\documentclass[fleqn,usenatbib]{mnras}

\usepackage{newtxtext,newtxmath}

\usepackage[T1]{fontenc}
\usepackage{ae,aecompl}

\usepackage{amsfonts,amssymb}
\usepackage{mathrsfs}
\usepackage{color}
\usepackage{enumerate}
\usepackage{booktabs}

\usepackage{float} 

\usepackage{xcolor}

\usepackage{needspace}
\usepackage{dcolumn}
\usepackage{bm}
\usepackage{graphicx}	
\usepackage{amsmath}	
\usepackage{amssymb}	
\usepackage{rotating, graphicx}
\usepackage{adjustbox}

\newcounter{qcounter}
\newcommand{\ra}[1]{\renewcommand{\arraystretch}{#1}}

\DeclareMathOperator\arcsinh{arcsinh}

\title[Galaxy rotation curves using a non-parametric regression method]{Galaxy rotation curves using a non-parametric regression method: 
core, cusp and fuzzy scalar field dark matter models}

\author[L. M. Fern\'andez-Hern\'andez et al.]{
Lizbeth~M. Fern\'andez-Hern\'andez,$^{1}$\thanks{E-mail: lfernandez@fis.cinvestav.mx}
Ariadna Montiel,$^{2}$\thanks{E-mail: amontiel@icf.unam.mx}
Mario A. Rodr\'{\i}guez-Meza$^{1}$\thanks{E-mail: marioalberto.rodriguez@inin.gob.mx}
\\
$^{1}$Departamento de F\'{\i}sica, Instituto Nacional de
       Investigaciones Nucleares, AP 18-1027, Ciudad de M\'{e}xico
       11801, M\'{e}xico\\
$^{2}$Instituto de Ciencias F\'isicas, Universidad Nacional
Aut\'onoma de M\'exico,  62210, Cuernavaca, Morelos, M\'{e}xico
}

\date{Accepted XXX. Received YYY; in original form ZZZ}

\pubyear{2018}

\begin{document}
\label{firstpage}
\pagerange{\pageref{firstpage}--\pageref{lastpage}}
\maketitle

\begin{abstract}
{We present a non-parametric reconstruction of the rotation curves (RC) for 88 spiral 
galaxies under the LOESS+SIMEX technique. 
In order to compare methods we also perform the parametric approach assuming core and cuspy 
dark matter (DM) profiles: PISO, NFW, 
Burkert, Spano, the soliton and two fuzzy soliton$+$NFW.} As result of this two approaches, 
a comparison of the RC obtained is carried out by computing the distance between central curves 
and  the distance between 1$\sigma$ error bands. {Furthermore, we perform a model selection according 
to two statistical criteria, the BIC and the value of $\chi^2_{red}$.}
We work with two groups. The first one is a comparison between 
PISO, NFW, Spano and Burkert showing that Spano is the most 
favored model satisfying our selection criteria. For the second group we select 
soliton, NFW and Fuzzy models, resulting the soliton as the best model. Moreover 
according to the statistical tools and non-parametric reconstruction we are able 
to classify galaxies as core or cusp. Finally, using an MCMC method, we compute
for each of the DM models the characteristic surface density, $\mu_{DM}=\rho_s r_s$, 
and the mass within 300 pc. We found that there is a common mass for spiral 
galaxies of the order of $10^7$ M$_\odot$, which is in agreement with results 
for dSph Milky Way satellites, independent of the model. This result is also consistent with 
our finding that there is a constant characteristic volume density of haloes. 
Finally, we also find that $\mu_{DM}$ is not constant, which is in 
tension with previous literature.

\end{abstract}




\section{\label{sec:level1}Introduction}
{Since the observation of the rotation curves in spiral galaxies (\cite{Rubin2001}) 
one of the main problems in galactic dynamics is how to determine the shape of the dark matter halo.
Currently, a variety of dark matter density profiles can be found in the literature, 
e.g.\cite{piso,NFW,Burkert,Spano,Einasto,Aspeitia-Rodriguez,Schive1,2018IJMPD..2750031H}).

{When used in $N$-body simulations, some of these models are able to reproduce the large scale structure of 
the universe. Nevertheless, at smaller scales, some problems arise such as the DM content in dwarf galaxies or the number of satellite galaxies, 
these problems are expected to be strongly related to the baryonic contribution to
the dynamics of structure formation.} 
On the other hand, there is the so called cusp-core problem consisting in the observation that in some galaxies the center of the dark 
matter halo features a core behavior $ \rho_ {DM} \sim \textrm {constant} $, which is in disagreement with the cuspy behavior predicted by the 
simulations, $\rho_ {DM} \sim r ^ {- 1}$ (for review see: \cite{2010AdAst2010E...5D,2014Natur.506..171P}).  
Furthermore, by cosmological simulations \cite{Schive2}, a density profile with inner soliton-like profile was derived,
called Fuzzy, wave or ultra-light axion with an asymptotic NFW declination in the outer points, see 
\cite{2018MNRAS.475.1447B} and references therein for a review.

It would be a step forward toward to the solution of the cusp-core problem, if we had data of galaxy rotation curves with higher spatial resolution at the center of the system. This later was one of the main concerns of the Spitzer Photometry and Accurate Rotation 
Curves (SPARC) database team \citep{Lelli:2016zqa}. They probed the inner regions at high spatial resolutions, and classify galaxies using a 
quality flag $Q$: 1 = high, 2 = medium and 3 = low. With more accurate data we may be able to fit core models and on the contrary, 
with low resolution data the cusp models, like the NFW \citep{2002A&A...385..816D,2015AJ....149..180O}.

On the other hand, in the standard approach one selects a particular density profile or model of rotation curves, with a specific functional form between dependent and independent 
variables and then constrain the parameters of the chosen theory employing bayesian methods together observational data of galaxies 
\citep{2018MNRAS.475.1447B,Aspeitia-Rodriguez,2019MNRAS.482.5106L}.
The main problem of this method is the susceptibility to bias if the data is not well-represented by the assumed parametric model. Since the goal is to determine the best-fit of the model parameters, this approach is called parametric. 

In contrast, the goal of a \textit{non-parametric} approach is to infer a global trend directly from data to assess sensitivity about the 
assumptions made in an specific parametric model and try to distinguish among competing cosmological models, for instance. Besides, it is 
also useful if a well justified functional form for the object of interest is not available or, even with a reasonable parametric model but there is 
a lack of data to infer other details of the theory. Specifically, in this kind of scenarios a correlation between each data point is assumed 
but prior information about the functional form of the observable is not required. So far it has been possible to gain useful information about the 
dark energy cosmic evolution directly from observations by using this kind of non-parametric methods, see for instance 
\cite{Huterer2003,Espana05,Bonvin06,Shafieloo06,Bogdanos09, Holsclaw2010,AlbertoVazquez2012,Montiel:2014fpa}.

In this work, the approach LOESS+SIMEX presented in \cite{Montiel:2014fpa}, which is a simple but powerful non-parametric regression technique 
that takes into account the observational errors, is employed. LOESS (Locally Weighted Scatterplot Smoothing) method originally was introduced in \cite{Cleveland79} and further developed in \cite{Cleveland88}. It recovers the global 
trend of data by using a weighted least squares and fitting a low-degree polynomial to a subset of the data, giving more weight to points near 
the point whose response is being estimated and less weight to points further away. Because LOESS method does not take into account the 
measurement errors while reconstructing the response parameter,  the effect of observational errors should be captured with another statistical 
technique: the SIMEX (Simulation and Extrapolation) method \citep{Cook94,Cook95}. These methods were first used together in cosmology 
in \cite{Montiel:2014fpa} with significant success. Later on, this technique was applied for reconstructing other cosmological 
quantities in \cite{Rani:2015lia,Rana:2015feb,Escamilla-Rivera:2015odt} providing also good results. Here we adapt the method to 
reconstruct the rotation curves of galaxies without assuming any prior or DM model.

We investigate the influence of the galactic DM halo on the rotation curves applying the LOESS+SIMEX method, 
for which the galaxy rotation curves from the SPARC database \citep{Lelli:2016zqa} are used, in particular a subsample
whose main feature is that the contribution of dark matter is greater than 50$\%$ such that the structure of the rotation curve comes 
mostly from dark matter and not from baryonic information (gas and stars). Another important characteristic of the subsample is that we 
only consider those galaxies with a quality factor of $Q=1$ or $2$, according to   \cite{Lelli:2016zqa}.

Furthermore, it is presented a comparison of the $\chi^2$ fitted velocity rotation curves with the non-parametric reconstruction method 
through the distance between central curves ($B_{DIST}$) and the distance between 1$\sigma$ errors bands ($B_{D1\sigma}$). 
Besides, in our analysis we use statistics tools such as $P$-value, the Bayesian information criterion (BIC), the Akaike information criterion 
(AIC) and minimization of the $\chi^2$  in order to perform a model selection according to the best values. In particular we work with 
two groups of models. The first group is aimed to contrast Core and Cusp models (Piso, NFW, Spano and Burkert) and the second 
one is aimed to compare Core, Cuspy, and Fuzzy models (Schive, NFW, 
and two hybrid Schive-NFW). Our criteria for model selection are: the best BIC value; the lowest $B_{DIST}$ value and the lowest $B_{D1\sigma}$ value. 
Additionally, from the statistical tools mentioned above and from the non-parametric reconstruction, a possible classification
of galaxies as core or cuspy is investigated.

Finally, the scaling relationship of the DM haloes and the luminosity of the galaxies 
using a bayesian Markov Chain-Monte Carlo (MCMC) method are analyzed. 
In particular we study the characteristic volume density $\rho_s$, the scale length $r_s$, the characteristic central surface density $\rho_s r_s$ 
and the central DM halo mass within 300 pc.
We show that the scaling relations help us to understand better the core-cusp problem.

The structure of this work is as follows: in Sec. \ref{GC} we review the main properties of the galaxy rotation curves and  
the selection criteria of galaxies from SPARC we have worked with; in Sec. \ref{Nonparametric} we briefly summarize 
the LOESS and SIMEX techniques; in Sec. \ref{analysis}, we present the statistical methodology to perform the model selection; 
in Sec. \ref{Results} we report the results; 
in Sec \ref{Discussion} the results are discussed
and finally we end up with the conclusions in Sec. \ref{Conclusions}.

\section{Galaxy rotation curves} \label{GC}
Let us start by pointing out that we shall consider rotation curves of galaxies within the weak gravitational field limit in order to investigate 
DM parameters. 

The rotation curve $V(r)$ is obtained from the absolute value of the effective gravitational force as:
\begin{eqnarray}
V^2(r)&=&r\left\vert \frac{d\Phi(r)}{dr}\right\vert\nonumber\\
&=&\sqrt{G\frac{ M_{\rm DM}(r)}{r} + \Upsilon_{\rm Disk}V_{\rm Disk}^2 + 
V_{\rm Gas}^2},  \label{rotvel}
\end{eqnarray}
where $\Phi(r)$ is the gravitational potential, $M_{\rm DM}(r)$ is the halo mass  of the DM distribution, $V_{\rm Disk}$ and $V_{\rm Gas}$ are 
the contributions of the stellar disk and gas disk velocities and $\Upsilon_{\rm Disk}$ is the stellar mass-to-light ratio of disk. 
In $V_{\rm Gas}$ is already included  the mass-to-light ratio $\Upsilon_{\rm Gas}$.

The models we select to fit each rotation curve can be reviewed in the appendix \ref{Ap}; there the explicit form of the density profile and 
the velocity function are included for each case considered in this work. Specifically PISO, NFW, Spano, Burkert  and the scalar field DM (SFDM) models.

Here we discuss only the less known SFDM cases (for a recent review see \cite{2018MNRAS.475.1447B}).
As was pointed out in the Introduction, large scale numerical simulations have provided two profiles. The first one is the NFW using 
CDM model, and, the second one, using SFDM, an inner soliton-like profile, that we called Schive Wave DM model (WDM), with an 
asymptotic NFW decline in the outer points, called Schive Wave+NFW DM (SNFW), \cite{Schive2}.
At cosmological scales these two models are equivalent. However, CDM presents some challenges at the galactic level that can 
be solved using the Wave DM model because of its core nature. Moreover, \cite{Schive2} derived an empirical density profile that 
described DM haloes with a soliton-like core embedded in an NFW density profile dominant at large radii. 
Such profile can be approximated by
\begin{eqnarray}
\label{SNFW}
\rho_{\rm SNFW}(r)&=& \Theta(r_\epsilon - r) \rho_{\rm WDM}(r)\nonumber\\
 &+& \Theta(r-r_\epsilon) \rho_{\rm NFW}(r), 
\end{eqnarray}
where $\Theta$ is the step function, $r_\epsilon$ the transition radius where density changes from the Schive profile
\begin{equation}
\rho_{\rm WDM}(r)=\frac{\rho_w}{(1+a_w(r/r_w)^2)^{8}}, \label{eq:WaveDM}
\end{equation}
to the NFW profile Eq. (\ref{NFW}). 
Here $a_w=0.091$, $\rho_w$ is the core density and $r_w$ is its scale radius.
Once $\rho_w$ and $r_w$ are fitted, the boson mass $m_w$ is found using $\rho_w= 1.9 (m_w/10^{-23}$ eV$)^{-2}(r_w/$kpc$)^{-4}$ $M_\odot$pc$^{-3}$.
The total density (\ref{SNFW}) has five parameters ($\rho_w$, $r_w$, $r_\epsilon$, $\rho_n$, $r_n$), the last two parameters are defined in Eq. (\ref{NFW}).
The last issue is related to the matching conditions at the transition radius $r_\epsilon$ in Eq. (\ref{SNFW}). There are two possibilities. Either use only 
the continuity in the density function, or apply a continuity condition to the density function and its derivative. The former case yields four parameters 
to fit, while the latter case leaves us with three parameters to fit. The last SFDM model to consider is where the core part dominates the galaxy. In 
such a case Eq. (\ref{SNFW}) reduces to only the first term, $\rho_{WDM}(r)$, with two parameters to fit. The complete expressions for all models are 
given in the Appendix \ref{Ap}.

Regarding the galaxy selection, we are considering the baryonic contribution for each rotation curve 
of SPARC (Disk and Gas), 
the same values of
$\Upsilon_{\rm Disk}=0.5$ $M_\odot/L_\odot$ at 3.6 $\mu$m
and 
$\Upsilon_{\rm Gas}=1.33$ $M_\odot/L_\odot$
\citep{schombert_mcgaugh_2014,Lelli:2016zqa,2017ApJ...836..152L}. 

Since for the fitting we shall assume a considerable contribution of dark matter, 
we choose galaxies from the SPARC catalogue \citep{Lelli:2016zqa} for which the percentage of dark matter contribution lies between 
$90> f_{ \rm DM} >60 $, where $f_{\rm DM}/100=M_{\rm DM}(r_{\rm max})/(M_{\rm Bar}(r_{\rm max}) + M_{\rm DM}(r_{\rm max}))$, $M_{\rm DM}$ 
corresponds to the dark matter contribution for each model, and $M_{\rm Bar}$ is the total baryonic mass, both evaluated at the outer 
value of the observed radius of the rotation curve $r_{\rm max}$.

Within our sample, we classify galaxies according to the dominant baryonic component, consequently, those dominated by gas are of the 
order of $83.3\%$, the ones dominated by the stellar disk, $14.6\%$ and $2.08\%$ where the gas and the stellar disk contribute with 
the $50\%$ from the total photometric information (NGC 0300, for example). 
However, according to the morphology of the galaxies and the Hubble classification, we have a galaxy selection where $\sim 33.3\%$ 
are Sm, $27.08\%$ Im, $20.83\%$ Sd, Scd = Sdm = $6.25\%$, $4.17\%$ Sc and $2.08\%$ BCD \citep{Lelli:2016zqa}. 

In the following section we briefly review the  non-parametric curve fitting method employed in our sample.

\section{Non-parametric reconstruction method}\label{Nonparametric}

\subsection{\textit{Basis of LOESS}\label{basics}}

LOESS is a non-parametric method in the sense that the fitting is performed without having to specify in advance the relationship between 
the dependent and independent variables. At each point of the data set a low-degree polynomial is fit to a subset of this data, the polynomial
is obtained by weighted least squares, giving more weight to points near to the point whose response is being estimated and less weight to points 
further away. The value of the regression function for the point is then obtained by evaluating the local polynomial using the independent variable 
value for that data point. The LOESS fit is complete after regression function values have been computed for each of the \textit{n} observational 
measurements. This whole process offers the possibility to get the full view of the global trend of the data, which is the original objective of the procedure. 
To obtain the graphical account of the relationship between dependent and independent variables, it suffices to join the reconstructed points with a line.

Here we work with observational data of rotation curves of galaxies, so the radius $R$, measured in 
kpc, will be our independent variable while the velocity $V$ of the stars or gas particles, in units of km/s, corresponds to the dependent variable.

In the following the main elements of the method are briefly discussed. For further details we encourage the reader to see \cite{Montiel:2014fpa} 
and references therein.

\begin{list}{\roman{qcounter}.}{\usecounter{qcounter}}

\vspace{1mm}

\item \textit{Smoothing Parameter} 

\vspace{1mm}

The \textit{smoothing parameter}, $s$, also called \textit{span} determines how many data points should be used in each weighted least squares fit. 
The \textit{span} ranges between $0$ and $1$ and controls the flexibility of the \textit{LOESS} regression function. 
Large values of $s$ produce the smoothest functions that wiggle the least in response to fluctuations in the data but small values of $s$ produce 
more irregular reconstructed curves, because the intrinsic noise and dispersion of data is fully captured. Choice of the span, the degree of polynomial, 
and the weight function can all affect the trade-off between the bias and variance of the fitted curve, however, the value of the span has the most 
important effect as can be seen in Fig. \ref{fig:span}, in which the differences in the fitted regression curves are associated to the choice of span. In order to minimize this effect, as in \cite{Montiel:2014fpa}, the election of the optimal value of the \textit{span} $s$ is done by using the 
\textit{cross-validation} method which aims to minimize the mean squared error of the fit.  

\begin{figure*}
\begin{center}$
\begin{array}{ccc}
\includegraphics[width=3.25in]{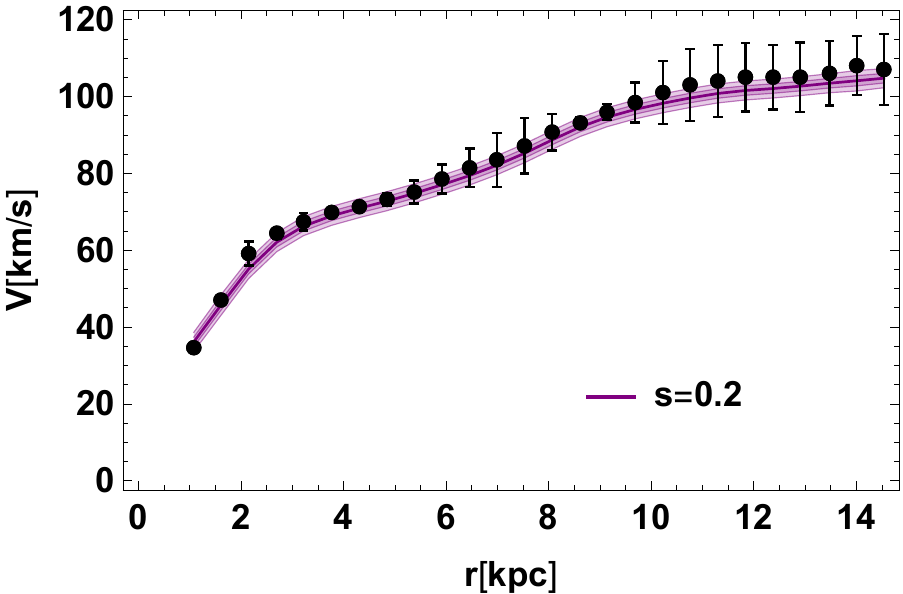}&
\includegraphics[width=3.25in]{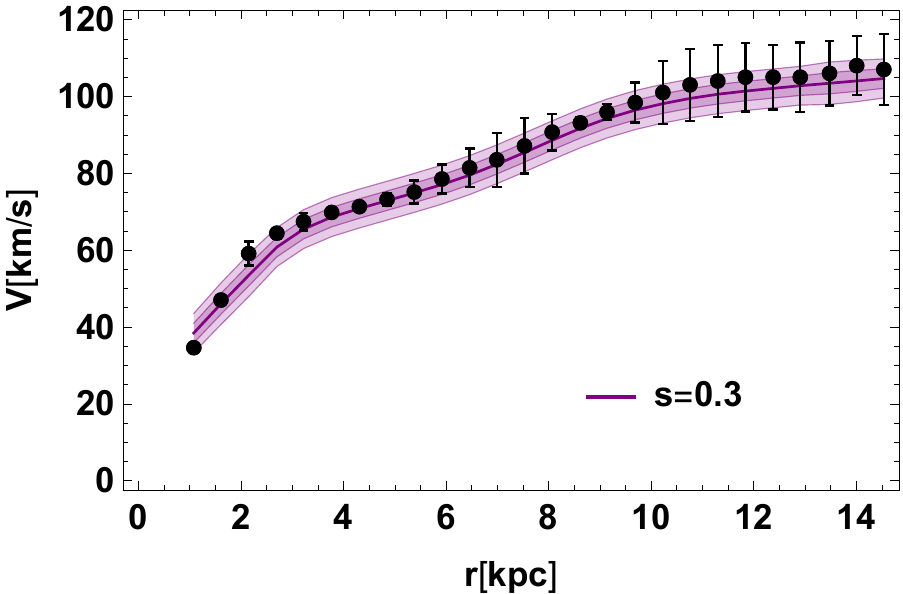}\\
\includegraphics[width=3.25in]{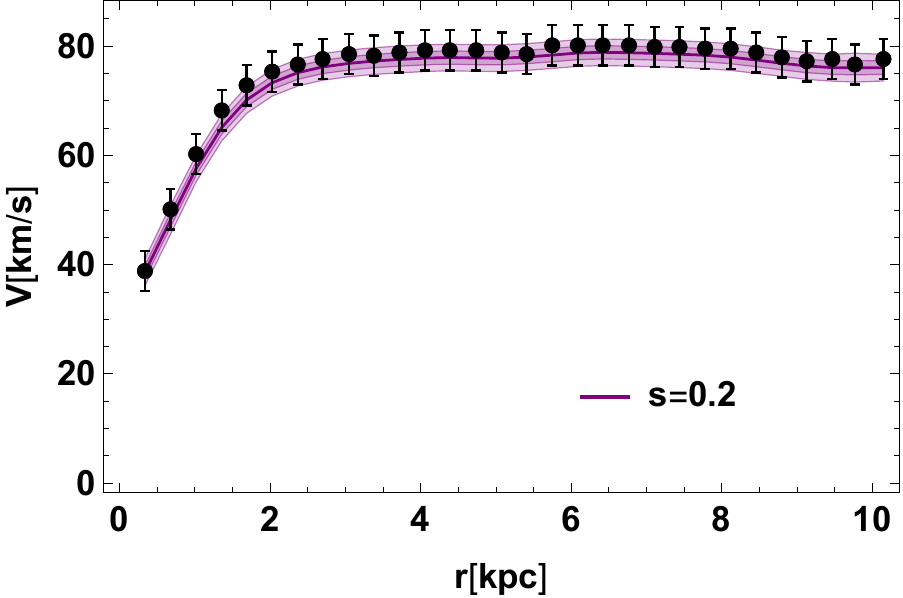}&
\includegraphics[width=3.25in]{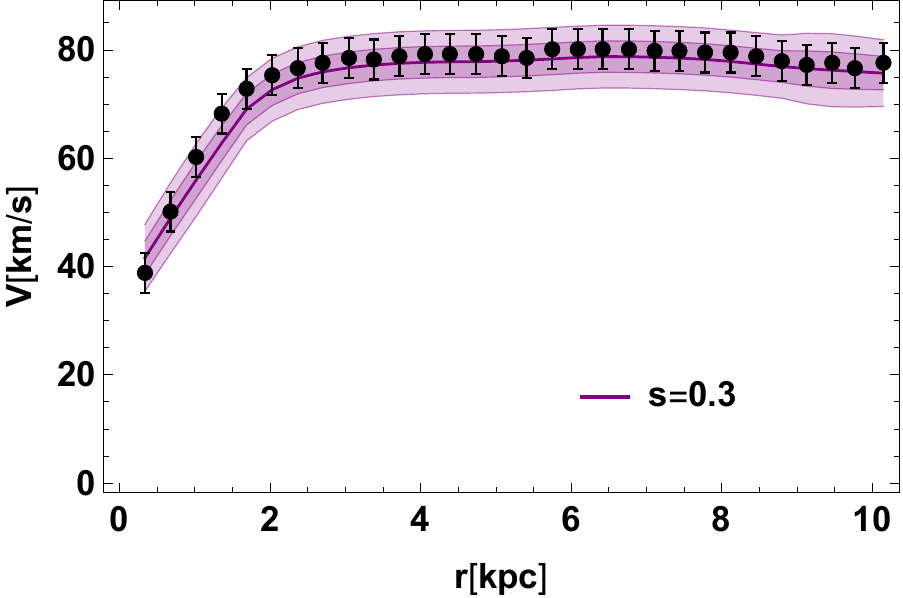} \\
\end{array}$
\end{center}
\caption{LOESS reconstruction for the NGC0247 (upper figures) and UGC08490 (lower figures) galaxies, obtained by using different values for the span, 
$s$, showing the impact of the choices of $s$ in the reconstructed curve (regardless the observational errors). }
\label{fig:span}
\end{figure*}

\vspace{2mm}

\item \textit{Weight Function}

\vspace{1mm}

As mentioned above, the weight function assigns a large weight to the data points nearest the point of estimation and a low weight to the data 
points that are furthest away. 

The usual weight function employed for LOESS is the tri-cube weight function,
\begin{equation}
w(x) = \begin{cases}
  \left(1-\vert x \vert^3 \right)^3 & \text{for $\vert x \vert < 1$} \\
  0 & \text{for $\vert x \vert \geq 1$}
\end{cases}\;.
\label{eq:funw}
\end{equation}
However, any other weight function that satisfies the properties listed in \cite{Cleveland79} is also useful. Such properties are a positive weight function and 
even a treatment of points left and right of each point $x_i$. In addition, it is desirable that $w(x)$ decreases smoothly to 0 as $x$ goes from 0 to 1. Among 
the weight functions that decrease to 0, normally the tricube function is chosen since it provides an adequate smoothing in almost all cases. 
Here, as in \cite{Montiel:2014fpa}, this weight function is used.

\vspace{2mm}
\item \textit{Degree of Local Polynomials}
\vspace{1mm}

The idea behind LOESS is to fit each subset of data using polynomials of a low degree after the weight function is applied. Most of the polynomials are 
of first or second degree and although higher-degree polynomials would work, it could cause overfit of the data in each subset, besides of higher 
computational cost without any significant improvement in the result.

\vspace{1mm}
\item \textit{Confidence regions}
\vspace{1mm}

In order to construct the confidence regions of the nonparametric regression we perform an analogy with the standard least squares regression by 
using the local polynomial estimate, $\hat{y}_i$, that results from the locally weighted least-squares regression of $y$ on the $x$ 
values in each chosen window $i$. Thus, by assuming normally distributed errors, the $68\%$-percent confidence interval and the 
$95\%$-percent confidence interval of the regression function can be constructed.  

It is worth mentioning that this procedure for constructing a confidence region is not completely accurate, due to the bias in $ \hat{y} _i $ 
as an estimate of the regression function \cite{Fox20}. The value of the smoothing parameter changes significantly 
the function estimate, and consequently the confidence bands. In this sense, such a bias can produce an overestimation of the error variance, 
thus making the confidence interval too wide. On the contrary, an inadequate small span (i.e. with insufficient data within the window fit) 
yields large variance. As we have pointed out, this can be alleviated by using the cross-validation procedure in order to choose the optimal value of $ s $ 
as is suggested in \cite{Montiel:2014fpa}. 
\end{list}

\subsection{Basis of \textit{SIMEX}\label{simex}}

SIMEX method is based on two resampling approaches. At the first stage, the simulation, additional errors 
are introduced in the data by using a controlling parameter $\lambda$. Next, a regression analysis is used on this new dataset to trace the effect of the measured error in the original dataset \cite{Cook94, Cook95}. More specifically, SIMEX work as follows: 

\begin{list}{\roman{qcounter}.}{\usecounter{qcounter}}
\item[{\bf 1.- }]  The \textit{simulation step}. The method starts by introducing additional errors on each observation $y_i$ in the data set, with $i=1,...n$ and $n$ 
the number of data points, following the rule 
\begin{equation}
\eta_i(\lambda)=y_i+\sqrt{\lambda} \sigma_i, ~~~~ \lambda > 0
\label{Eq:simex}
\end{equation}
where $\sigma_i$ is the measurement error variance associated to the observed data $y_i$ and the parameter $\lambda$ acts as the 
controlling parameter for the amount added of $\sigma_i$. We take, as in \cite{Montiel:2014fpa}, $\lambda = \lbrace 0.5, 0.6, 0.7,..., 2.0 \rbrace$ 
to simulate the \textit{new} datasets.
\\

\item[{\bf 2.- } ] The \textit{extrapolation step}. After introducing the variable $\lambda$, the final measurement error variance 
associated with the simulated data points, $\eta_i(\lambda)$, is $(1+\lambda) \sigma^{2}_{i}$. So, it is necessary to take $\lambda \rightarrow -1$, 
in order to return back to the original data without uncertainties and to trace the effect of the measured error in the original data. 
This is done via a regression analysis, using a quadratic polynomial, on the \textit{new} datasets. 

\end{list}

Up to this point, we have addressed briefly the features and free parameters of the \textit{LOESS+SIMEX} method, however many 
more details can be found in \cite{Fox20, Andersen, Handbook,Cosma,Larry,Carroll95,Cook94,Cook95,Carroll99,Montiel:2014fpa} and references 
therein.

\section{Parametric and non-parametric statistical methodology} \label{analysis}

We have selected 88 galaxies from SPARC with an important contribution of DM to the rotation curve in order to study, from the 
non-parametric and parametric point of views, the seven theoretical models that we have pointed out before. This section reviews how 
we perform the estimation of the best fit parameters as well as the statistical tools used to perform the model selection.

Regarding the parametric approach, which focuses on estimating the model parameters, we use a fitting code to minimize the 
reduced $\chi^2_{red}$
function,
\begin{equation}
\label{chired}
\chi^2_{red}= \frac{1}{N_d -N_p} \sum_{i=1}^N{\frac{\left(V_{\rm obs} ({r_i}) - V_{\rm model} (r_i,\hat{\theta})\right)^2}{\delta V^2_{\rm obs} ({r_i})}}\, ,\\
\end{equation}
where $V_{\rm obs}$ and $\delta V_{\rm obs}$ are the observed galaxy velocity rotation curve and its uncertainty at the observed radial 
distance $r_i$, $\hat{\theta}$ are the $N_p$ fitting parameters of every model studied and
$N_d$ is the total number of data. Remember that $\chi^2_{red} \sim 1$ is desirable for a good fit \citep{numrecip}.

We also compute the $P$-value ($\chi^2$ test) which is an indicator between the compatibility of the distribution data and 
the fitting model. $P > 0.95$ indicates that we have a ($>95\%$) chance of finding a result less close to the central 
data; while if $P <$ 0.05,  the data and the fitting model are incompatible and then we can reject the fitting 
model; in case that $P > 0.05$ we can not reject the model \citep{numrecip}. 

Since we are comparing models with a different number of parameters, we compute the AIC (Akaike information criterion)  
and the BIC (Bayesian information criterion) in order to perform the model selection. In these statistical methods there is a penalty, 
which is applied to compensate for the obligatory difference in likelihoods due to the different number of parameters \citep{2012msma.book}. 
That is, if we have a model $M_j$ with a $p_j$ parameters $\hat{\theta}_j$, then
\begin{equation}
\label{AIC}
AIC=-2l(\hat{\theta}_j) +2 p_j\, ,\\
\end{equation}
where $2l(\hat{\theta}_j)$ is the goodness-of-fit term and $2p_j$ is the penalty of number of parameters. 
On the other hand, the Bayesian information criterion is defined as
\begin{equation}
\label{BIC}
BIC=-2l(\hat{\theta}_j) +2 p_j \ln{N_d}\, ,\\
\end{equation}
where, as before, $N_d$ is the number of data points.

The goodness-of-fit definition used in both cases is 
\begin{equation}
\label{goodnes-of-fit}
l(\hat{\theta}_j)= \prod_{i}\frac{1}{\sqrt{2\pi\delta V^2_{\rm obs} ({r_i})}} \exp\left(-\chi^2_j\right)\, ,
\end{equation}
where $\chi^2_j$ and $\delta V^2_{\rm obs} ({r_i})$ describe the galaxy studied, Eq. (\ref{chired}), and are associated to 
the $j$-model \citep{Cousineau2015}.

AIC penalizes free parameters less strongly than does the BIC. BIC imposes a greater penalty for larger datasets 
while the AIC is independent of the sample size.

The galaxies we have analyzed are the ones that fulfill the quality condition (see \cite{2016PhRvL.117t1101M}). 
We used a Markov Chain-Monte Carlo (MCMC) method \citep{Gamerman:1997} implemented in {\it Mathematica} 
by one of the authors (MARM),\footnote{Public at https://github.com/rodriguezmeza/MathematicaMCMC-1.0.0} 
in order to constrain the free parameters of the DM models through a maximization of 
the likelihood function $\mathcal{L}(\mathbf{p})$ given by

\begin{equation}
	\mathcal{L}({\hat{\theta}_j}) = \frac{1}{(2 \pi)^{N_d/2}
	|{\bf C}|^{1/2}} \exp{\left ( - \frac{{\bf \Delta}^{T}
	{\bf C}^{-1} {\bf \Delta}}{2} \right )} ,
\label{eq:likelihood}
\end{equation}
where $\hat{\theta}_j$ is the vector of parameters, 
${\bf \Delta} = V_\mathrm{obs}(r_i) -
V_\mathrm{model}(r_i,\hat{\theta}_j)$ 
and $V_\mathrm{model}$ the derived total velocity for one of the seven models
computed in the same position where $V_\mathrm{obs}$ was
measured, and $\mathbf{C}$ is a diagonal matrix.

We sample, using the Metropolis Hastings algorithm, the parameter space from uniform prior ranges with two Markov
chains and tested the convergence of the fit with the Gelman-Rubin convergence
criterion ($\mathcal{R}-1 < 0.01$) \citep{Gelman-Rubin:1992}. The fitting
parameters and the $1$ $\sigma$ and $2$ $\sigma$ confidence levels (CL) are computed from the
Markov chains with 30\% of burn-in.
We have tested the MCMC code against the results obtained with the minimization method described above.
The Markov chain was analyzed using GetDist code that is included in the CosmoMC code \citep{Lewis:2002ah}.

On the other side, regarding the nonparametric approach, we would like to recall we are using a technique which focuses 
on the fitted curve such that the fitted points and their errors are estimated with respect to the whole curve rather than a 
particular estimate and then, the overall uncertainty is measured on the basis of how well the estimated curve fits the data.

Since LOESS+SIMEX reconstruction has been used in an informal graphical way to assess the relationship between variables in 
\cite{Montiel:2014fpa,Rani:2015lia,Escamilla-Rivera:2015odt,Rana:2015feb}, in order to check the validity of a specific theoretical model 
by comparing with the nonparametric regression curve, we compute the distance between the best $\chi^2$ fitted velocity curve and the 
velocity curve obtained by reconstruction and also the distance between the $1\sigma$ band from the best $\chi^2$ fitted curve and the 
$1\sigma$ band from LOESS+SIMEX one.

We define the distance (area) between curves as:
\begin{equation}
\label{eq:distances}
D= \frac{ \int\displaylimits_{r_{min}}^{r_{max}} \big{|}V_{\rm LOESS}({r}) - V_{\rm model}(r, \hat{\theta}) \big{|} d{r} } {A_{\rm DATA}}\, ,\
\end{equation}
where $V_{\rm LOESS}({r})$ is the reconstruction velocity rotation curve from the data distribution, $V_{\rm model}(r, \hat{\theta})$ is the 
velocity function of each model characterized with their parameters $\hat{\theta}$. Notice we are normalizing the area between curves 
with $A _{\rm DATA}$, which is the area enclosing the velocity rotation data, including the error bars.

For the distance (area) between 1$\sigma$ bands (model and LOESS+SIMEX) we take into account the following cases:  
\begin{itemize}
\label{areas}
\item If there is not overlap between bands at a given $R$ and the LOESS $1\sigma$ band is above the model $1\sigma$ band, there 
we compute its contribution to the distance as 
\begin{equation}
\label{eq:distancesBandsR}
\delta D=\frac{1}{A_{\rm DATA}} \int\displaylimits_{R}^{R+\delta R} \big{|}V_{\rm LOESS}^{L}({r}) - V_{\rm model}^{U}(r, \hat{\theta}) \big{|} d{r}\, ,
\end{equation}
where $V_{\rm LOESS}^{L}(r)$ is the lower curve of the LOESS $1\sigma$ band and $V_{\rm model}^{U}(r, \hat{\theta})$ is the upper curve 
of the model $1\sigma$ band.
When the model $1\sigma$ band is above the LOESS $1\sigma$ band, we use a similar expression.

\item If there is an overlap between the $1\sigma$ bands but 
one band is inside the other there $\delta D =0$. On the opposite case, if
$V_{\rm LOESS}^{U}(r) > V_{\rm model}^{U}(r, \hat{\theta})$
the contribution to the distance
is given by
\begin{equation}
\label{eq:distancesBandsROverlap}
\delta D= \frac{1}{A_{\rm DATA}} \int\displaylimits_{R}^{R+\delta R} \big{|}
V_{\rm model}^{U}(r, \hat{\theta}) - V_{\rm LOESS}^{L}({r}) \big{|} d{r}\, .
\end{equation}
If
$V_{\rm model}^{U}(r, \hat{\theta}) > V_{\rm LOESS}^{U}(r)$
the contribution to the distance
is given by a similar expression.
\end{itemize}

\section{RESULTS} \label{Results}

In Table \ref{tab:PNSB} we show the statistical results for PISO (1), NFW (2), Spano (3) and Burkert (4) models 
and for the galaxy selection of the SPARC catalog where DM is dominant.

\begin{figure*}
\begin{center}$
\begin{array}{ccc}
\includegraphics[width=3.6in]{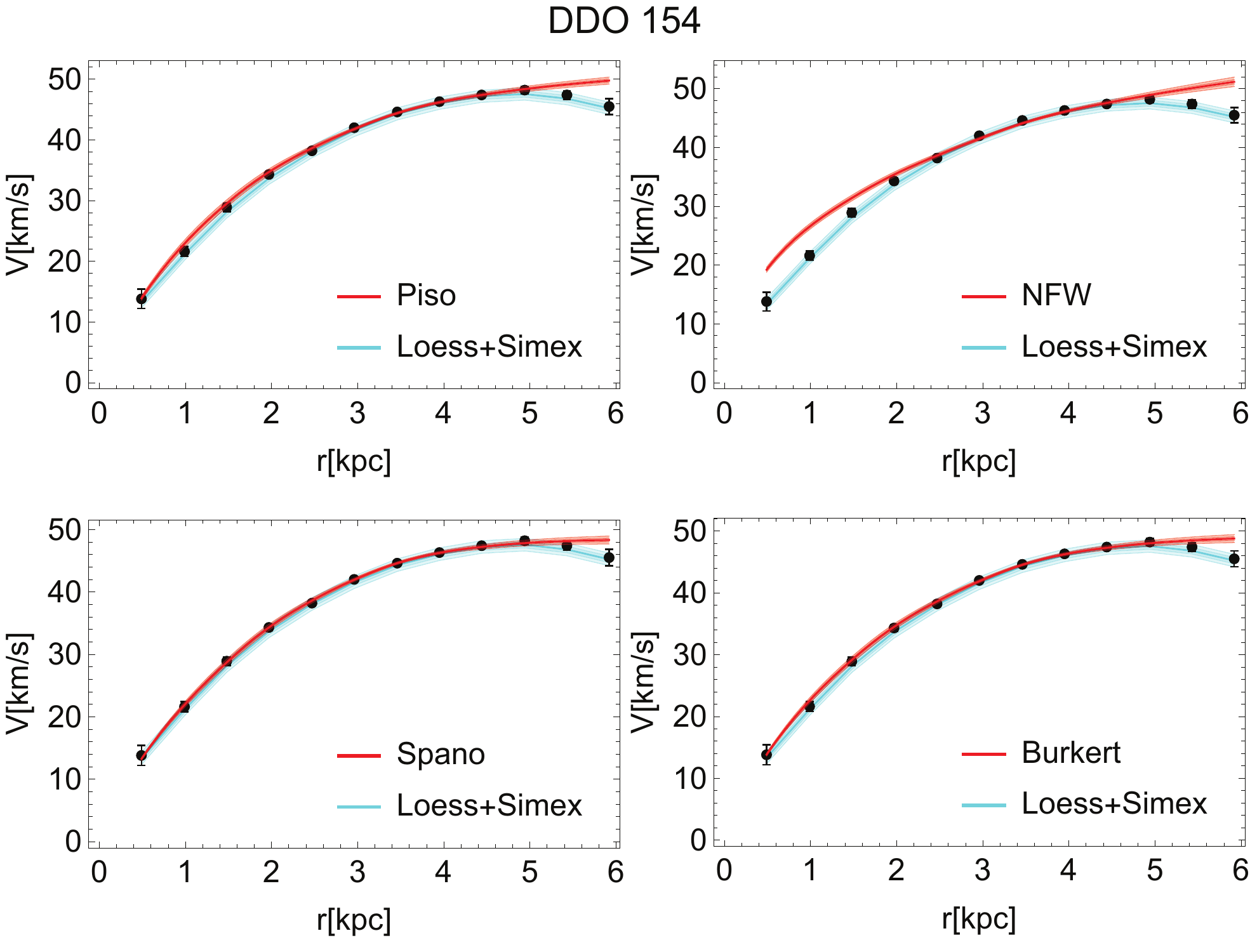}&
\includegraphics[width=3.6in]{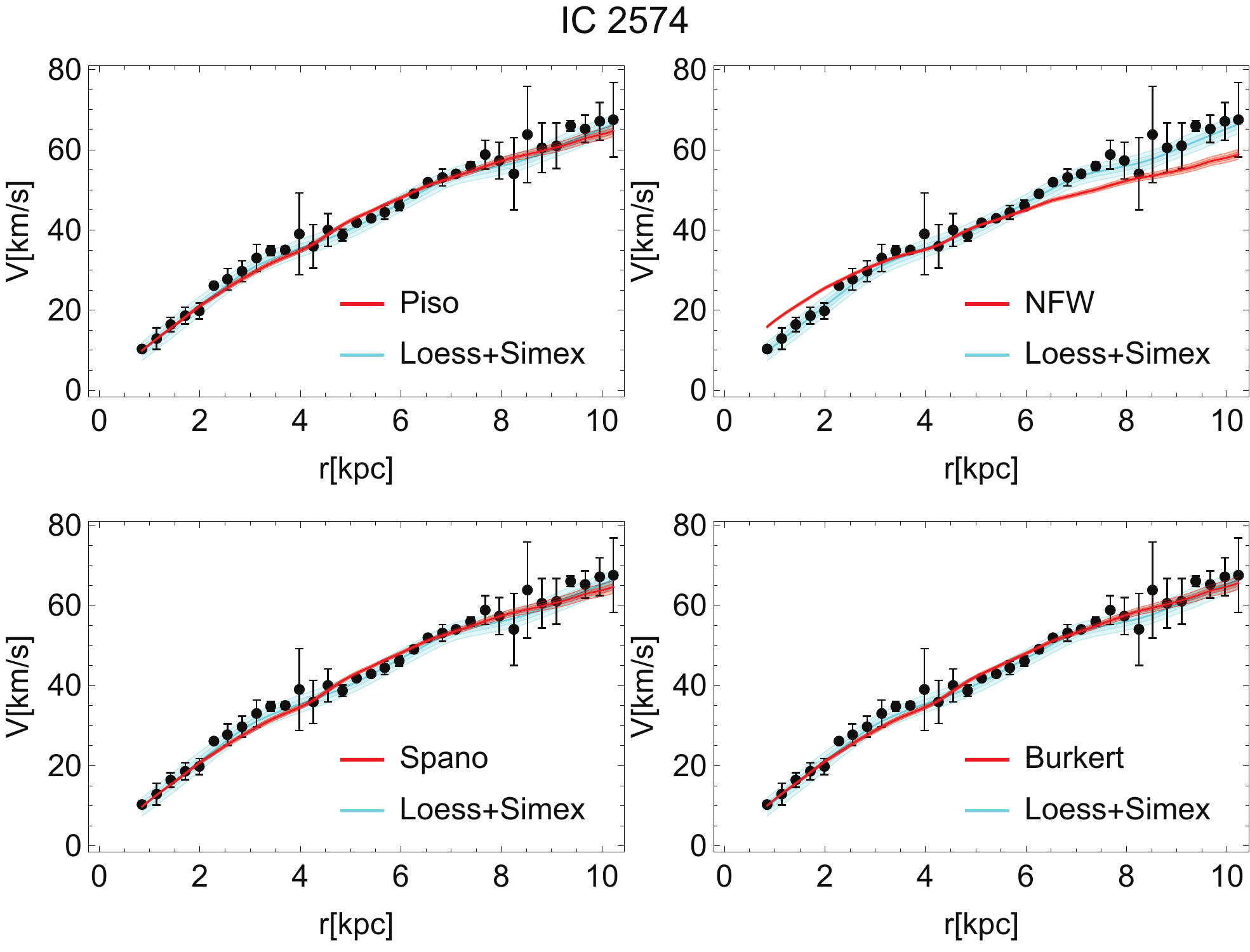}\\
\includegraphics[width=3.6in]{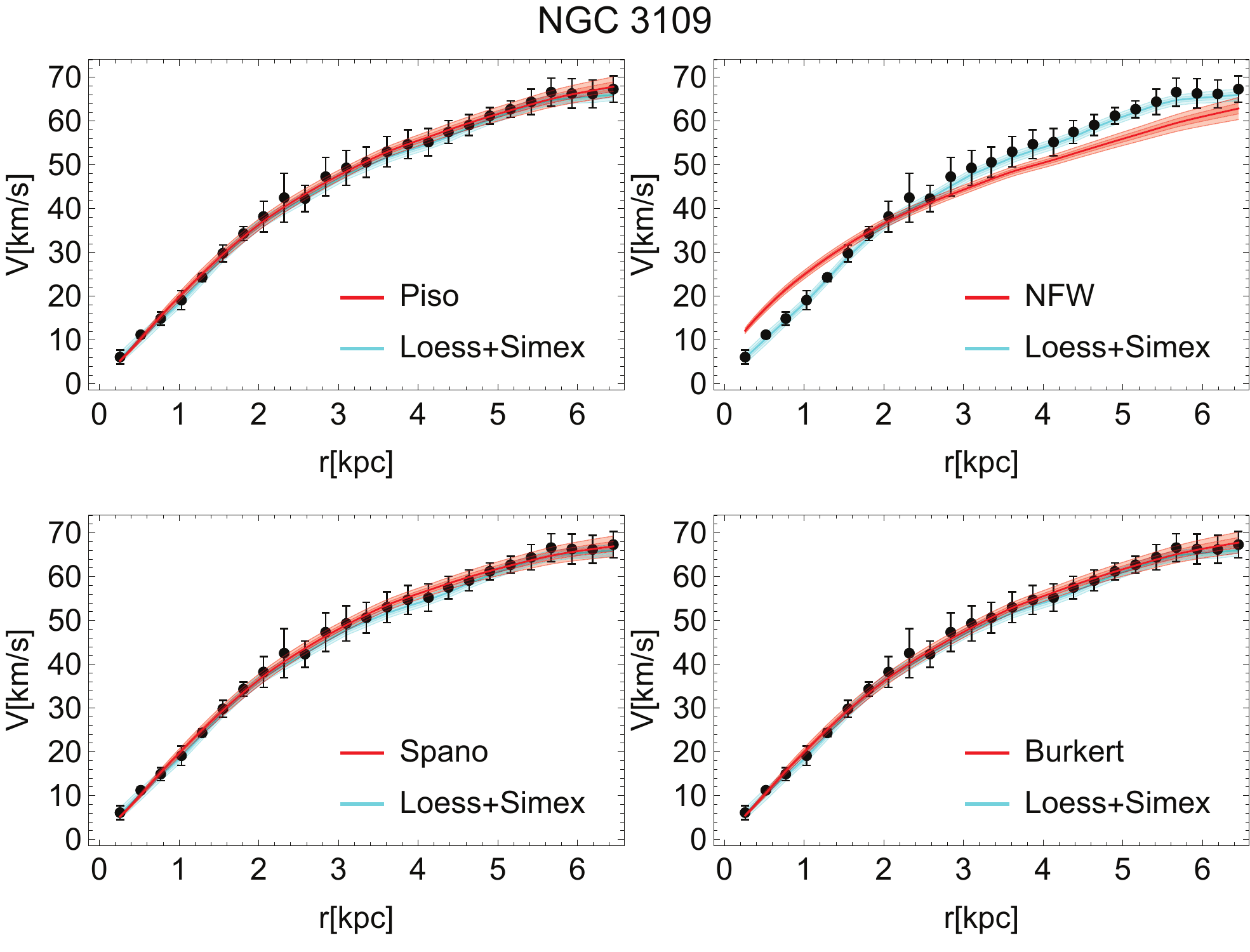}&
\includegraphics[width=3.6in]{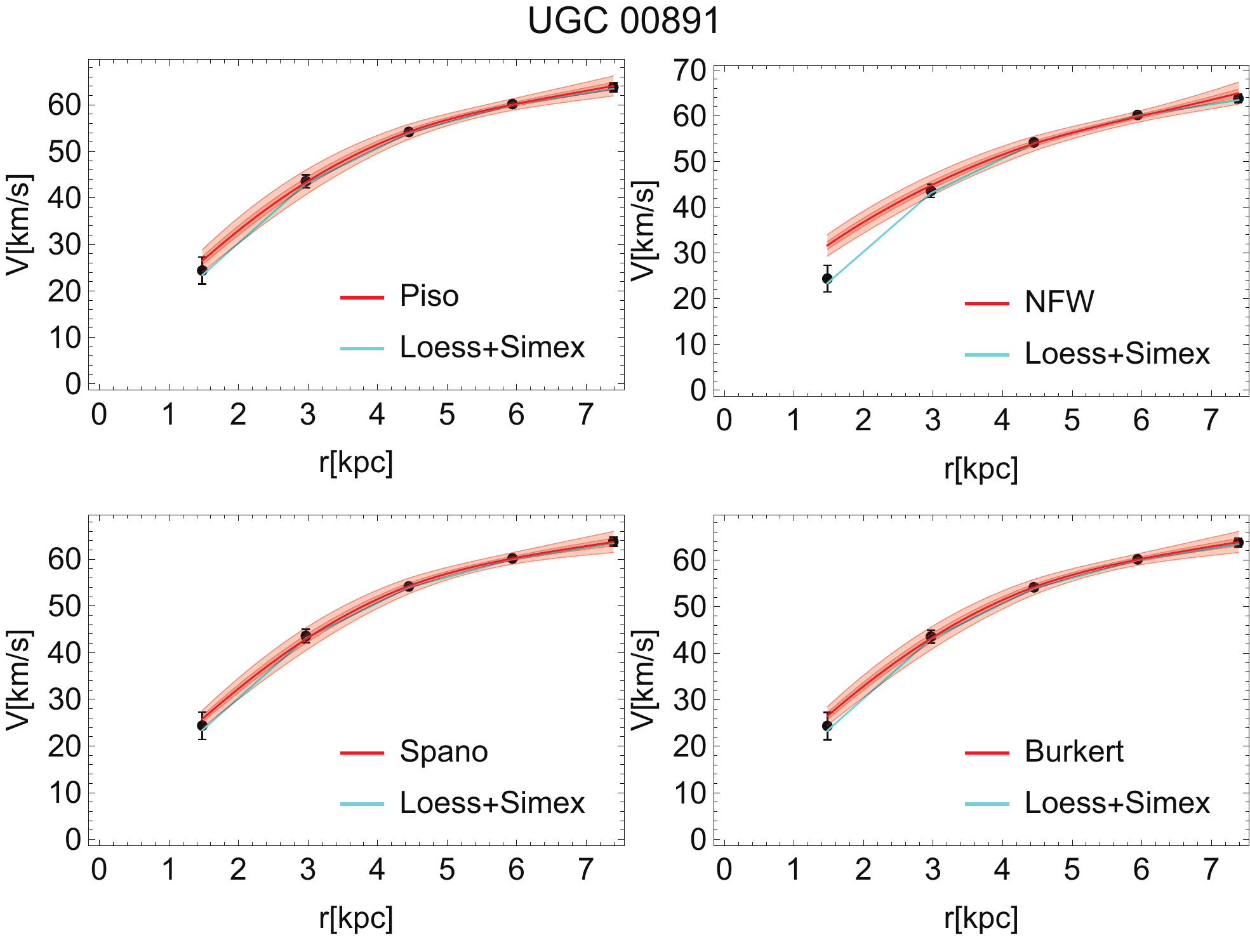}
\end{array}$
\end{center}
\caption{Galaxies where the distance between the best fit and 1$\sigma$ bands for NFW has a notable separation 
with the LOESS+SIMEX band, according with the results from Table \ref{tab:PNSB}. In the figure we show the reconstruction 
LOESS+SIMEX 1 and 2$\sigma$ bands in cyan color, the best fit for the four studied models in the second group including 
1 and 2$\sigma$ bands in red.}
\label{fig:Models1}
\end{figure*}

\begin{figure*}
\begin{center}$
\begin{array}{ccc}
\includegraphics[width=3.6in]{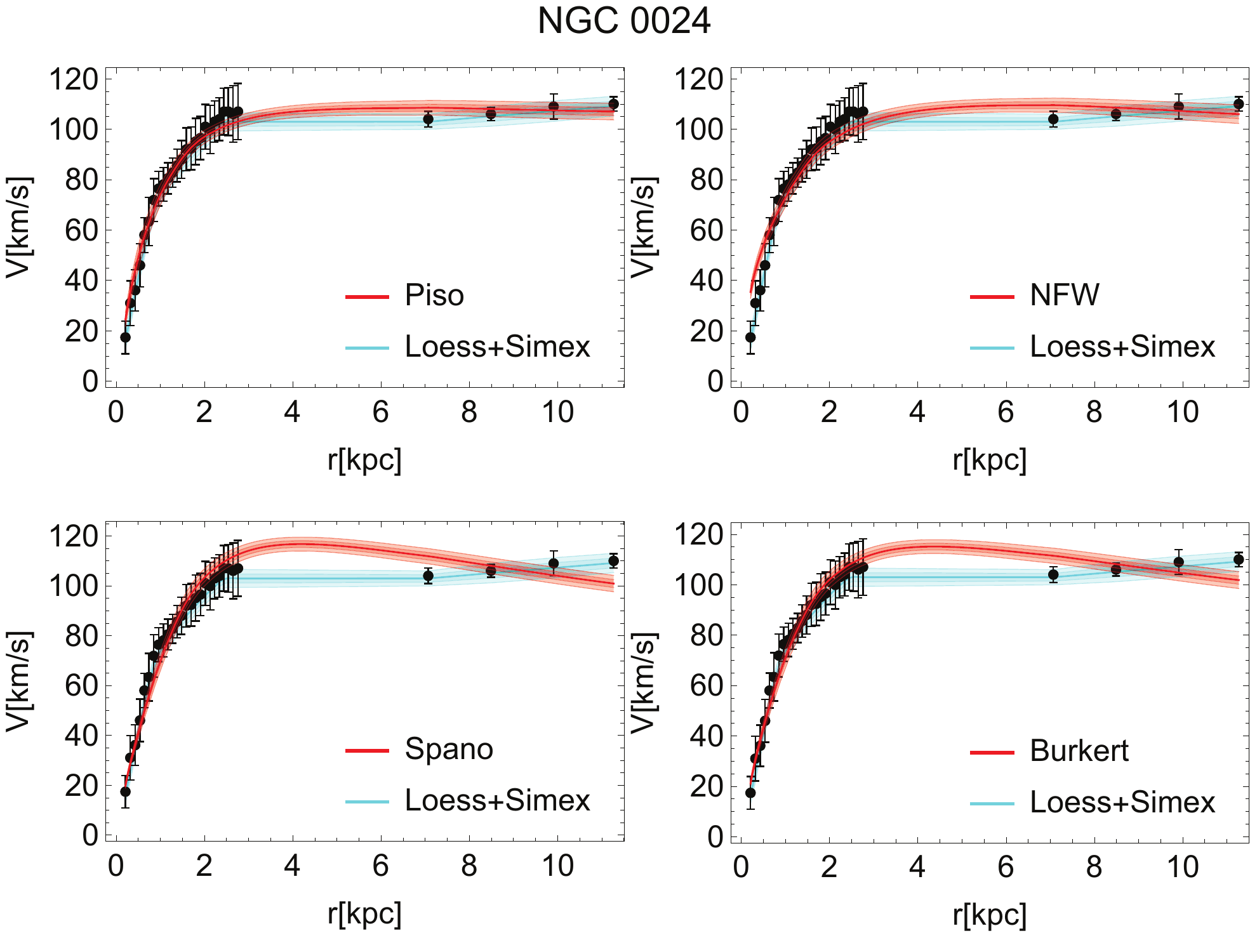}&
\includegraphics[width=3.6in]{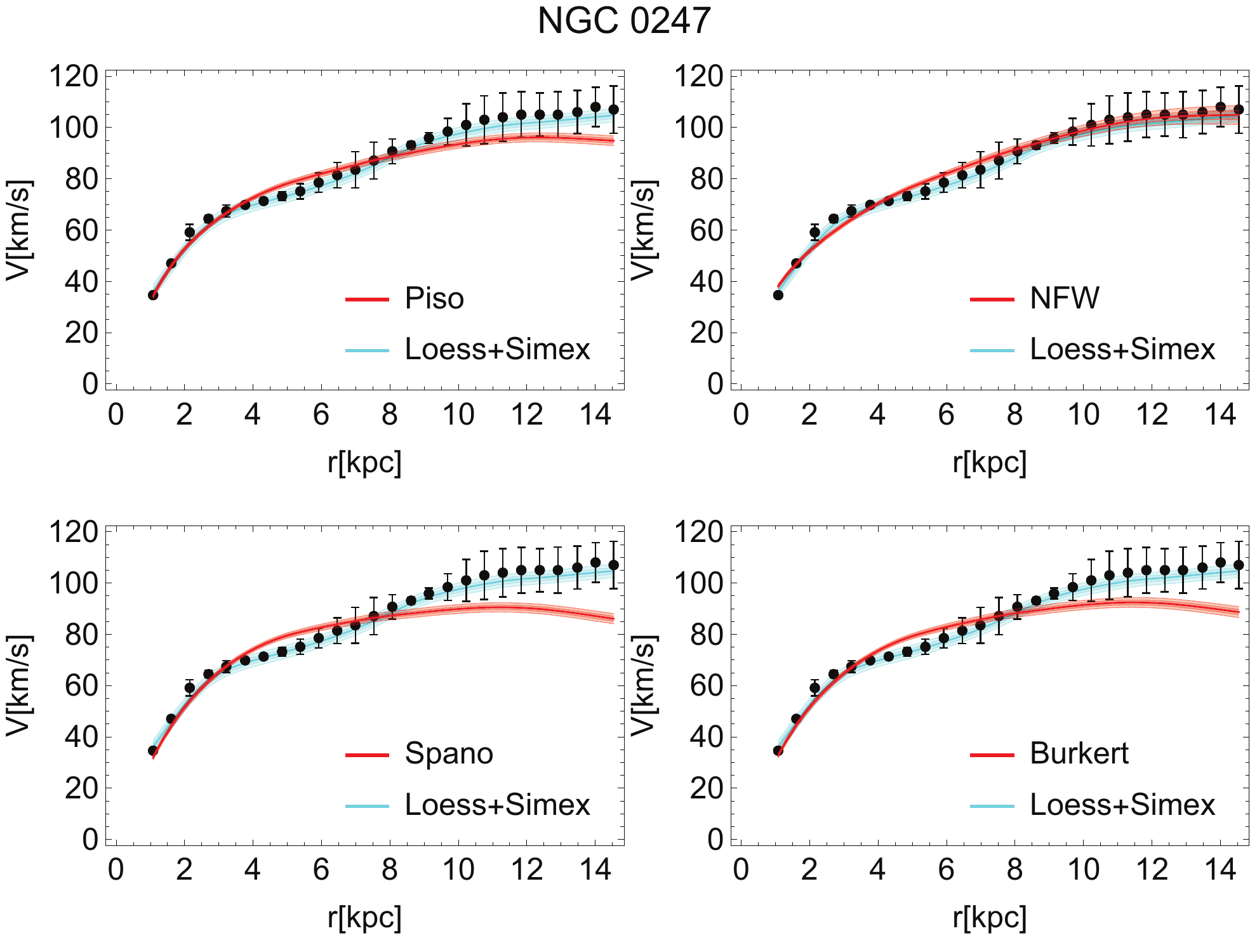}\\
\includegraphics[width=3.6in]{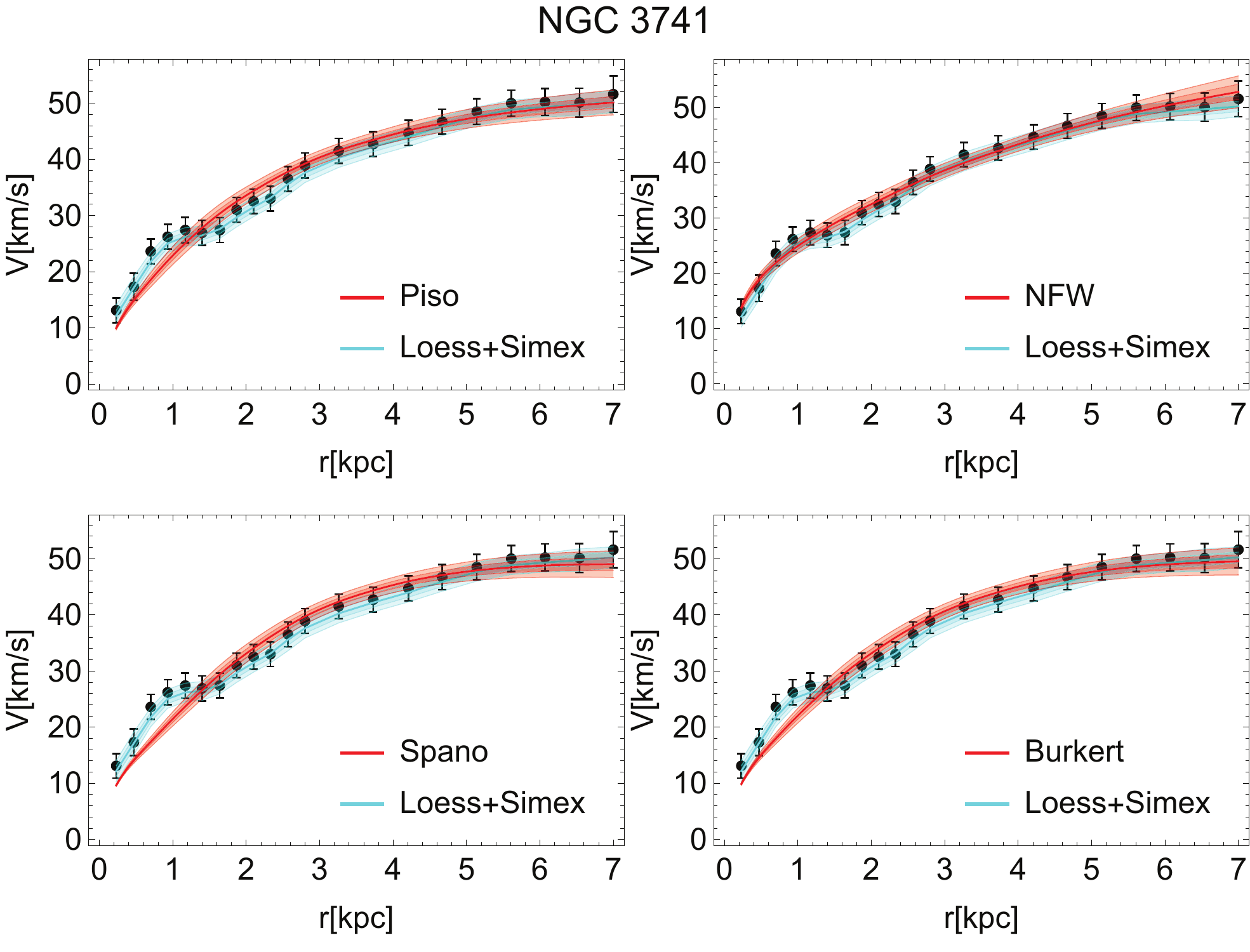}&
\includegraphics[width=3.6in]{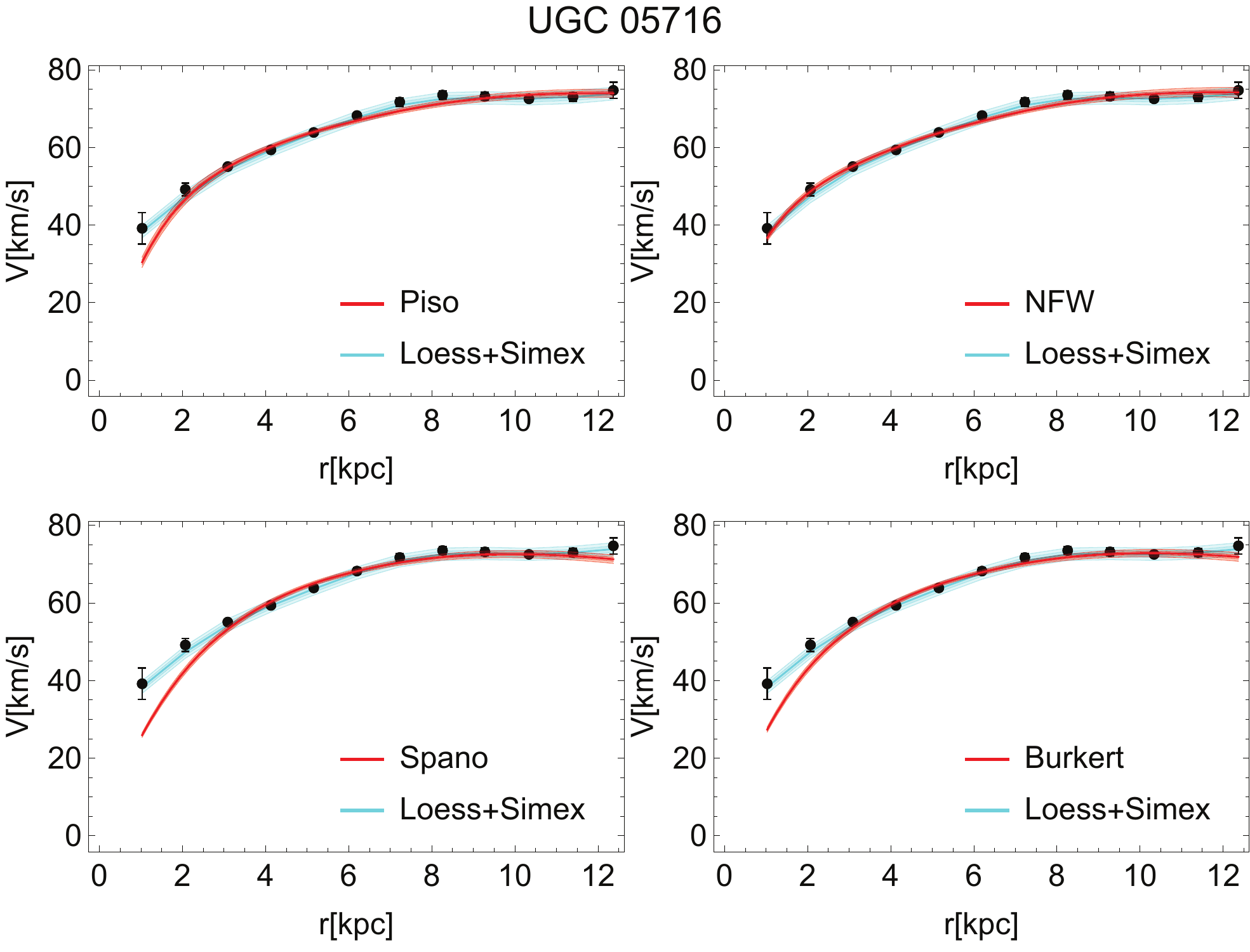} \\
\end{array}$
\end{center}
\caption{Galaxies where the distance between the best fit and 1$\sigma$ bands for Spano and Burkert have a notable 
separation with the LOESS+SIMEX band, according with the results from Table \ref{tab:PNSB}. In the figure we show the reconstruction 
LOESS+SIMEX 1 and 2$\sigma$ bands in cyan color, the best fit for the four models studied in the second group 
including 1 and 2$\sigma$ bands in red.}
\label{fig:Models2}
\end{figure*}

For each model we report the $\chi^2_{red}$, $P$-value, AIC, BIC, distance between the best fit velocity curve and the 
reconstruction with LOESS+SIMEX (DIST), the distance between the $1\sigma$ band from the best fit and the $1\sigma$ 
band from LOESS+SIMEX ($D1\sigma$). Also in columns 26--29, the Best Model columns are shown according to 
the BIC value ($B_{\rm BIC}$), the distance between the LOESS+SIMEX reconstruction and the best fit curves ($B_{\rm DIST}$), 
the distance between $1\sigma$ bands $B_{D1\sigma}$ and $\chi^2_{red}$ value. Columns 26--28 with models ordered according 
to the best values allow us to establish the three conditions as a criterion to accept or reject models. The ordering is as follows.

In the table we associate a number (1--4) to each model, and for column 26 and according to the best BIC (the model with the 
minimum value in the BIC), we write in the first place the number of the model to point out the best one. Followed by the second 
one that has the best BIC among the rest, and so on.  We found from the comparison of the best BIC's that the success for PISO is 
$ \sim 15.9 \% $, for NFW $\sim 18.1 \%$, Spano $\sim  46.6 \%$ and for Burkert $\sim 18.2 \% $.

In columns 27 and 28 of Table \ref{tab:PNSB} we show, in ascending order, the 
models (1--4) according to the distance between LOESS+SIMEX and best fit velocity curve ($B_{\rm DIST}$) and the distance 
between the $1\sigma$ confidence bands for LOESS+SIMEX and  best fit velocity rotation curve, ($B_{D1 \sigma}$),
respectively. 

From this table and according to the lower BIC, the lower distance $B_{\rm DIST}$ and lower distance $B_{\rm D1\sigma}$, the percentage 
of galaxies satisfying the three conditions, i. e., galaxies that have the best values in all the cases, 
is the $44.32\%$. From this $44.32\%$ of the analyzed galaxies, approximately the $53.85\%$ points out to Spano as the most favored 
model and for NFW $\sim 20.51\%$, for Piso we found a percentage value of $\sim 10.25\%$ and Burkert $\sim 12.82\%$.

It is important to mention that  BIC and AIC have the same penalty term, therefore the best AIC is compatible 
$100\%$ with
the best BIC value.

Finally, as an additional information, we show in column 29 the best $\chi^2_{red}$, in this case the first number indicates 
the model with the lower $\chi^2_ {red}$. 
It is important to mention that, in this group, the number of parameters of each model is the same, therefore there is a consistency between 
AIC and BIC definitions where the selection of the most favorable model according to the AIC has a coincidence of 100$\%$ with the selection 
of the best BIC (in this case the model order for each galaxy according to AIC is the same according to BIC).

As a way to illustrate the considerable differences between the models studied in this work, by taking a look to the Best Model columns of 
Table  \ref{tab:PNSB} (columns 26--28), we choose four galaxies where the distance between the fitting velocity curve and the nonparametric 
reconstructed curve  (central lines and 1$\sigma$ bands) is the biggest for the NFW model. Also in agreement with the respective BIC 
results and $\chi^2_{\rm red}$, the four columns of the Best Model select NFW in the last place.

In Figs. \ref{fig:Models1} and \ref{fig:Models2} we show the observed rotation curves for four galaxies and
the rotation curves that result from the fitting procedure with four models, Piso, NFW, Spano and Burkert. 
In these figures are shown the curves resulting from the nonparametric LOESS-SIMEX procedure (in cyan).
The red colors are for the best fit curves, 1$\sigma$ and 2$\sigma$ error bands, respectively.
For the galaxies 
DDO 154, IC 2574, NGC 3109 and UGC00891 (Fig. \ref{fig:Models1}), 
and for NFW model the distance between the best fit and 1$\sigma$ bands has a noticeable separation from the LOESS-SIMEX band 
and in agreement with results shown in Table \ref{tab:PNSB}.
And in Fig. \ref{fig:Models2} we show that for the rotation curves NGC 0024, NGC 0247, NGC 3741 and UGC 05716,
Spano and Burkert models are having the biggest distance between the 1$\sigma$ confidence  and the reconstruction LOESS+SIMEX 
confidence bands, even the columns of the Best Model place Spano in the last place of the selection.

\begin{figure*}
\begin{center}$
\begin{array}{ccc}
\includegraphics[width=3.45in]{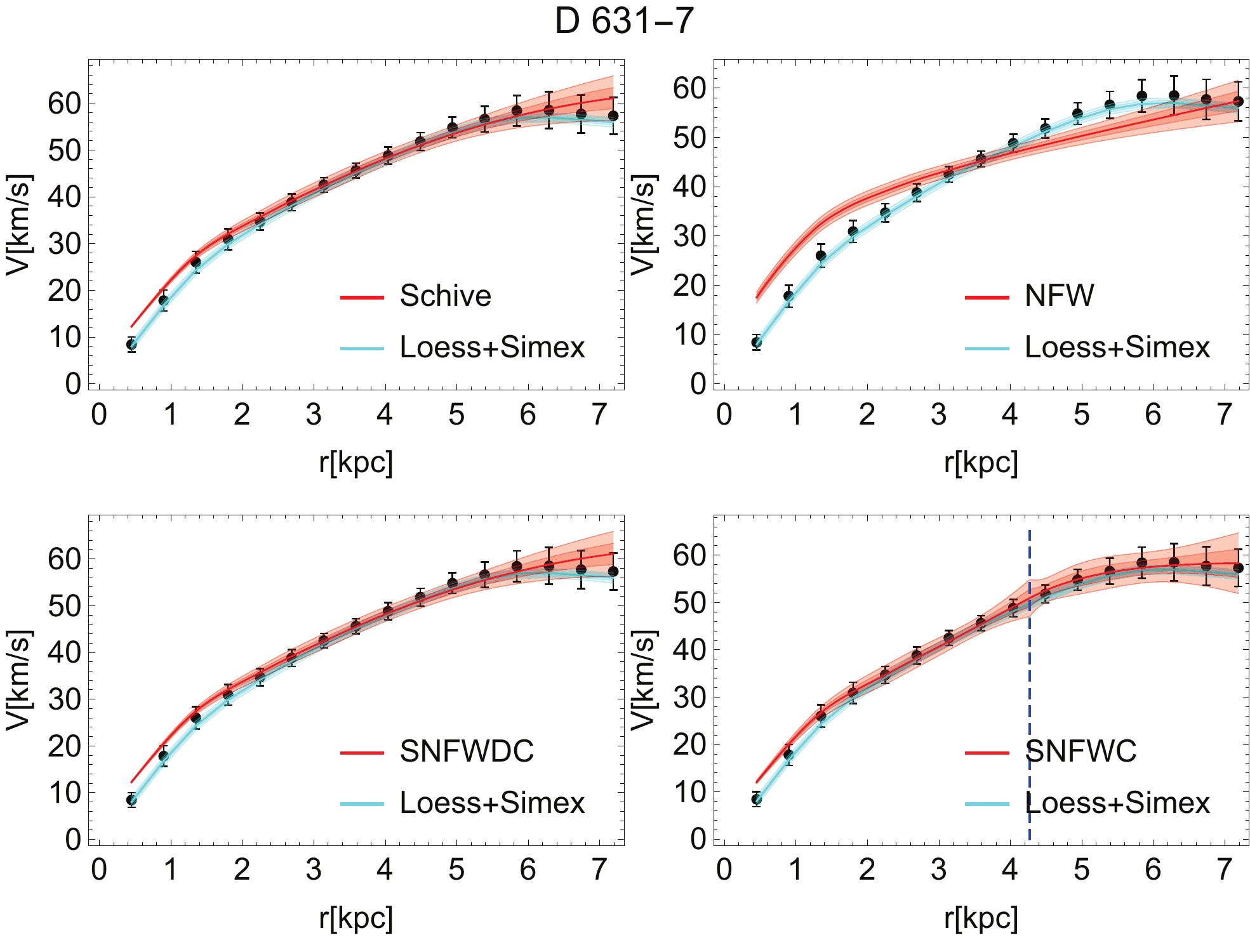}&
\includegraphics[width=3.45in]{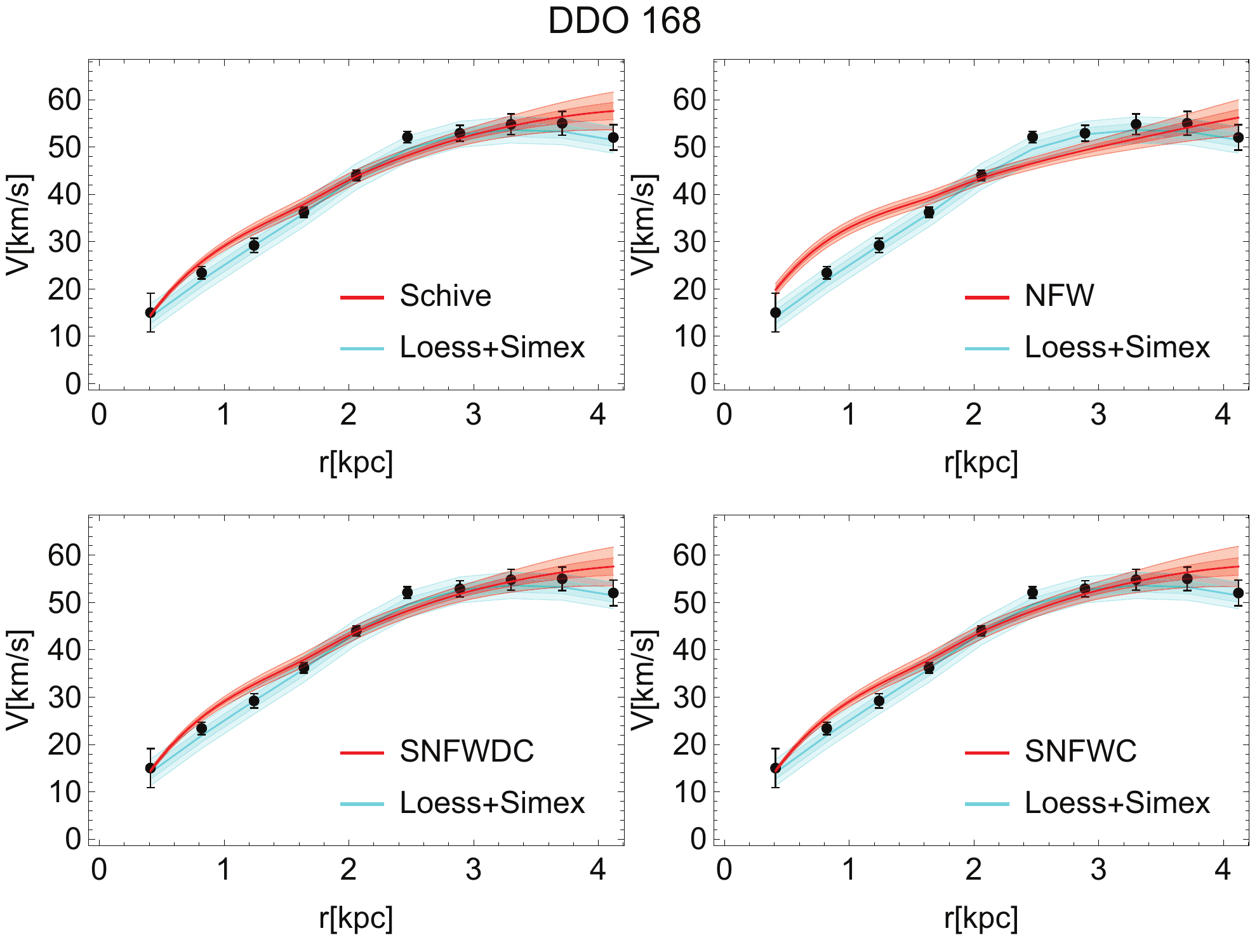}\\
\includegraphics[width=3.45in]{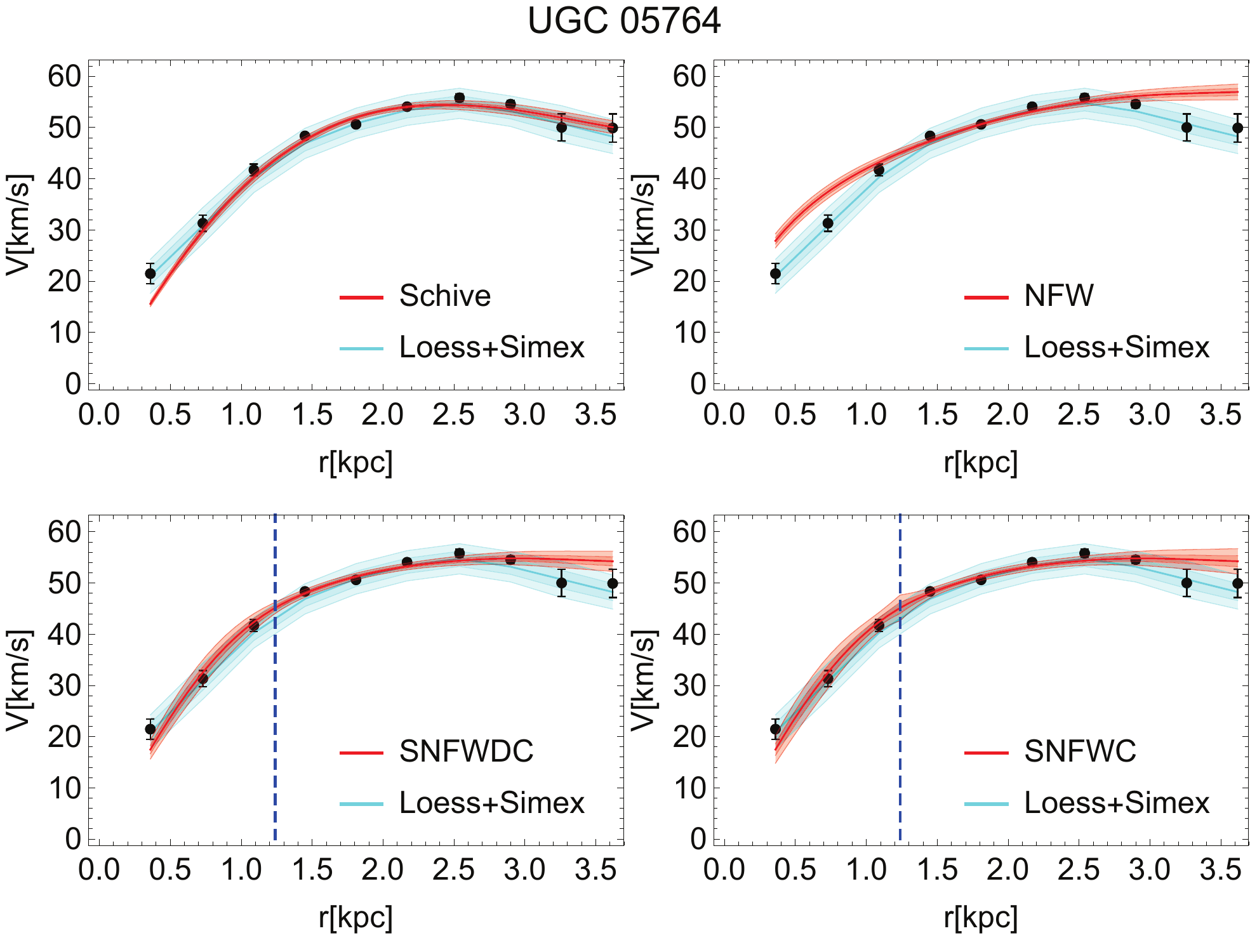}&
\includegraphics[width=3.45in]{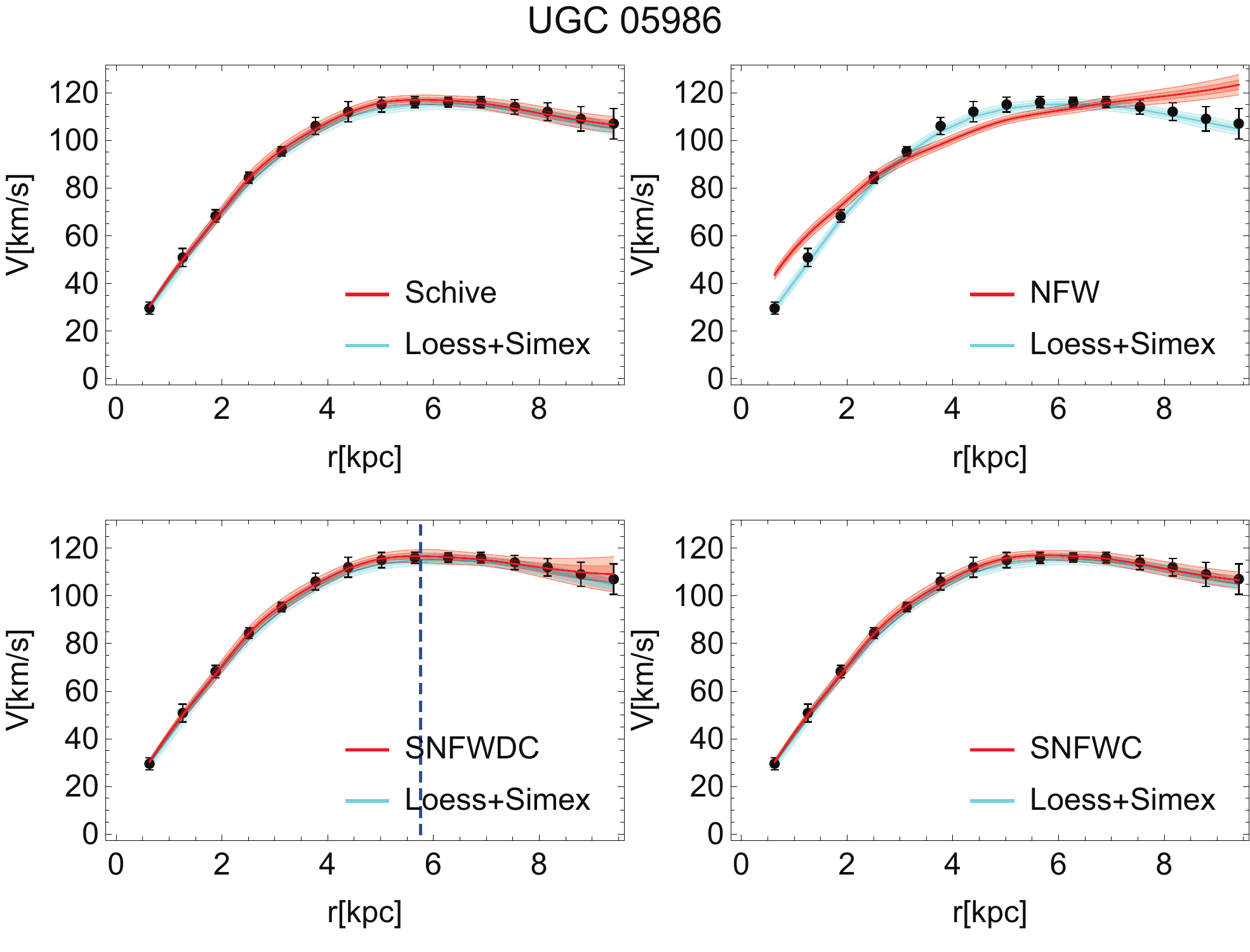} \\
\end{array}$
\end{center}
\caption{Galaxies where the distance between the best fit and 1$\sigma$ bands for NFW has a notable separation with the 
LOESS+SIMEX band, according with the results from Table \ref{tab:SNSS}. In the figure we show the reconstruction LOESS+SIMEX 1 and 
2$\sigma$ bands in cyan color, the best fit for the four models studied in the second group including 1 and 2$\sigma$ bands in red and the 
vertical lines correspond to $r_\epsilon$ (transition radius for the fuzzy models, Eq. (\ref{WDM-density})).}
\label{fig:Models3}
\end{figure*}

\begin{figure*}
\begin{center}$
\begin{array}{ccc}
\includegraphics[width=3.45in]{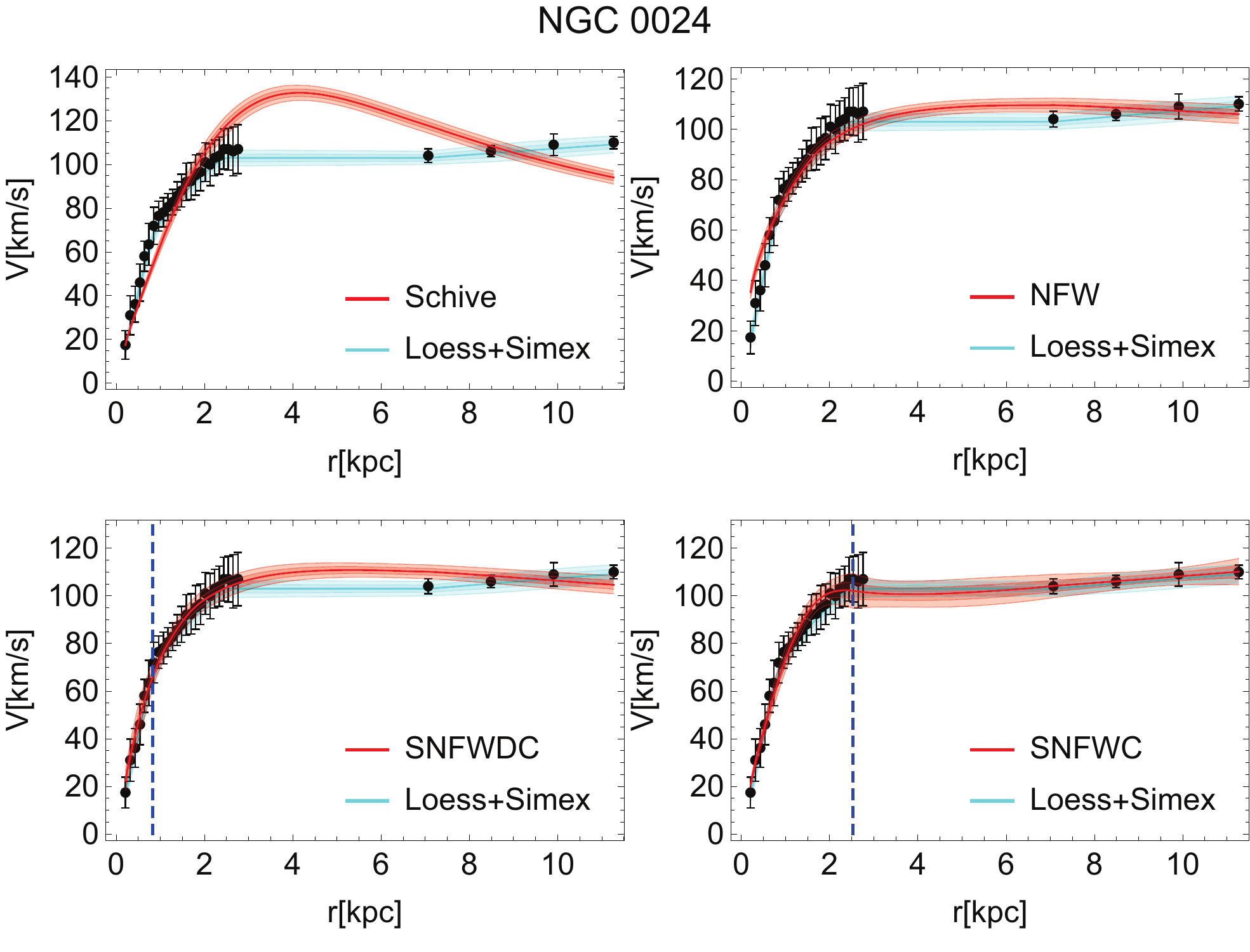}&
\includegraphics[width=3.45in]{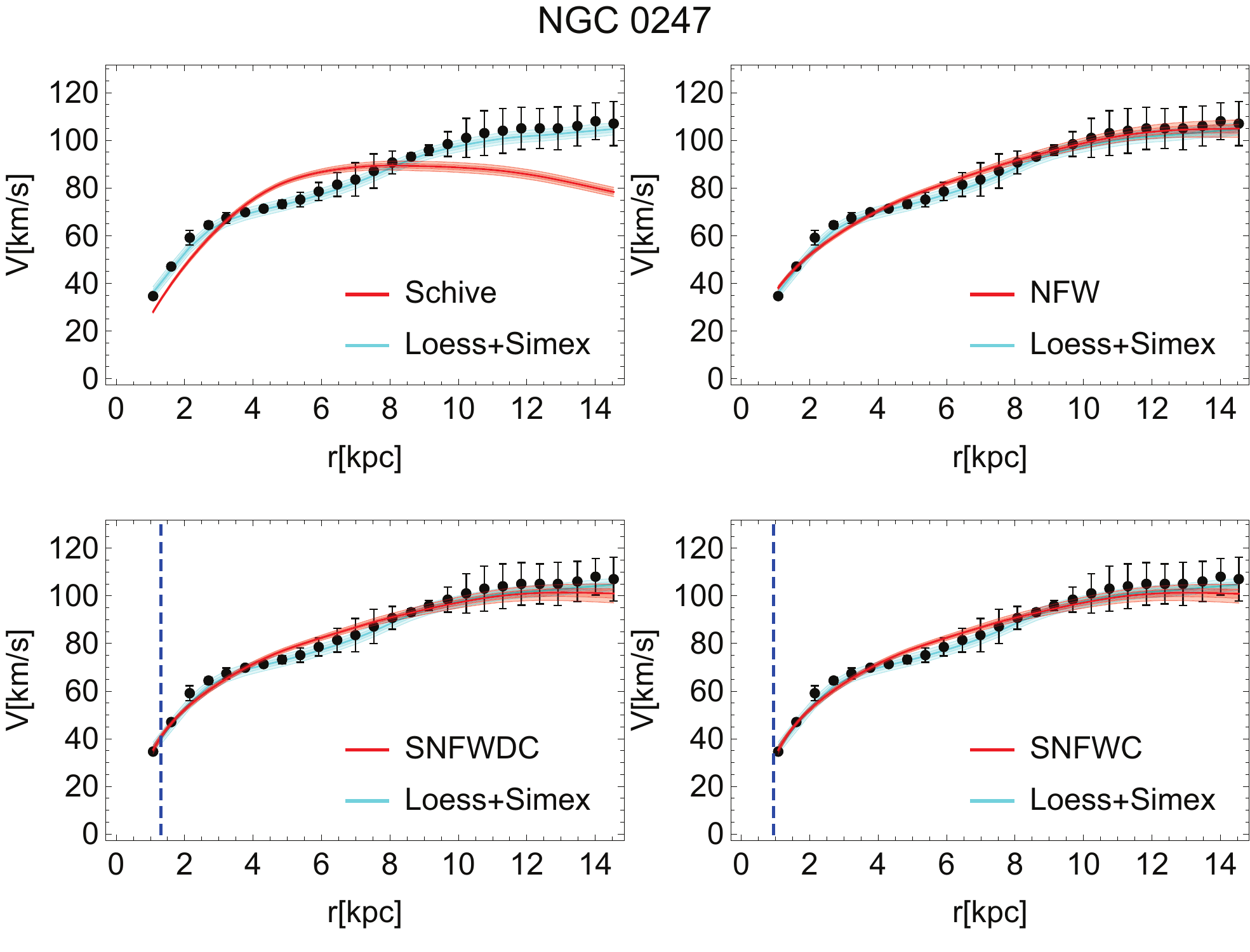}\\
\includegraphics[width=3.45in]{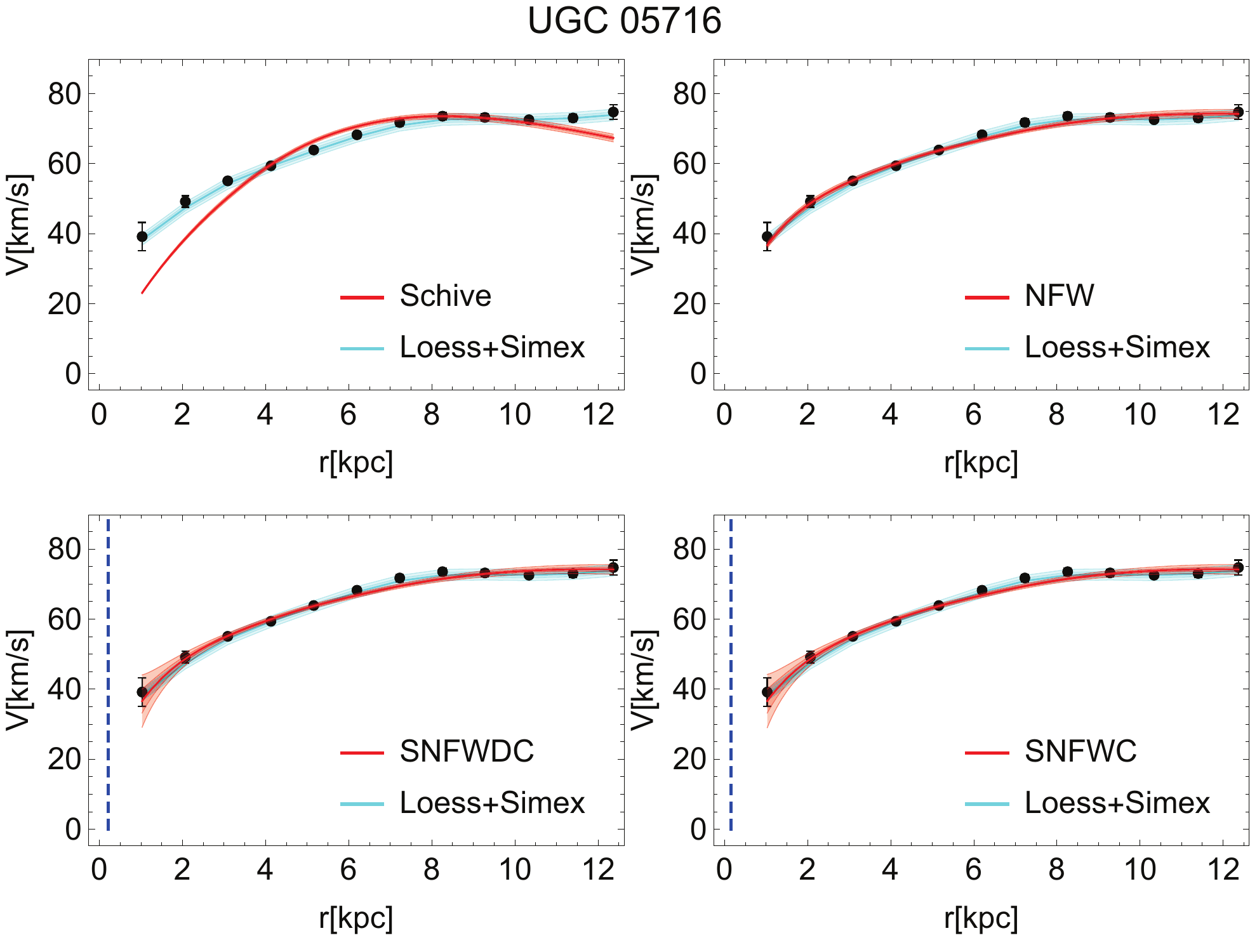}&
\includegraphics[width=3.45in]{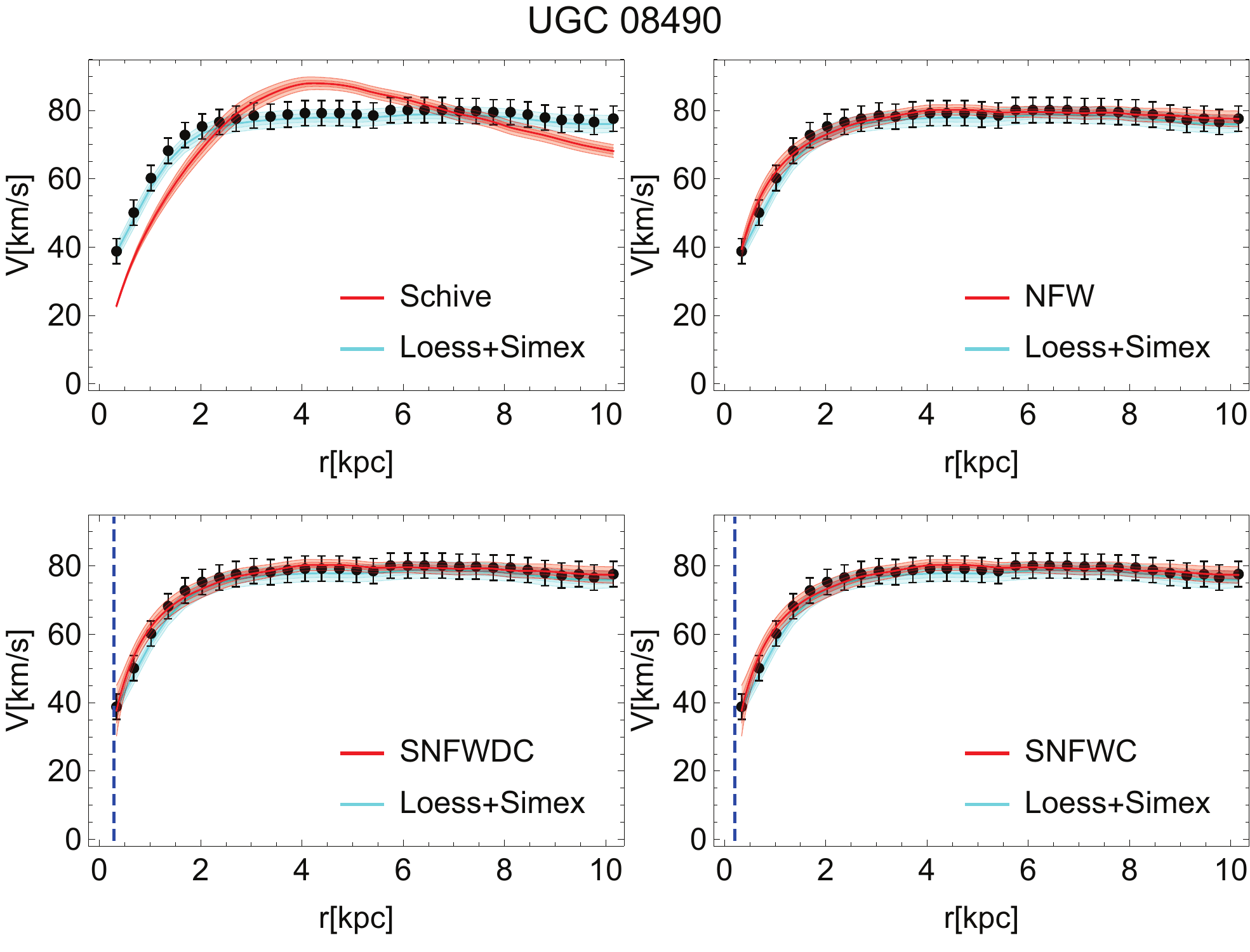} \\
\end{array}$
\end{center}
\caption{Galaxies where the distance between the best fit and 1$\sigma$ bands for Schive has a notable separation with the 
LOESS+SIMEX band, according with the results from Table \ref{tab:SNSS}. In the figure we show the reconstruction LOESS+SIMEX 1 
and 2$\sigma$ bands in cyan color, the best fit for the four models studied in the second group including 1 and 2$\sigma$ bands in red 
and the vertical lines correspond to $r_\epsilon$ (transition radius for the fuzzy models, Eq. (\ref{WDM-density})).}
\label{fig:Models4}
\end{figure*}

On the other hand, keep in mind that LOESS+SIMEX can provide great assistance in model selection, particularly when the 
data are very noisy or have other features that make patterns difficult to see. Indeed, following \cite{10.2307/2336363}, a very 
simple way to assess the goodness of fit of a parametric model is by determining if the nonparametric fit falls within the parametric 
error bars. So, if we take a look at Fig. \ref{fig:Models1} and the respective plots for the NFW model, for all the galaxies showed 
there, it is noticeable that the reconstructed curve lies outside the error bands of NFW model which indicates that this model 
does not appear to be a good fit to these data which gives support to our 
results that point out NFW is not the most favored model. The same argument can apply to other figures. 

Taking as reference the NFW model, we studied the Schive and the Fuzzy models (SNFWDC and SNFWC) to compare the quantitative 
distances between the best fitting and the LOESS+SIMEX method. 
From this group of models we compare core, cusp and fuzzy models, where in addition, these fuzzy models have the freedom to adapt 
themselves to the structure of each galaxy and reproduce either core or cusp models. The value of $r_\epsilon$, 
Eq. (\ref{WDM-density}), can provide us the three different structures of a galaxy: Core ($r_\epsilon > r_{\rm max}$, being $r_{\rm max}$ 
the last observed radius of the galaxy), Cusp ($r_\epsilon \sim r_{\rm min}$ with $r_{\rm min}$ is the first observed 
radius of the galaxy) and Core+Cusp ($r_{\rm min}<r_\epsilon <r_{\rm max}$).

The results of the analysis of this group of models are shown in Table \ref{tab:SNSS}. Its description is similar to Table \ref{tab:PNSB}. 
Columns 27--30 show the Best Model values, from 27--29 we apply our selection criteria. In order to obtain an 
alternative classification value for the baryonic contribution in each galaxy, we compute $\delta A_V = 100\frac{ A_{V\rm Bar}}{A_{V\rm Obs}}$, 
where $A_{V\rm Bar}$ is the total area enclosed by the baryonic velocity contribution, $\sqrt{\Upsilon_{\rm Disk}V_{\rm Disk}^2 + V_{\rm Gas}^2}$ and
$A_{V\rm Obs}$, is the area under the velocity rotation curve of the observed data (in the radius range observed), see column 2 of Table  \ref{tab:SNSS}. 

Here, it is important to mention that we have three galaxies where the number of observed data is the same as the number of free parameters 
for the SNFWC model (we use the * symbol, next to the galaxy name to identify these galaxies, see Table \ref{tab:SNSS}); in these cases 
we are not reporting the $\chi^2_{red}$ and the $P$-value, nevertheless it is possible to calculate the other statistic quantities.

In Figs. \ref{fig:Models3} and \ref{fig:Models4}
we show the observed rotation curves for four galaxies
and
the rotation curves that result from the fitting 
procedure with the four models, 
NFW, Schive and the Fuzzy models. 
The resulting curves from the nonparametric LOESS-SIMEX procedure are displayed as well.
The cyan bands are the nonparametric reconstruction LOESS+SIMEX and the red colors are for the best fit curves, 
1$\sigma$ and 2$\sigma$ error bands, respectively.

From Table \ref{tab:SNSS} we studied the relationship between the lower BIC and the lower value from $B_{\rm DIST}$ and 
$B_{\rm D1\sigma}$, and we found that 
$\sim 28.41\%$ of galaxies satisfy the three conditions. In this case Schive has been selected with the $44\%$ to be the most 
favored model, NFW with $ 36\%$, SNFWDC $12\%$ and SNFWC obtained $8\%$ of model selection coincidence. 

It is important to mention that even when BIC and AIC have different penalty terms, the best AIC is compatible 
$\sim 76.1\%$ with
the best BIC value, 
which gives us a strong support for the Best Selection Model techniques used in this work.

According to the Best Model columns from Table \ref{tab:SNSS}, we notice that the $B_{\rm DIST}$ selection model is the same 
for the Fuzzy models (3 and 4) in 24 galaxies, and 11 galaxies where fuzzy models have the same $B_{\rm D1\sigma}$. If we considered 
the cases where $B_{\rm DIST}$ and $B_{\rm D1\sigma}$ have the same value for the Fuzzy models (3 and 4) we found that only 8 
galaxies fulfill this condition. For example, let us consider the galaxy NGC 0247 and compare the two columns $B_{\rm DIST}$ 
and $B_{\rm D1\sigma}$ (from the Best Model column) and we found the combination (34) in first and second place of the numerical 
combination of models, respectively. This is because $r_\epsilon ^{(3)} \sim r_{min}$ and  $r_\epsilon ^{(4)} < r_{min}$, which tell us that 
NFW model fits better the structure for the rotation curve. 
For the other cases, galaxies DDO 064, DDO 168, KK 98-251 we obtain $r_\epsilon ^{(3)} > r_{max}$ and  $r_\epsilon ^{(4)} > r_{max}$, 
where Schive is the dominant model for the rotation curves; for UGC 05716, UGC 07399 and UGC 08490 $r_\epsilon ^{(3)} < r_{min}$ 
and  $r_\epsilon ^{(4)} < r_{min}$; and for UGC 07524 $r_\epsilon ^{(3)} = r_\epsilon ^{(4)} = 3.378$ kpc.

Additionally, we notice that Schive model has problems in fitting the final data points while NFW in the initial data. On the other hand, Fuzzy models solve the two problems, the transition radius $r_\epsilon$ for each SNFW model is the same for 13 galaxies, 
while, for the other 35 galaxies we have for SNFWDC (3), $0.11< r_\epsilon^{(3)} < 9.4$ and for SNFWC (4), $0.15 < r_\epsilon ^{(4)} < 19.7$. 
In the cases where SNFW models are the same, we use the $B_{\rm BIC}$ in order to perform the model selection because it considers the 
penalty term for the number of parameters instead of taking the $B_{\rm D1\sigma}$ which can lead to misleading results due to the error 
propagation in the LOESS+SIMEX technique.

So far we have compared the results obtained from parametric versus non-parametric analysis. 
The SPARC subsample consists of 88 rotation curves of galaxies with quality factor $Q=1$ or $2$. To find any indication of spatial 
resolution effect on the results we have divided the subsample into two: the first one with only galaxies with a $Q$ value of $1$ (50 galaxies) 
and the second one with $Q=2$ (38 galaxies). The analysis results are as follows:

\begin{list}{\roman{qcounter}.}{\usecounter{qcounter}}
 \item [{\bf Q= 1}] When the RC is extracted from the first group Table \ref{tab:PNSB}: we found that 36\% satisfied the three conditions and 
 of which 55.56\% prefer Spano model while 16.67\% avored NFW.  And for the RC extracted from the second group 
 Table \ref{tab:SNSS}: 24\% satisfied the three conditions and 33.33\% support Schive as  the best model according to the criteria 
 and 25\% favored NFW and 25\%  were also in favor of  SNFWDC.
 \\

 \item[{\bf Q= 2}]  If the RC is extracted from the second group Table \ref{tab:PNSB}: 
we found that 55.26\% satisfied the three conditions and from them 52.38\% favored Spano model and 
23.81\% favored NFW. Table \ref{tab:SNSS}: for $Q=2$, 34.21\% satisfied the three conditions and 53.85\% of these favored 
Schive as the best model according to the criteria and 46.15\% set to the NFW model in second place.
\end{list}

\section{DISCUSSION} \label{Discussion}

From Table \ref{tab:SNSS} and according to the three columns of the Best Model it is possible to classify the galaxies 
DDO 064, DDO 168, F 571-v1, KK 98-251, NGC 6789, UGC 05986 and UGC 06399 as core type, where this criterion 
points out to Schive as the best model.

In the same way we select the cuspy models based on Tables \ref{tab:PNSB} and \ref{tab:SNSS}, observing that the Galaxies 
UGC 05716 and UGC 08490 point out to NFW (with the three conditions fulfilled in both cases) as the most favored model. 
On the other hand, UGC 02259, UGC 05918 and UGC 12732 are selected to be cuspy from Table \ref{tab:SNSS}, 
NGC 0247 and NGC 3741 from Table \ref{tab:PNSB}, satisfying again the three Best Model columns.

Comparing the two tables (seven models) for NGC 0247, NGC 3741, UGC 02259, UGC 05918 and UGC 12732 we found:

\begin{itemize}

\item For NGC 0247 comparing the seven models it is found that $B_{\rm BIC}$ and $B_{\rm DIST}$ point out 
to SNFWDC (3) and $B_{\rm D1\sigma}$ selects the NFW model. Analyzing the fuzzy models it is found that $r_\epsilon^{(3)} = 1.31$ kpc 
and the minimum radius observed for this galaxy is $r_{min} = 1.08$ kpc. In this case the major contribution in fitting the velocity of the galaxy 
comes from NFW ($r_\epsilon ^{(3)} \sim r_{\rm min}$, Fig. \ref{fig:Models4}), therefore the fuzzy model is in agreement with LOESS+SIMEX and 
the other conditions to classify it as a cuspy galaxy.

\item For NGC 3741, the NFW model satisfies $B_{\rm BIC}$ and $B_{\rm D1\sigma}$ and while the $B_{\rm DIST}$ points out to SNFWDC, 
but $r_\epsilon^{(3)} = 0.268$ kpc and  $r_{\rm min} = 0.23$ kpc which NFW gives the biggest contribution to the galaxy structure, 
as an additional condition, we compare the $B_{\chi^2_{red}}$ for the seven models and found that this points out to NFW, 
therefore, we can classify NGC 3741 as a cuspy galaxy.

\item For UGC 12732 we found that the NFW profile satisfies $B_{\rm BIC}$ and $B_{\rm DIST}$, while PISO satisfies $B_{\rm D1\sigma}$ 
and the $B_{\chi^2_{red}}$ points out to SNFWDC, where $r_\epsilon ^{(3)} = 0.8$ kpc and $r_{\rm min} = 0.96$ kpc. We can conclude that 
this galaxy is classified as a cuspy one.

\end{itemize}

The remaining galaxies (UGC 02259, UGC 05918) can be excluded of being cuspy because of NFW satisfies only $B_{\rm DIST}$ 
for both of them while PISO satisfies the rest of the conditions from Best Model column.

Aditionally, in Fig. \ref{fig:Models3}, we show four galaxies
(D 631-7, DDO168, UGC 05764 and UGC 05986)
with the four models (Schive, NFW, SNFWDC and SNFWC) from there, one can see that NFW is at a far distance 
from the confidence bands. From Table \ref{tab:SNSS} we notice that these galaxies have in common that the NFW  
is in the last place from the selection number code of models. 
In contrast we observe in Fig. \ref{fig:Models4} 
(for galaxies, NGC 0024, NGC 0247, UGC 05716 and UGC 08490)
that the Schive model best fit is far from the reconstruction. In these cases, Schive comes in
the last place for the selection of the best model in Table \ref{tab:SNSS}.

To complement the comparison between non-parametric versus parametric methods of data analysis and in order to set some basic 
concepts to make an interpretation of our results, we have analyzed with the MCMC method the 152 galaxies that satisfy the quality condition of
\cite{2016PhRvL.117t1101M}  in the SPARC catalog of rotation curves. 

For each halo DM model in the core-cusp group (see below) we have computed four important quantities that should give us information about galaxy 
formation and evolution: the central characteristic volume density $\rho_s$, the scale length $r_s$, the characteristic surface density 
$\mu_{DM} \equiv \rho_s r_s$ and the DM mass within 300 pc, $M_{300}\equiv M_{DM}(300$ pc$)$.

Fig. \ref{fig:MCMCtriangles} illustrates the analysis of the Markov chains of the galaxy IC 2574 by using GetDist code that is included 
in CosmoMC bayesian analysis code \citep{Lewis:2002ah}. It is shown the posterior distributions of fitting ($\rho_s$, $r_s$) and derived parameters 
($\mu_{DM}$, $M_{300}$) for the PISO, NFW, Spano, Burkert and Schive models. We will refer to these group of models as core-cusp group. 
Also, in Fig. \ref{fig:MCMCtriangles}, it is displayed the confidence region at $1\sigma$ and $2\sigma$. 
We can notice the correlation between $\mu_{DM}$ and $r_s$
and the strong correlation between $M_{300}$ and $\rho_s$.
The pair $M_{300}$ and $\rho_s$ is anti correlated with the pair $\mu_{DM}$ and $r_s$.
We notice good Gaussian posteriors for the fitting parameters for all the models except for the NFW profile. 
As shown in Fig. \ref{fig:Models1} NFW is the worst fitting case for galaxy IC 2574.
\begin{figure*}
\begin{center}$
\begin{array}{ccc}
\includegraphics[width=3.0in]{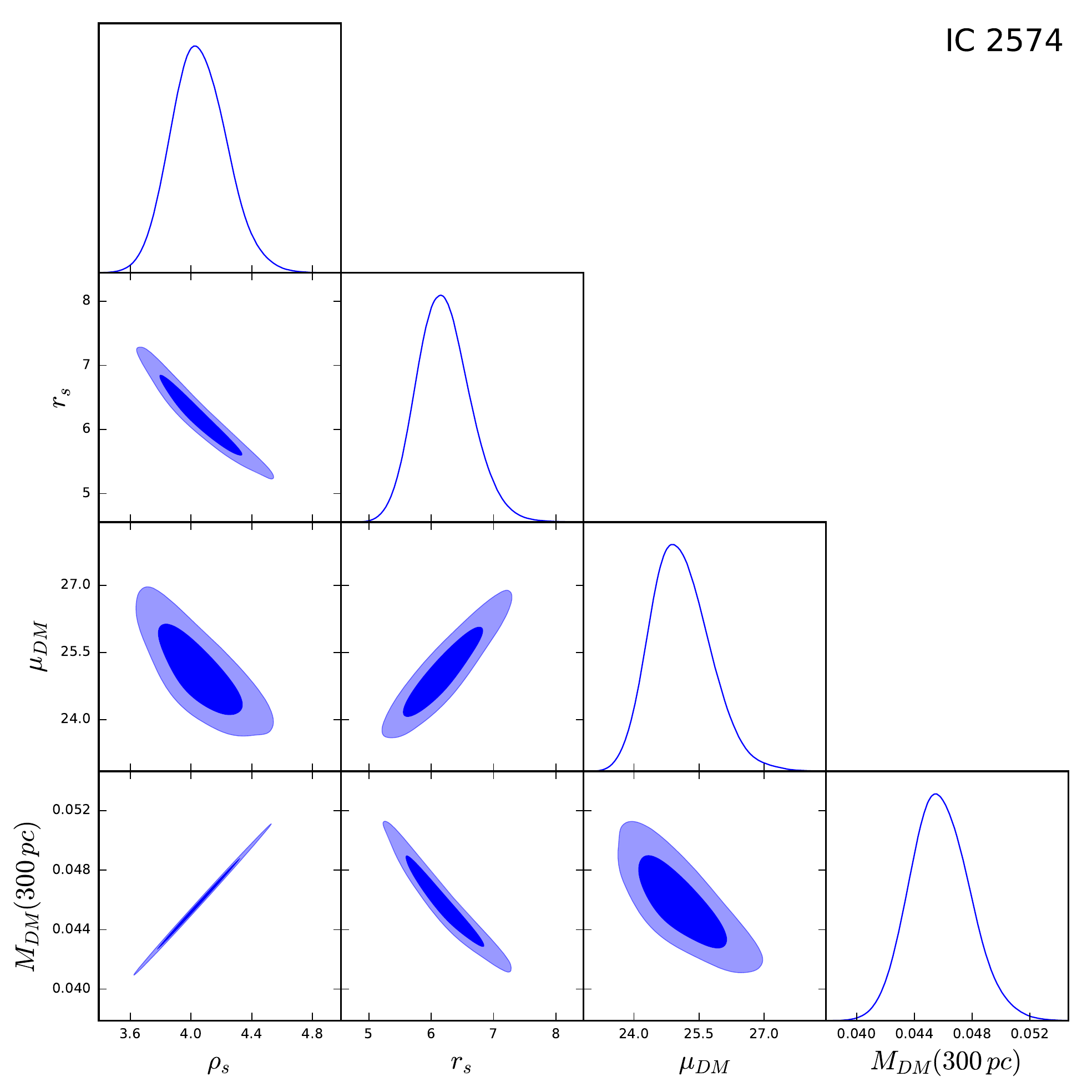} &
\includegraphics[width=3.0in]{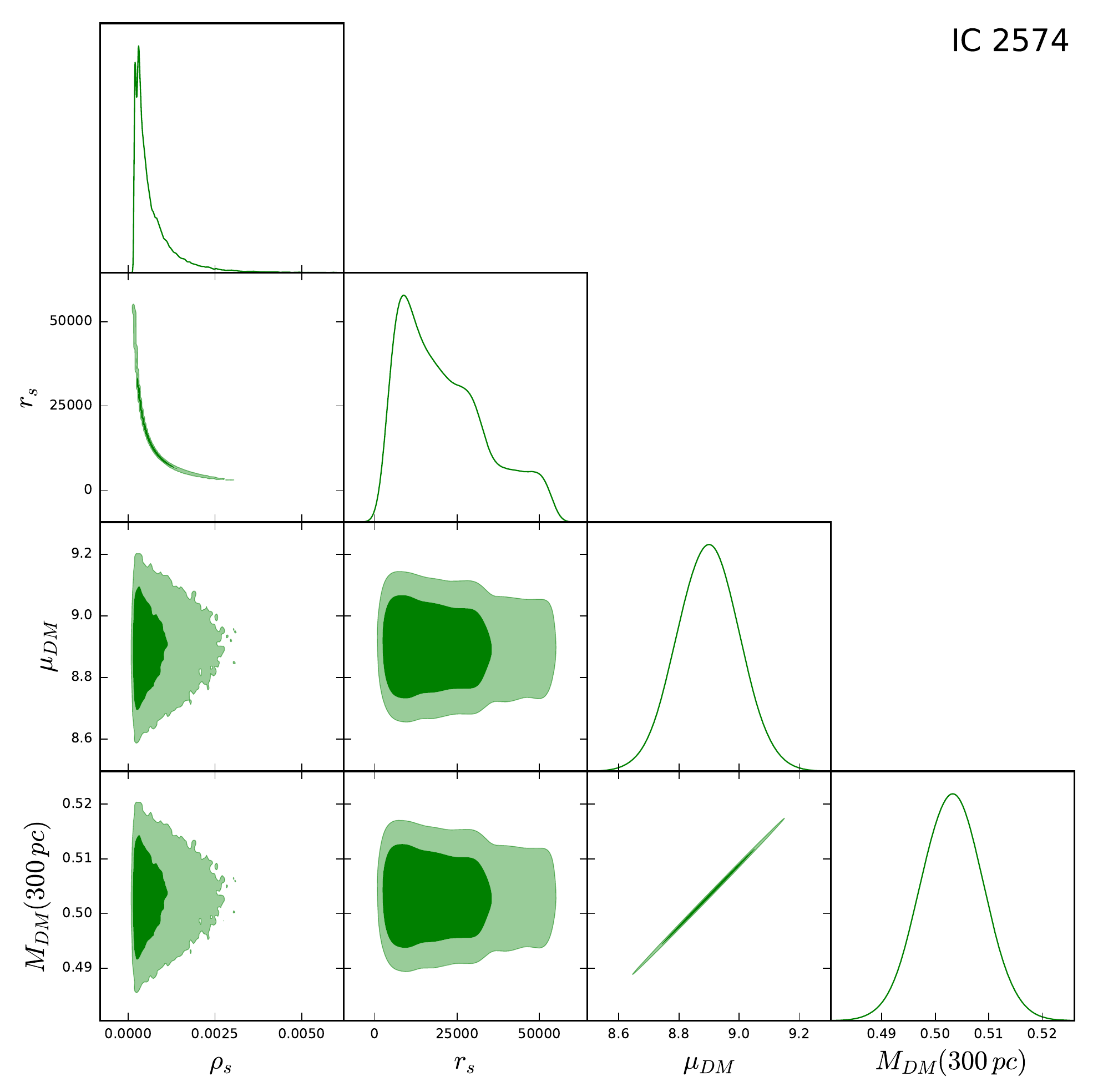} \\
\includegraphics[width=3.0in]{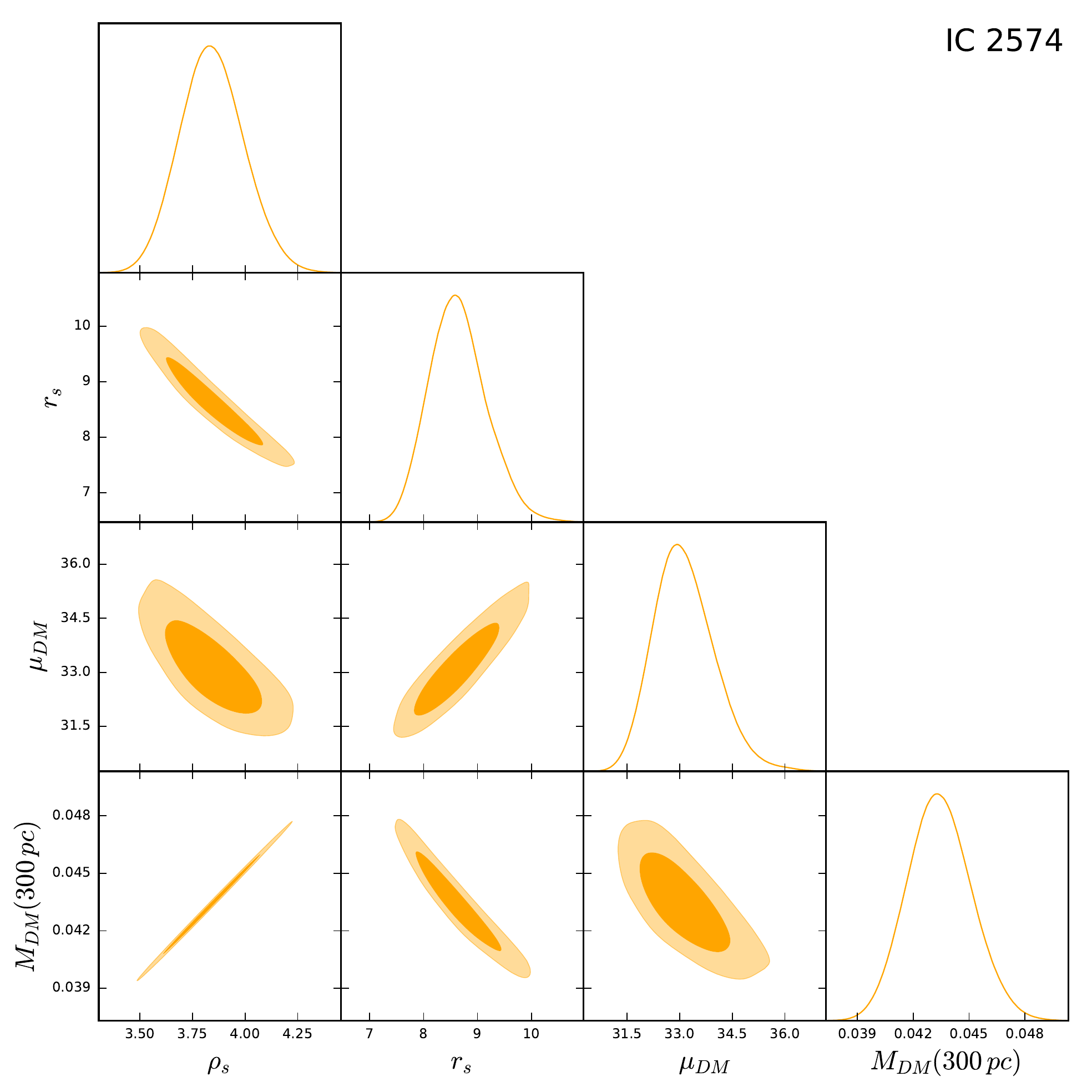} &
\includegraphics[width=3.0in]{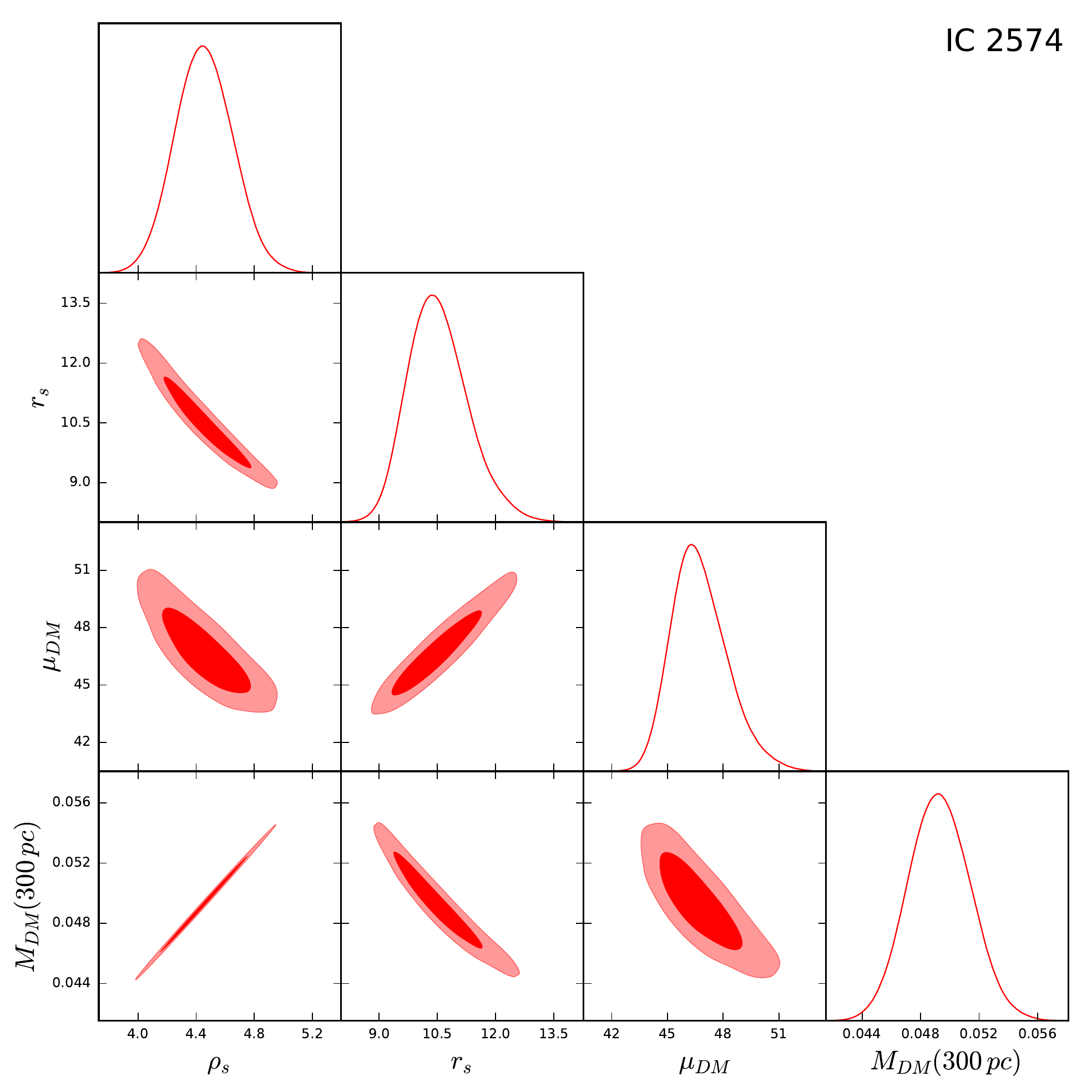} \\
\includegraphics[width=3.0in]{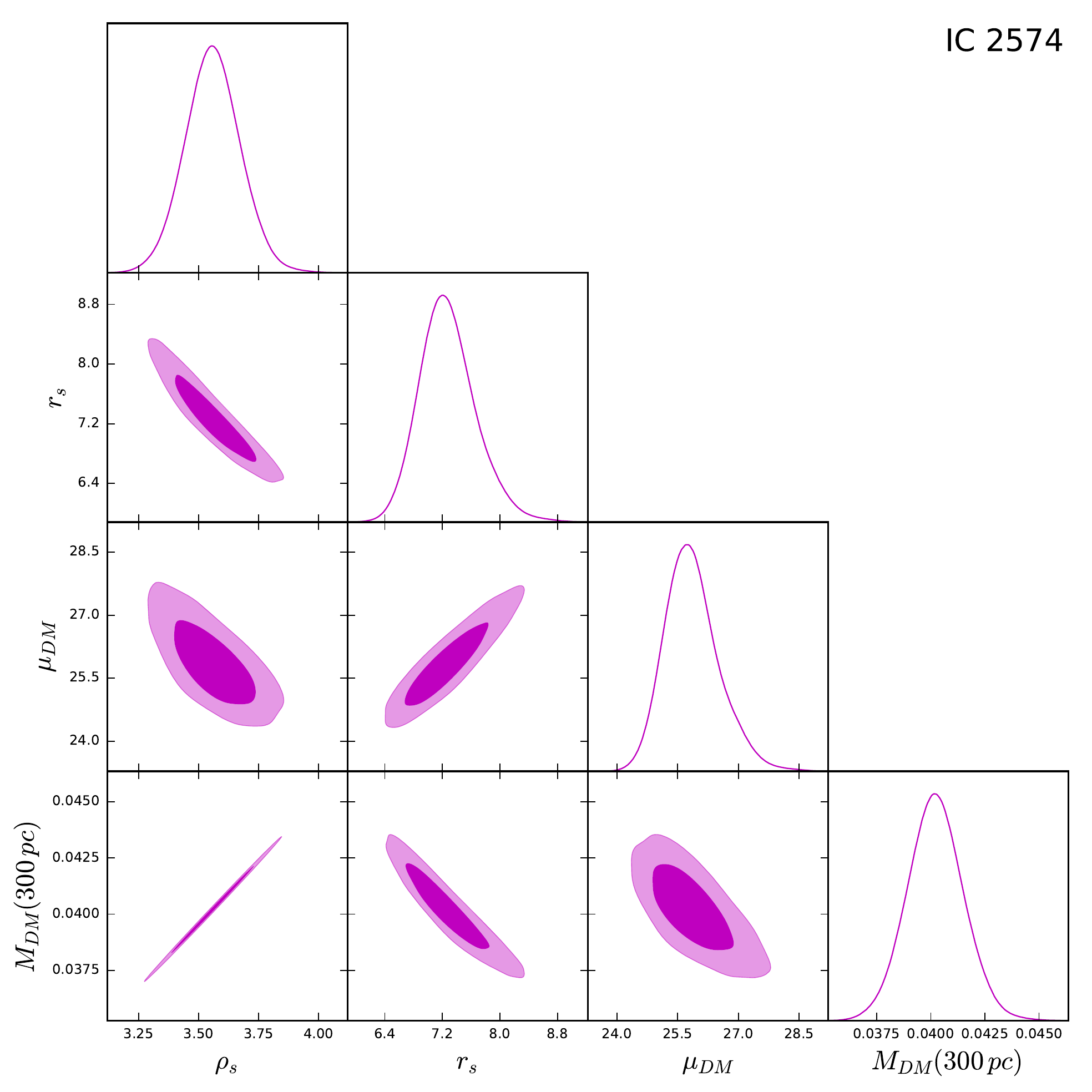} &
\end{array}$
\end{center}
\caption{Posterior distributions for parameters $\rho_s$, $r_s$, $\mu_{DM}$ and $M_{300}$.
The contour lines are $1\sigma$ and $2\sigma$ confidence regions.
From the top to right and to the bottom shown are PISO (blue), NFW (green), Spano (yellow), Burkert (red) and Schive (violet) models.
}
\label{fig:MCMCtriangles}
\end{figure*}

In Figs. \ref{fig:ModelsPISO_NFW}--\ref{fig:ModelsSchive} are shown these four astrophysical observables 
($\rho_s$, $r_s$, $\mu_{DM}$ and $M_{300}$) as a function of galaxy luminosity, $L_{[3.6]}$.
From other studies it was expected that $\mu_{DM}$ and $M_{300}$ were nearly constant, independent of the 
absolute magnitud of the galaxies
\citep{2017PhRvD..96d3005U,2004IAUS..220..377K,2008Natur.454.1096S}.
The general tendency we found for the above four quantities is that $\rho_s$ and $M_{300}$ are roughly constant independently 
of the galaxy luminosity in the SPARC catalog except for the NFW model. We have found that a constant characteristic volume density 
of DM haloes is in agreement with \cite{2019MNRAS.482.5106L}.  In agreement with \citep{2008Natur.454.1096S}, we obtained that 
for dSph galaxies, the value of the mass is the same for spiral galaxies within 300 pc and with the same order of magnitud, 
$\sim 10^7$ M$_\odot$. This would give a central density for DM haloes of $\sim 0.1$ M$_\odot$pc$^{-3}$ regardless the model, 
result not shared with NFW. Moreover, this result is consistent with both, non-parametric and parametric analysis, 
that are in favor of core models. 

Theoretical and observational studies have given the result that the central surface density $\mu_{DM}$ is constant independently of the absolute 
magnitude of the galaxies \citep{2017PhRvD..96d3005U,2004IAUS..220..377K}. We have obtained that this is not the case, 
and that $\mu_{DM}$ grows linearly in the Log-Log scale as a function of $L_{[3.6]}$, fact that is in tension with previous results. 
Therefore, there is not a strong correlation between the density $\rho_s$ and the scale $r_s$ parameters for the density profiles we studied here, 
on the contrary, $\rho_s$ is constant.

In Table \ref{tab:muM300} we show the compilation of the $a$ and $b$ parameters in the linear fit, $a +b L$,
of the Log-Log scale of each one of the plots $\rho_s$, $r_s$, $\mu_{DM}$ and $M_{300}$ as a function of the luminosity $L_{[3.6]}$.
For instance, we found that $M_{300} \sim L^{0.03\pm 0.04}$ for PISO model, that is the functional relationship found in 
\cite{2008Natur.454.1096S}. Similar behaviour is obtained for Spano, Burkert and Schive models. For the characteristic volume density 
we found a similar relation, $\rho_{s} \sim L^{0.04\pm 0.04}$ for PISO model and almost the same relations are found for Spano, Burkert and Schive. 
However, the relations for the scale length and the central surface density are 
$r_s \sim L^{0.17\pm 0.02}$ and
$\mu_{DM} \sim L^{0.18\pm 0.02}$
for PISO and the same for the other models, except NFW which is the worst case as can be appreciated 
in Fig. \ref{fig:allModels} where we are illustrating the collection of all models in each plot. The $1\sigma$ bands of the linear fit are 
also shown for comparison.

In Fig. \ref{fig:allModels_massvel} are displayed the expressions for the dimensionless mass and velocity functions of
the core-cusp group (included in the Appendix \ref{Ap}) and for a fixed scaling length $r_s=1$ and using the same value for 
$r_s^3 \rho_s$ which is in units of mass. Figs. \ref{fig:allModels} and \ref{fig:allModels_massvel} show clearly the difference 
between core and cusp models. Additionally, from Fig. \ref{fig:allModels_massvel} it was shown that Spano and Schive models
almost agree with each other even when Figs.  \ref{fig:allModels}(a) and (b) show clear differences between them for high luminosities. 
Up to $2\sigma$ in the linear fit we may find an agreement between all core models nevertheless, there are clear differences at $1\sigma$. 
A detailed analysis following the cosmological method in \cite{2019MNRAS.482.5106L} should clarify on this matter.

\begin{center}
\begin{table*}
\ra{1.3}
\begin{center}
\begin{tabular}{@{}l l r r r r r r r r r r r @ {}}\toprule
    \hline \hline
    \multicolumn{11}{c}{Results compilation for the linear fit ($a + b \log(L_{[3.6]})$) to SPARC galaxies observables} \\
    \hline 

\multicolumn{1}{c}{} &
\multicolumn{1}{c}{} &
\multicolumn{1}{c}{PISO }&
\multicolumn{1}{c}{NFW} &
\multicolumn{1}{c}{Spano} &
\multicolumn{1}{c}{Burkert}&
\multicolumn{1}{c}{Schive}
\\
 
\cmidrule[0.4pt](r{0.125em}){1-1}
\cmidrule[0.4pt](lr{0.125em}){2-2}
\cmidrule[0.4pt](lr{0.125em}){3-3}
\cmidrule[0.4pt](lr{0.125em}){4-4}
\cmidrule[0.4pt](lr{0.125em}){5-5}
\cmidrule[0.4pt](lr{0.125em}){6-6}
\cmidrule[0.4pt](lr{0.125em}){7-7}
\cmidrule[0.4pt](lr{0.125em}){8-8}
\cmidrule[0.4pt](lr{0.125em}){9-9}
\cmidrule[0.4pt](lr{0.125em}){10-10}
\cmidrule[0.4pt](lr{0.125em}){11-11}
$\rho_s$ 	& $a$ 		& $1.50966\pm 0.07482$ 	    & $0.77526\pm 0.09496$ & $1.39718\pm 0.06294$  & $1.55027\pm 0.06738$ 	& $1.275\pm 0.05622$ \\
	  	& $b$ 		& $0.03908\pm 0.04412$ 	    & $0.11217\pm 0.05117$  & $0.05472\pm 0.03611$  & $0.04798\pm 0.0386$ 	& $-0.0315\pm 0.03317$ \\
$r_s$  	& $a$ 		& $0.22215\pm 0.04073$ 	    & $0.94325\pm 0.04649$  & $0.51082\pm 0.03099$  & $0.5889\pm 0.21369$ 	& $0.55206\pm 0.02854$ \\
  		& $b$ 		& $0.16663\pm 0.02328$ 	    & $0.13417\pm 0.02419$  & $0.17885\pm 0.01701$  & $0.31454\pm 0.11695$ 	& $0.235\pm 0.0161$ \\
$\mu$  	& $a$ 		& $1.83778\pm 0.03038$ 	    & $1.44945\pm 0.04341$  & $1.96427\pm 0.03031$  & $2.10946\pm 0.03356$ 	& $1.86379\pm 0.02744$ \\ 
  		& $b$ 		& $0.18341\pm 0.01857$ 	    & $0.28382\pm 0.02636$  & $0.20559\pm 0.01802$  & $0.20571\pm 0.02016$ 	& $0.18314\pm 0.01668$ \\ 
$M_{DM}(300$ pc$)$ & $a$ & $-0.40992\pm 0.07378$ & $0.21738\pm 0.04243$  & $-0.54155\pm 0.06218$ & $-0.40713\pm 0.06306$ 	& $-0.66961\pm 0.05572$ \\ 
 		& $b$ 		& $0.03284\pm 0.04383$ 	    & $0.27863\pm 0.02588$  & $0.05115\pm 0.03576$   & $0.05376\pm 0.03647$ 	& $-0.03203\pm 0.03291$ \\ 

 \bottomrule
\hline \hline
\end{tabular}
\end{center}
\caption{Compilation of the results 
of the linear fit (log-log scale) parameters of
$\rho_s$, $r_s$,
$\mu_{DM}$ and $M_{DM}(300$ pc$)$.
 }\label{tab:muM300}
\end{table*}
\end{center}

\begin{figure*}
\begin{center}$
\begin{array}{ccc}
\includegraphics[width=3.25in]{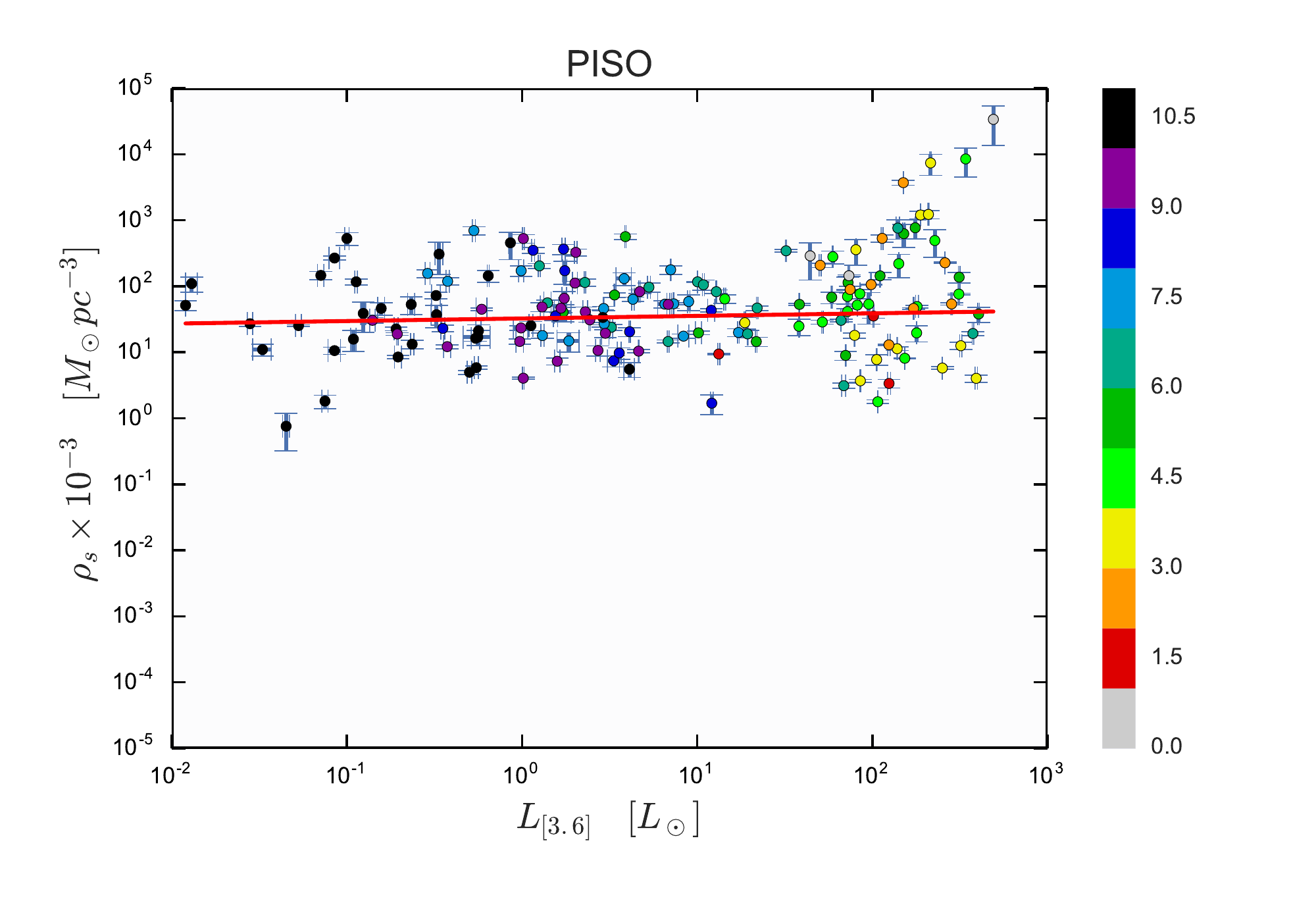}&
\includegraphics[width=3.25in]{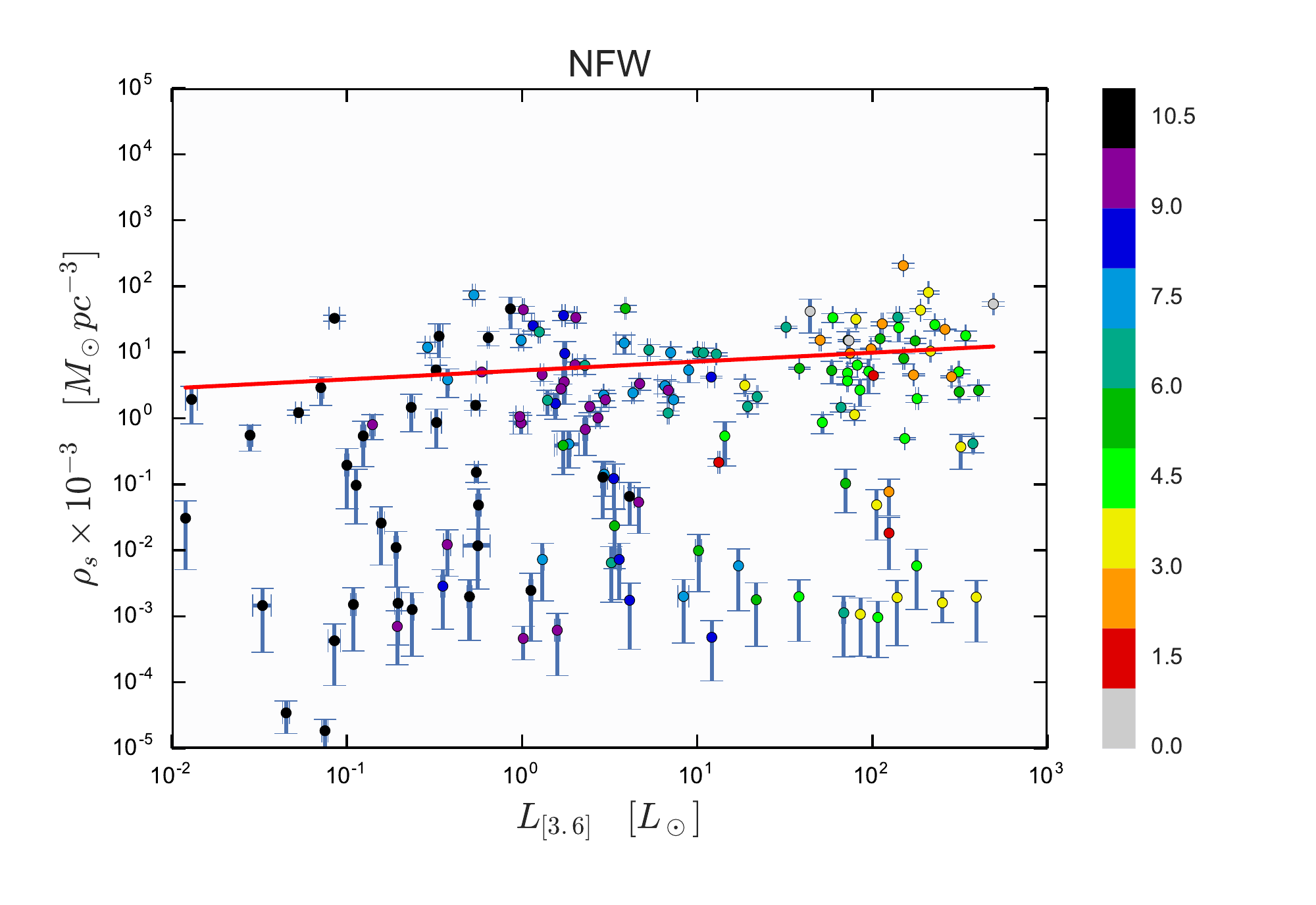}\\
\includegraphics[width=3.25in]{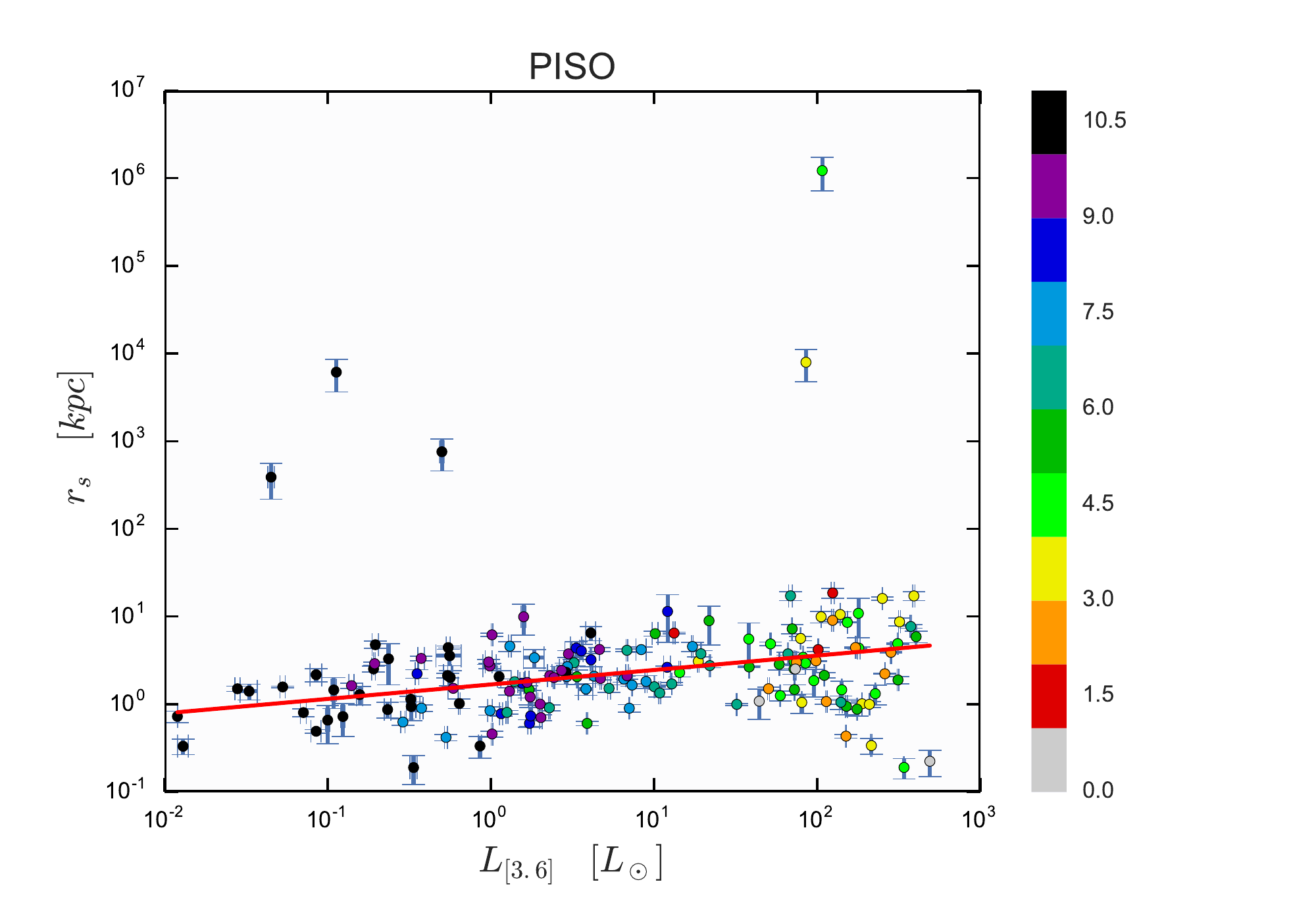}&
\includegraphics[width=3.25in]{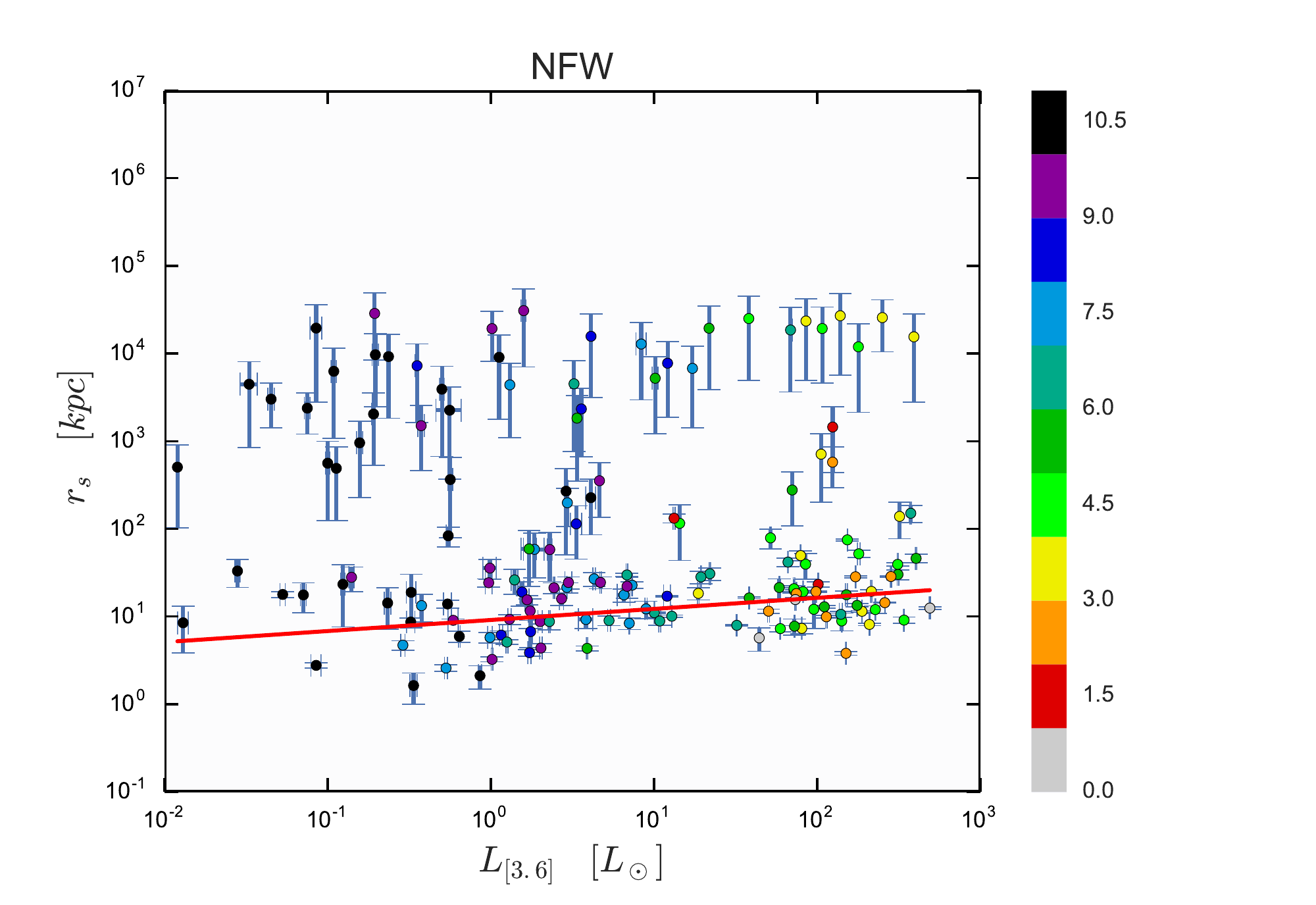} \\
\includegraphics[width=3.25in]{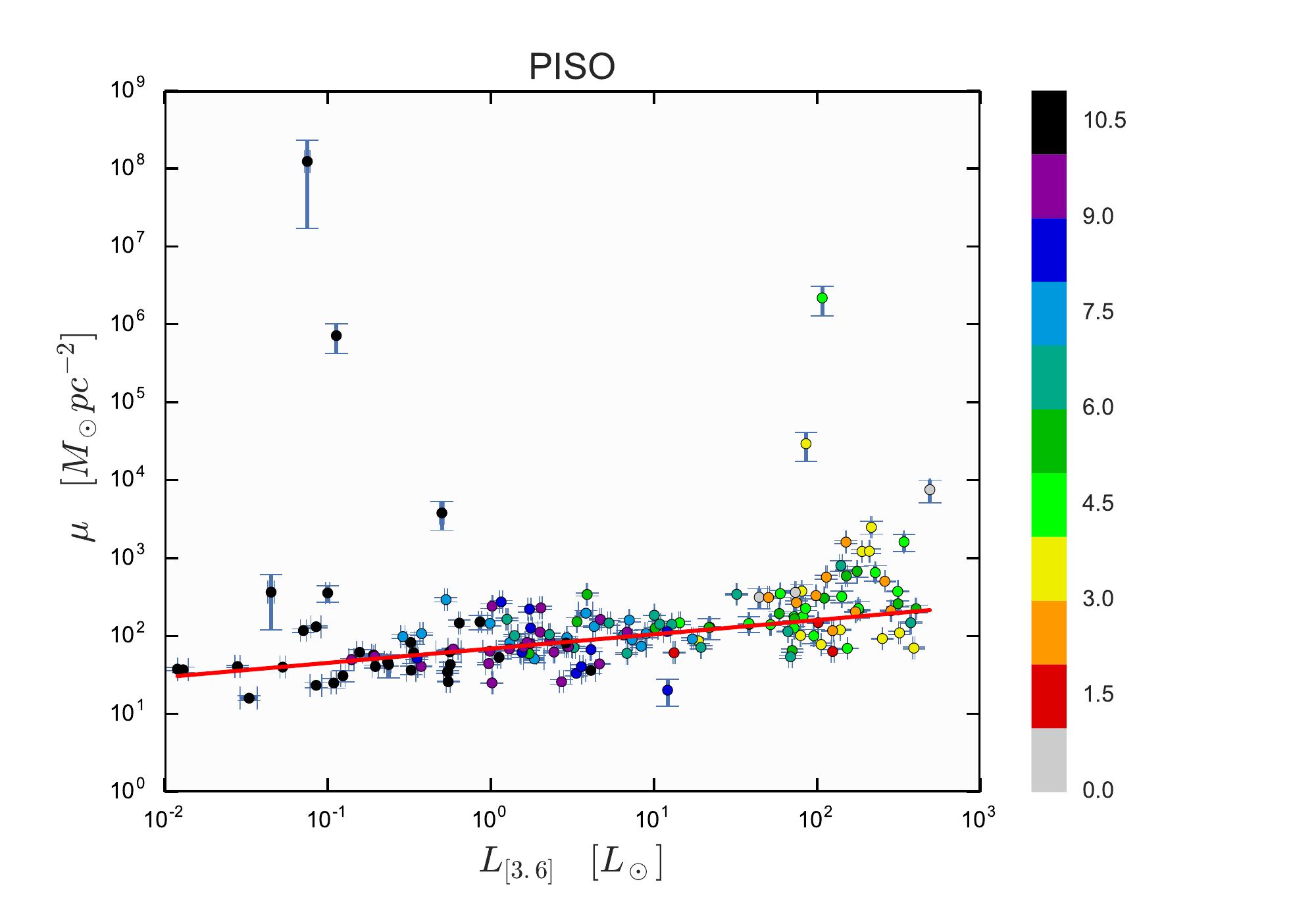}&
\includegraphics[width=3.25in]{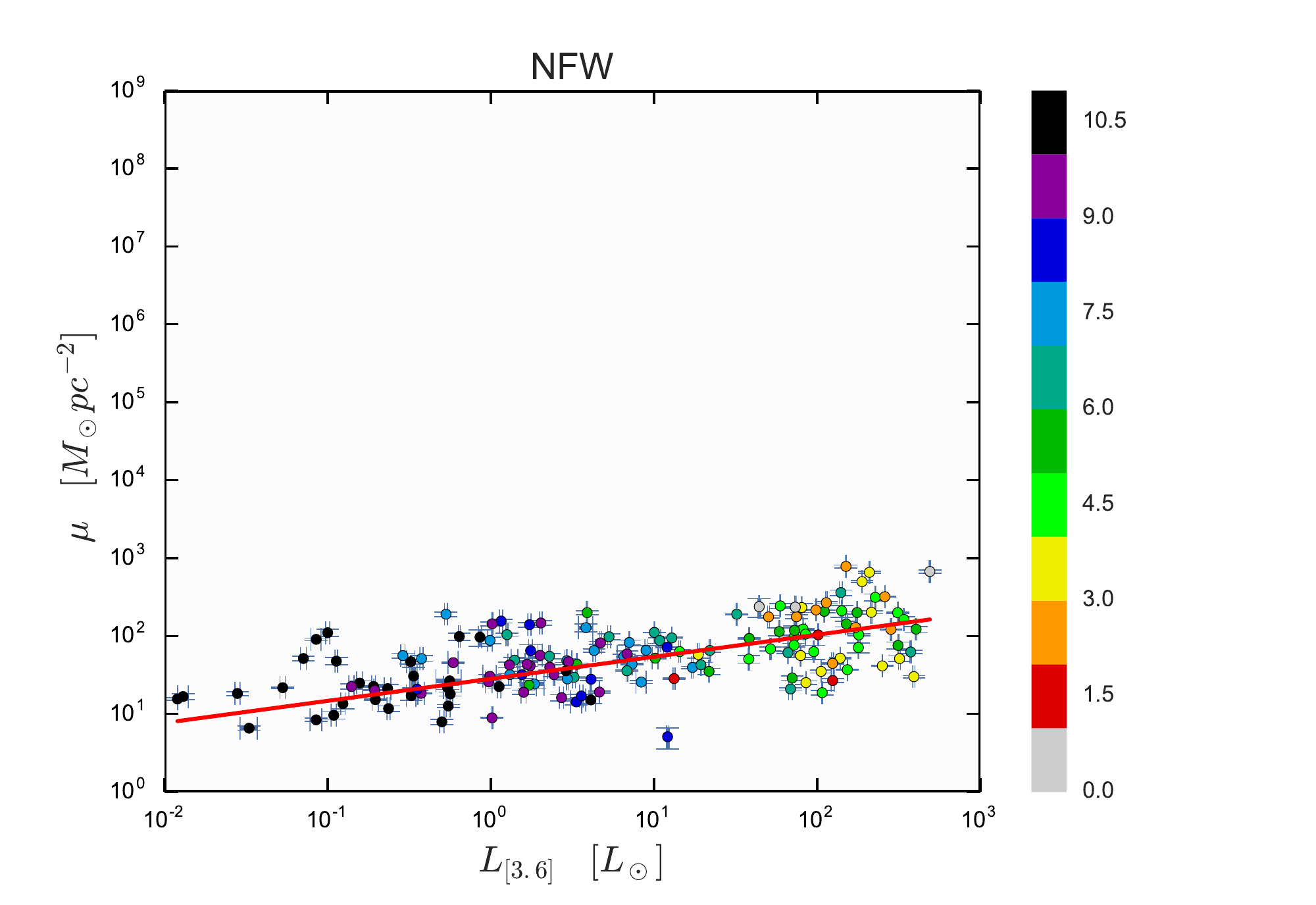} \\
\includegraphics[width=3.25in]{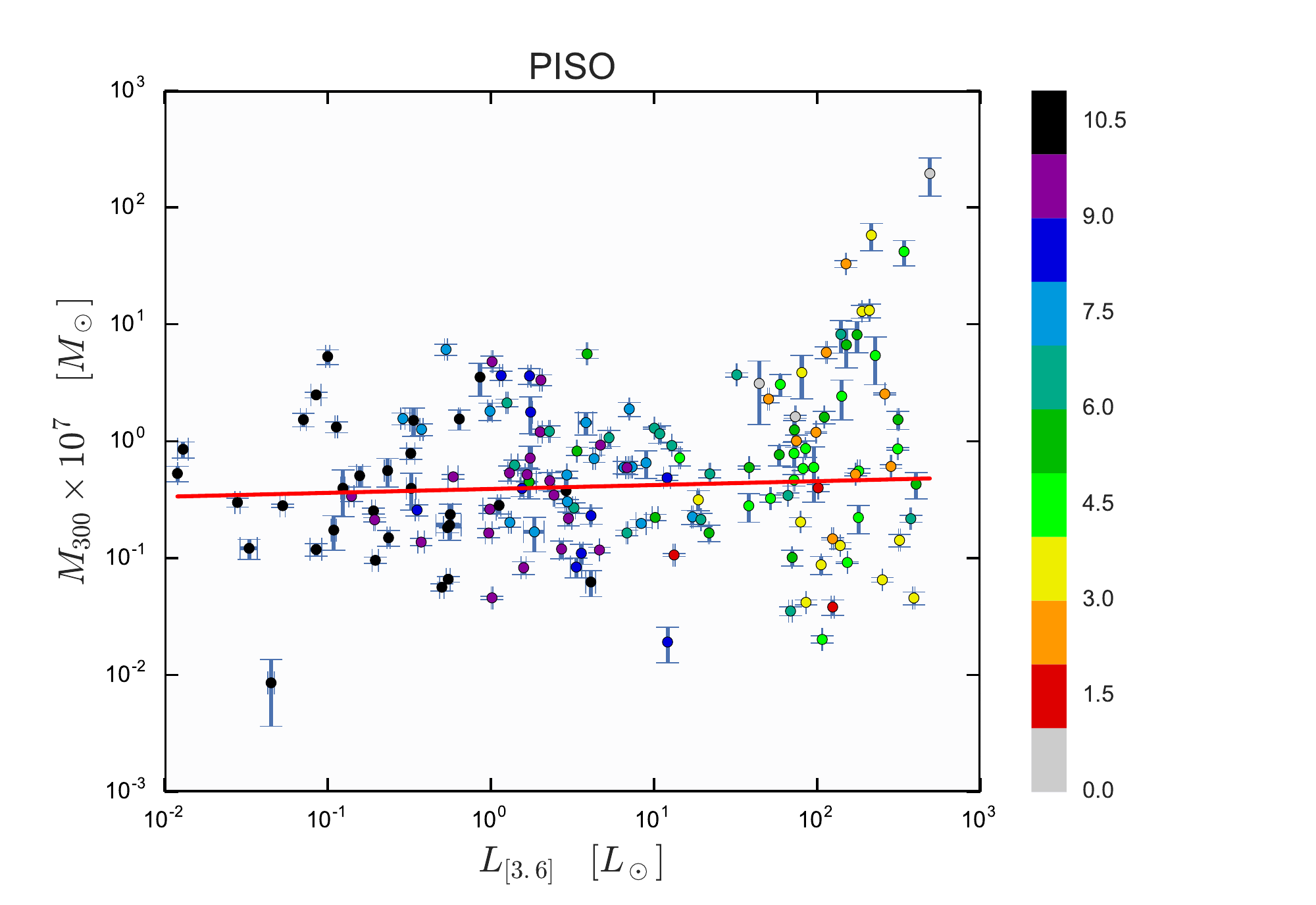}&
\includegraphics[width=3.25in]{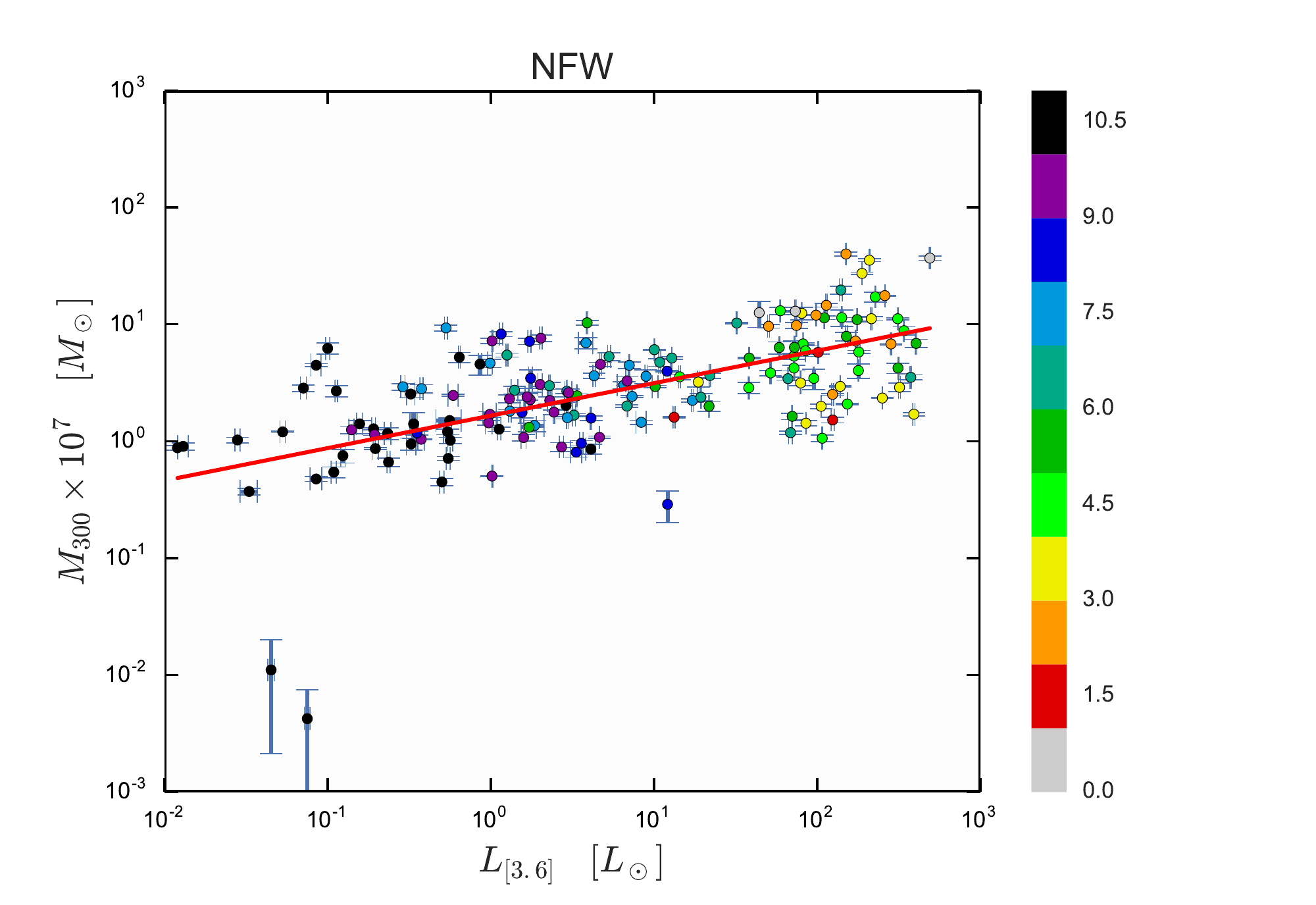} \\
\end{array}$
\end{center}
\caption{Shown are the characteristic volume density, the scale length, the characteristic central surface density and the mass within 300 pc as function of luminosity at 3.6 $\mu$m. Left panel PISO, right panel NFW.
Galaxies are colored by Hubble type with numbers from 0 to 11 corresponding to S0, Sa, Sab, Sb, Sbc, Sc, Scd, Sd, Sdm, Sm, Im, BCD, respectively. In all panels, solid lines show linear fits.}
\label{fig:ModelsPISO_NFW}
\end{figure*}

\begin{figure*}
\begin{center}$
\begin{array}{ccc}
\includegraphics[width=3.25in]{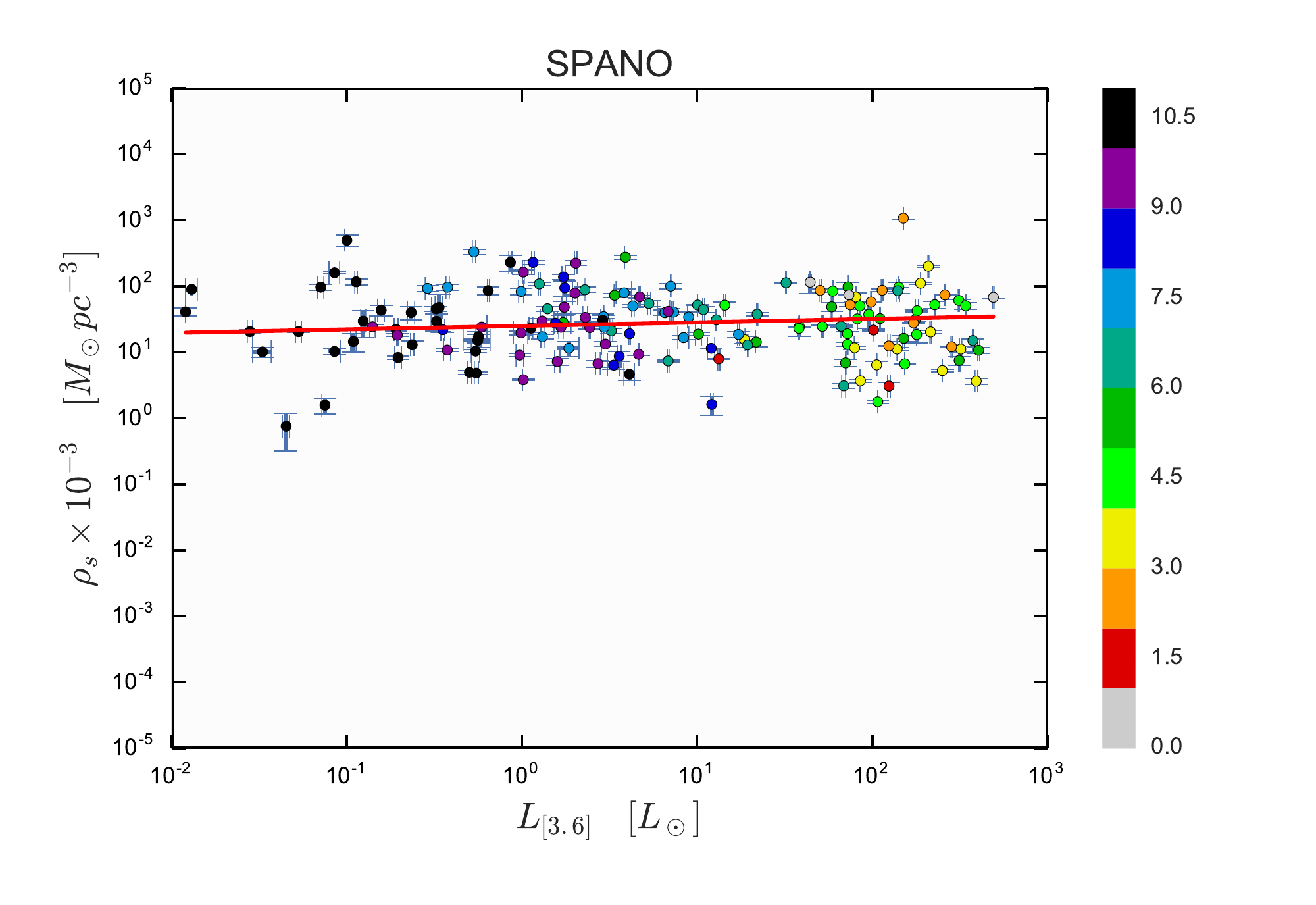}&
\includegraphics[width=3.25in]{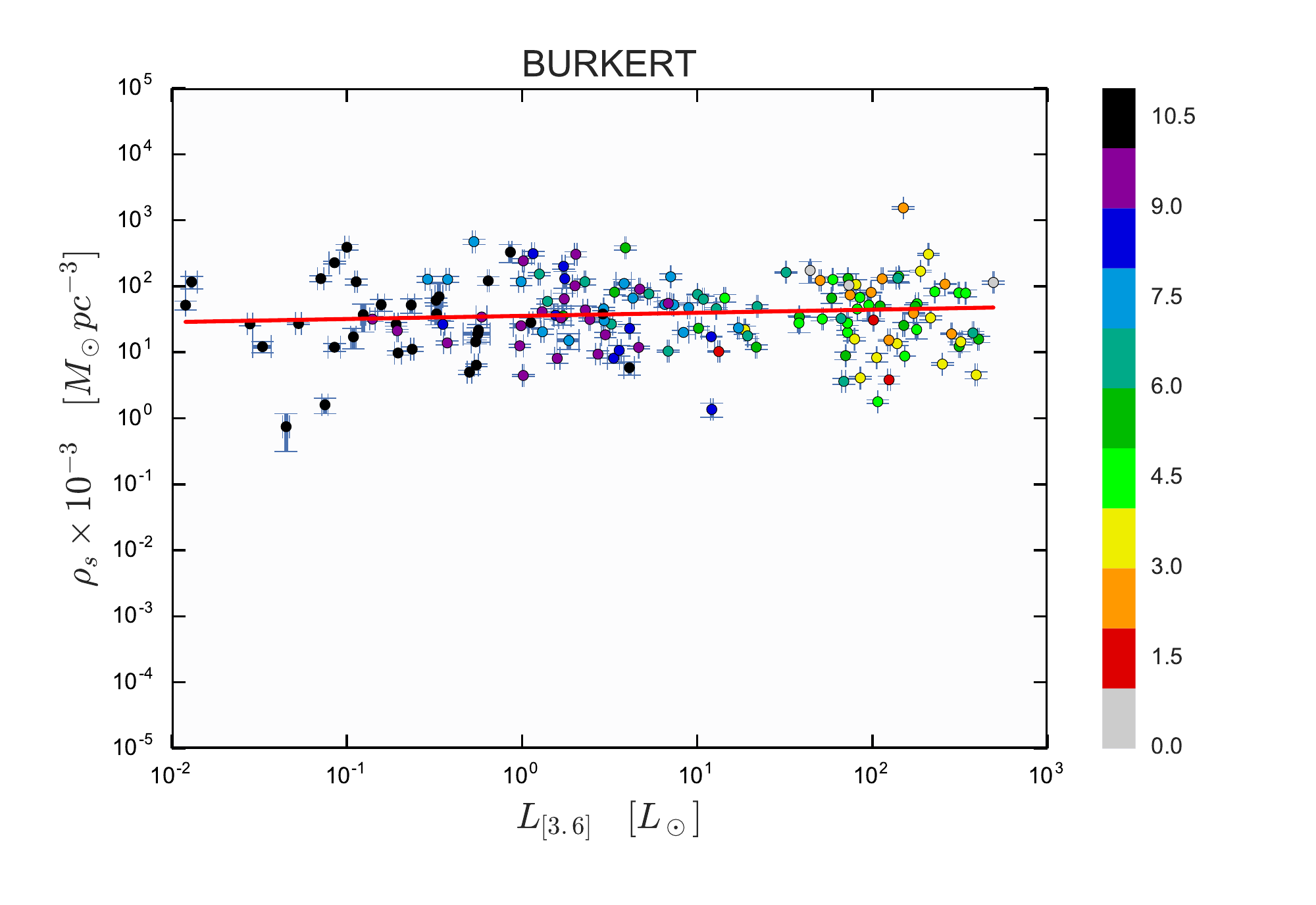}\\
\includegraphics[width=3.25in]{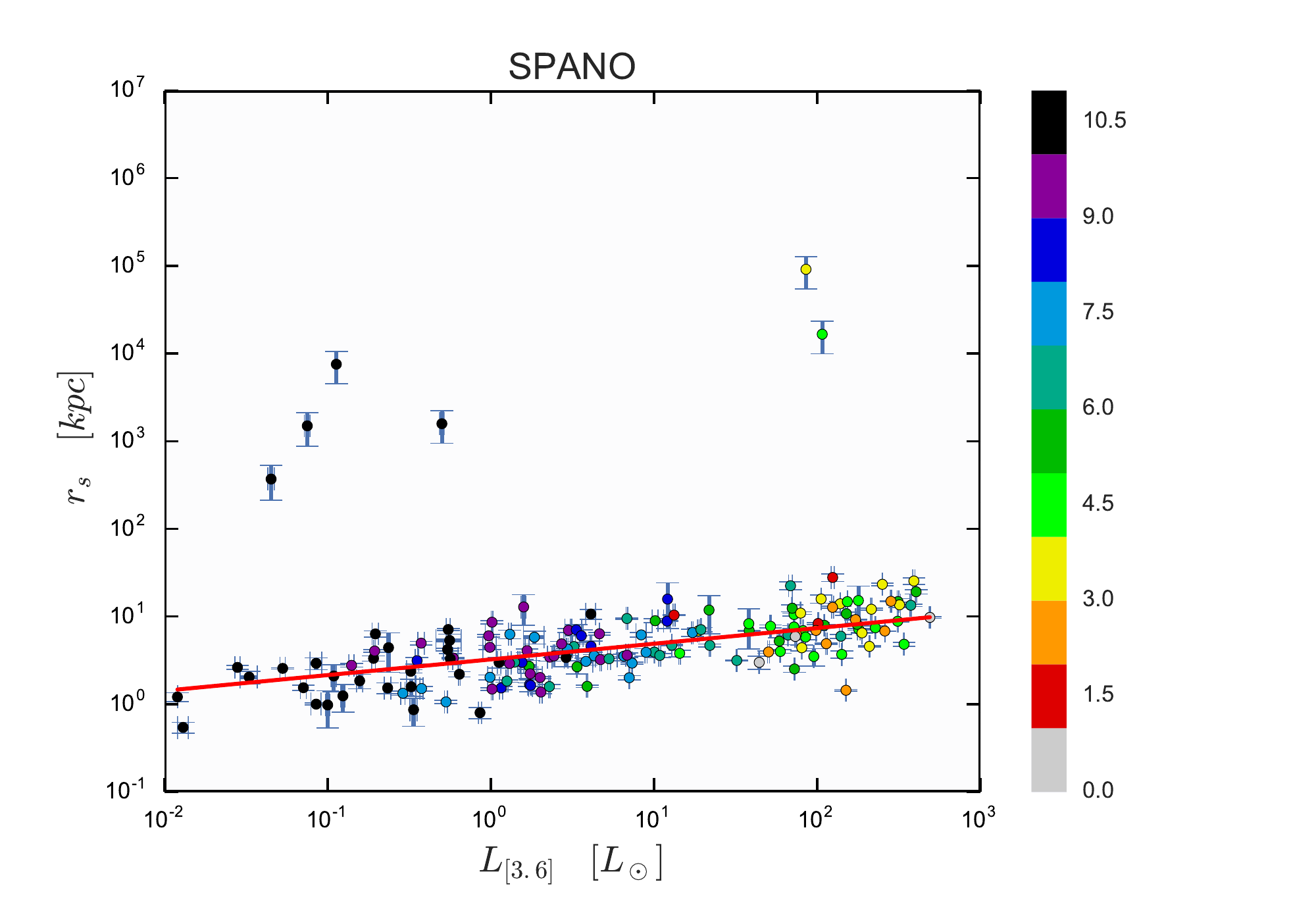}&
\includegraphics[width=3.25in]{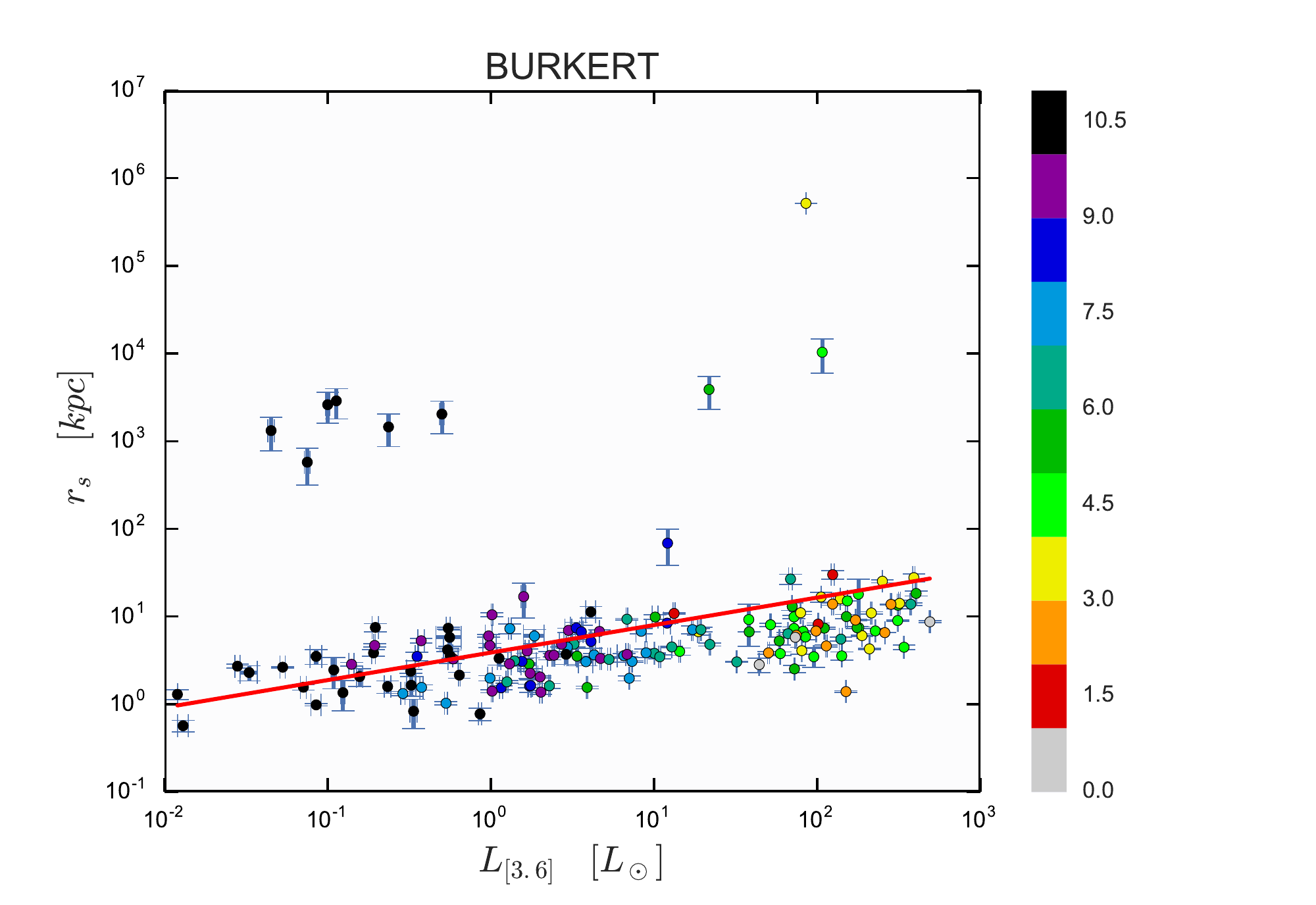} \\
\includegraphics[width=3.25in]{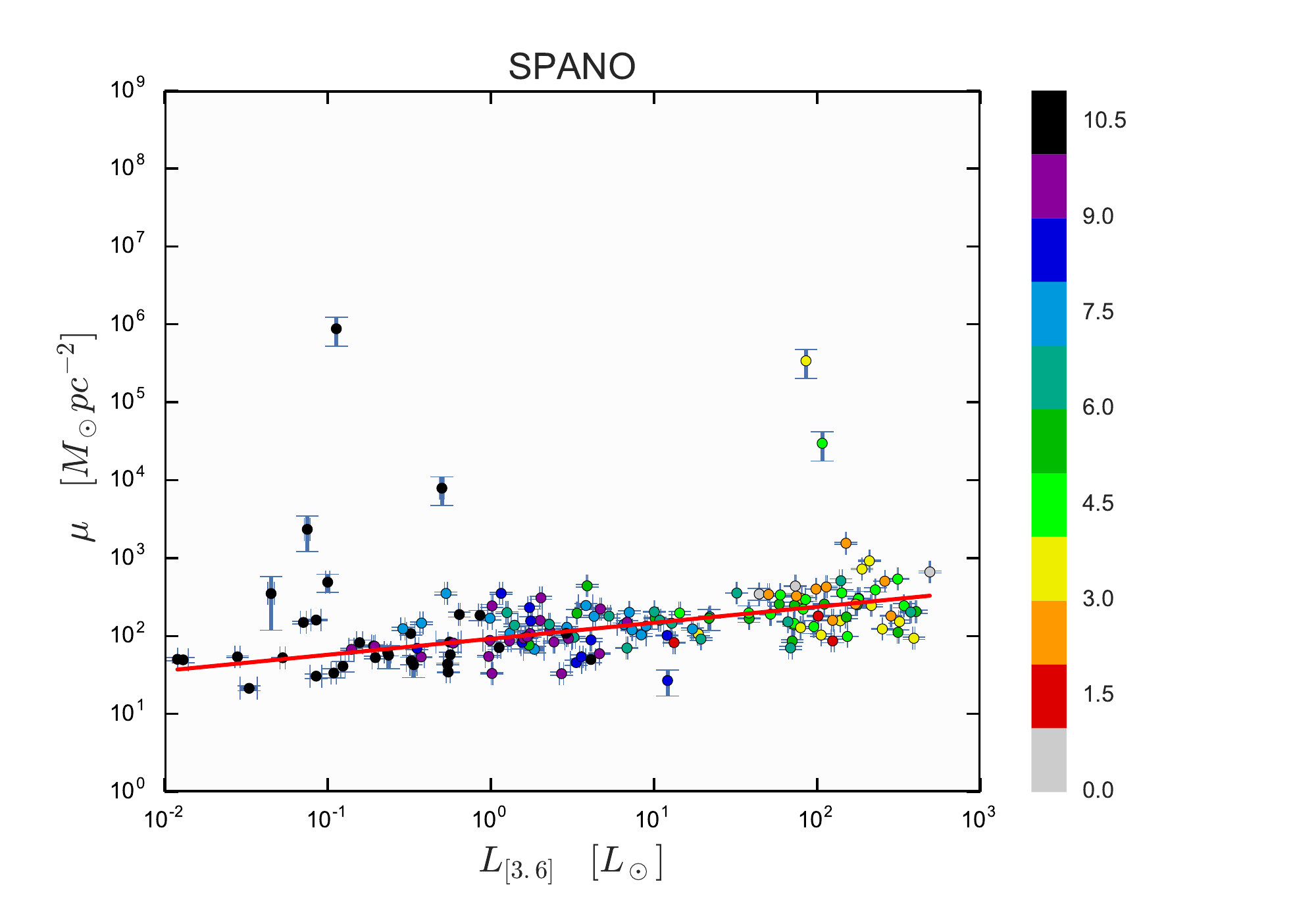}&
\includegraphics[width=3.25in]{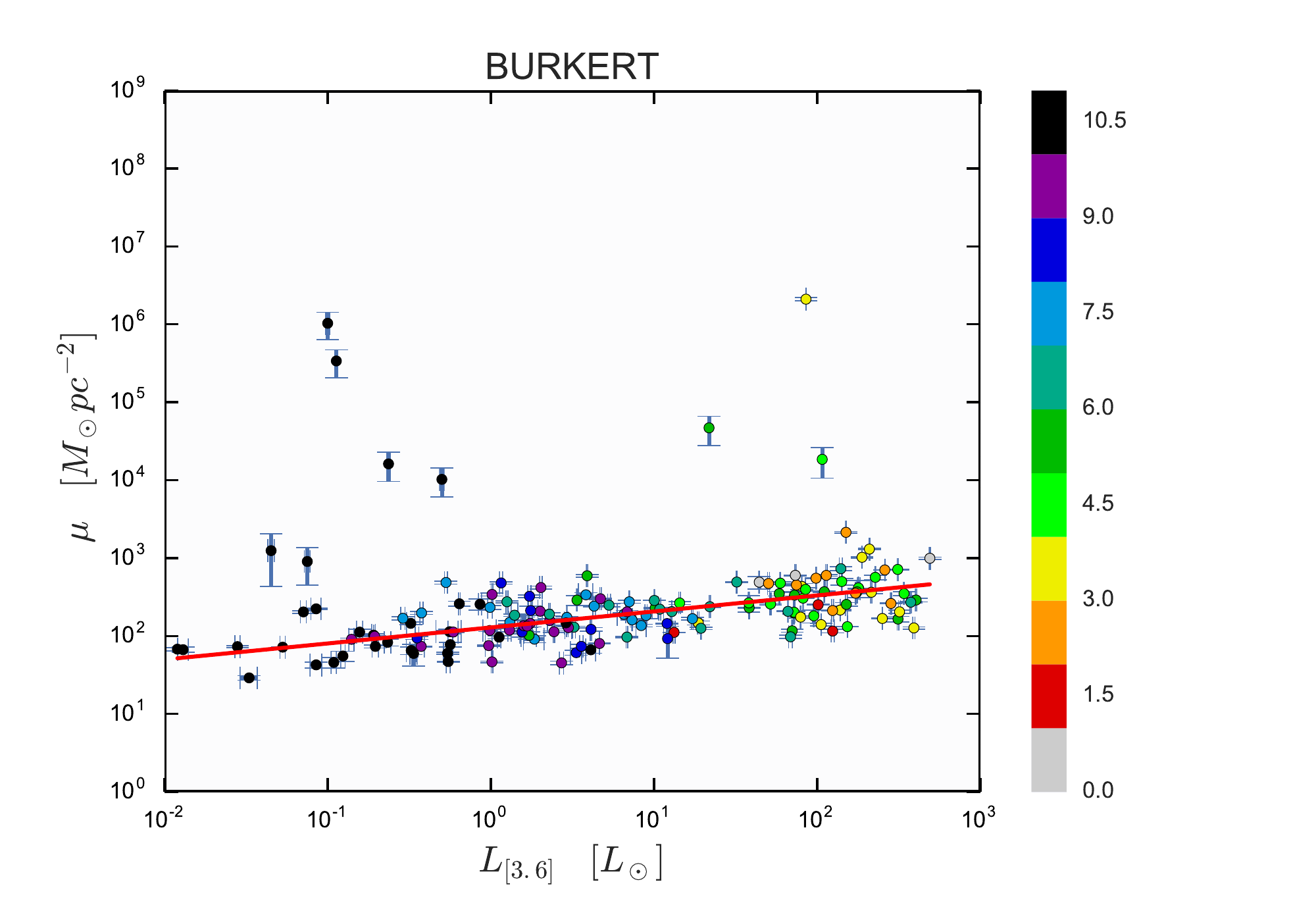} \\
\includegraphics[width=3.25in]{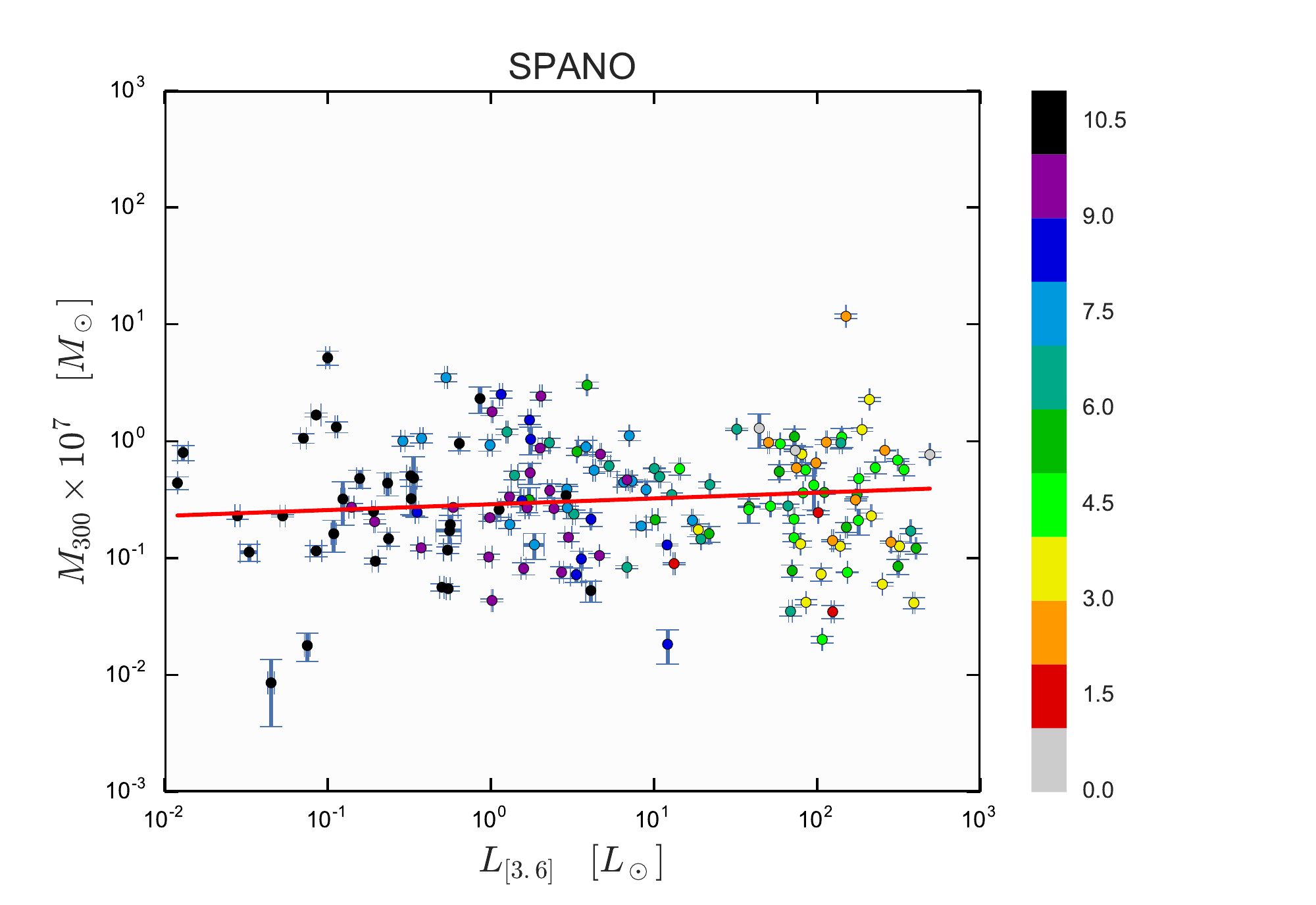}&
\includegraphics[width=3.25in]{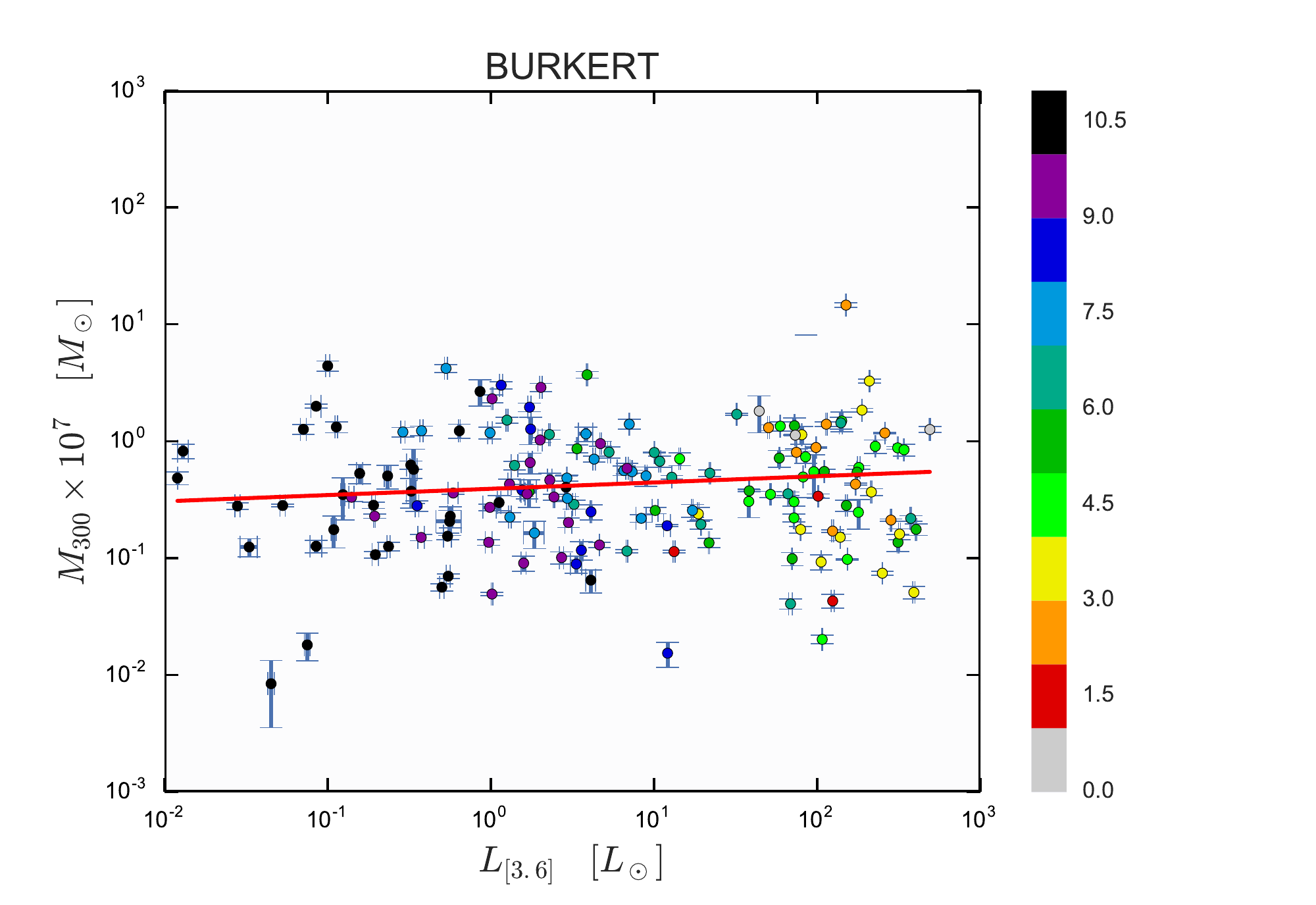} \\
\end{array}$
\end{center}
\caption{Shown are the characteristic volume density, the scale length, the characteristic central surface density and the mass within 300 pc as function of luminosity at 3.6 $\mu$m. Left panel Spano, right panel Burkert.
Galaxies are colored by Hubble type. In all panels, solid lines show linear fits. See caption of Fig 
\ref{fig:ModelsPISO_NFW}.}
\label{fig:ModelsSpano_Burkert}
\end{figure*}

\begin{figure}
\includegraphics[width=3.25in]{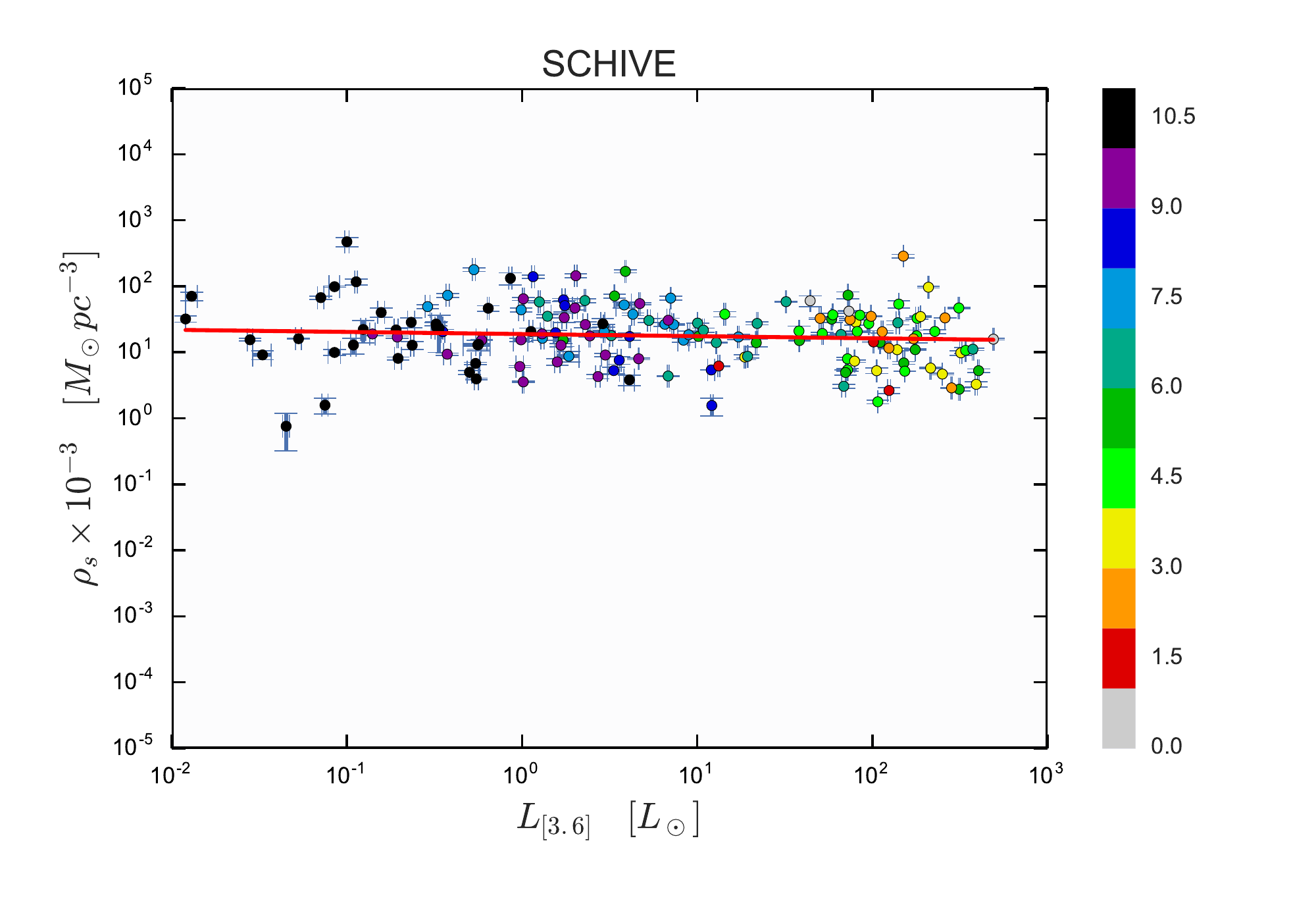} \\
\includegraphics[width=3.25in]{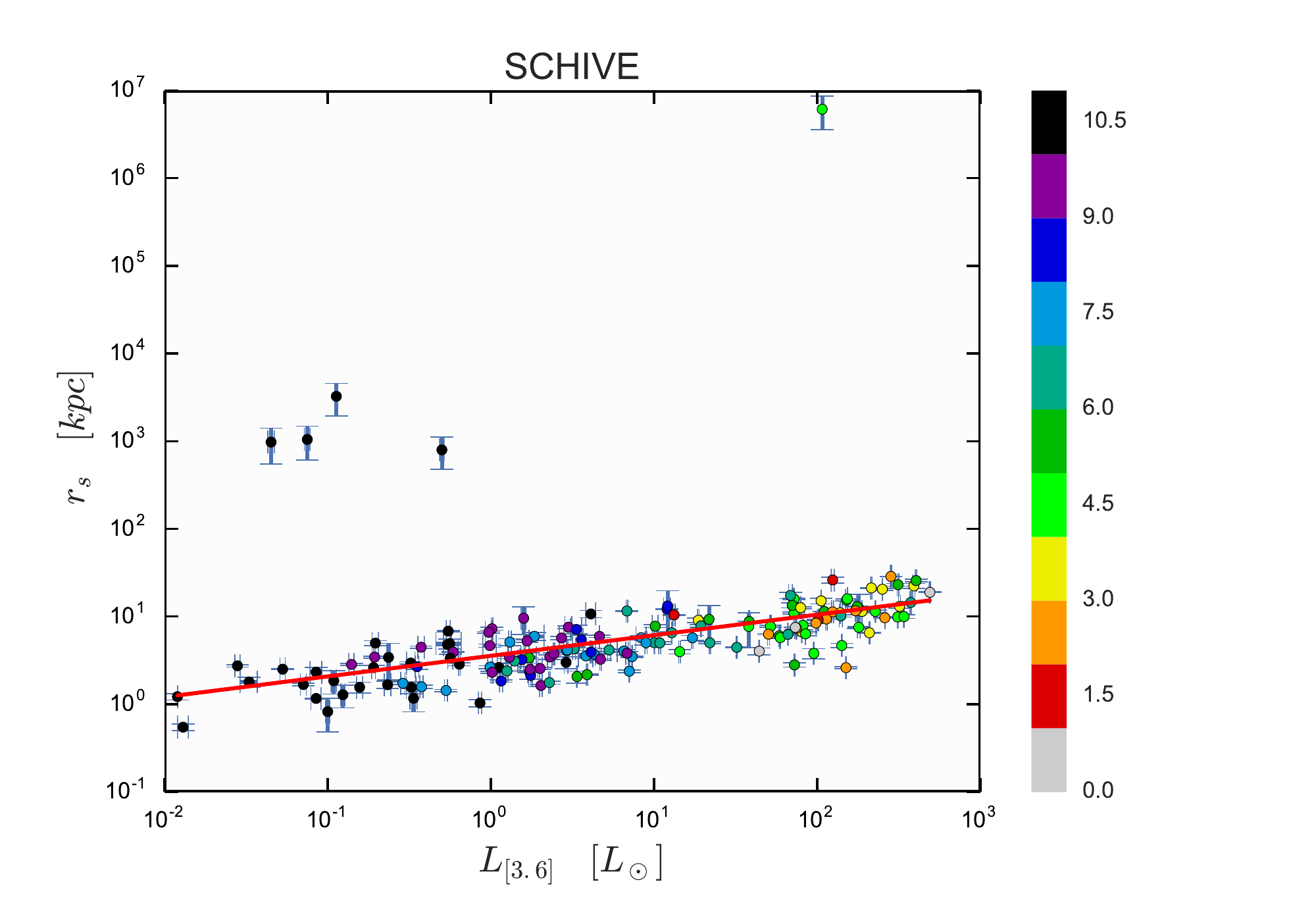} \\
\includegraphics[width=3.25in]{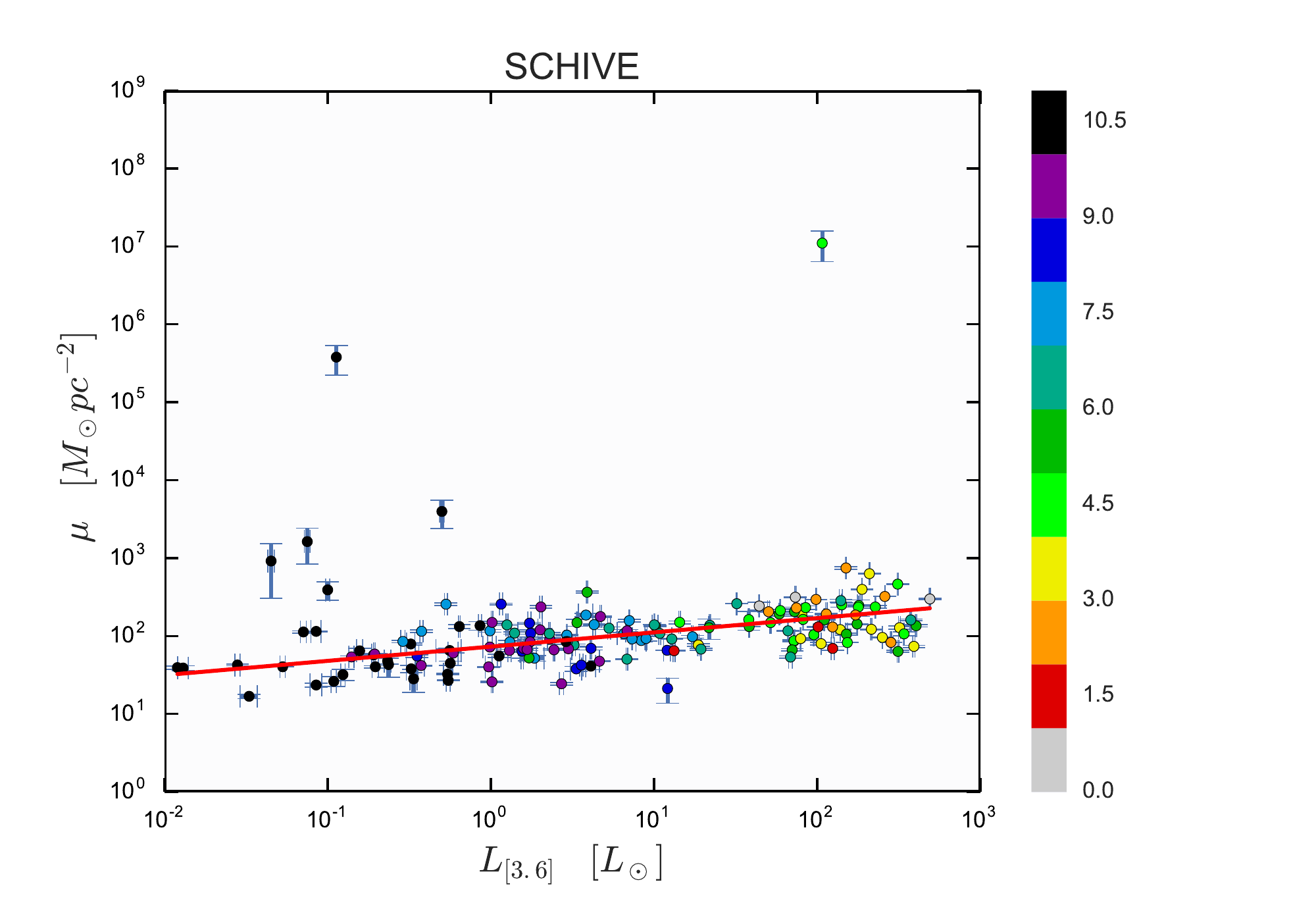} \\
\includegraphics[width=3.25in]{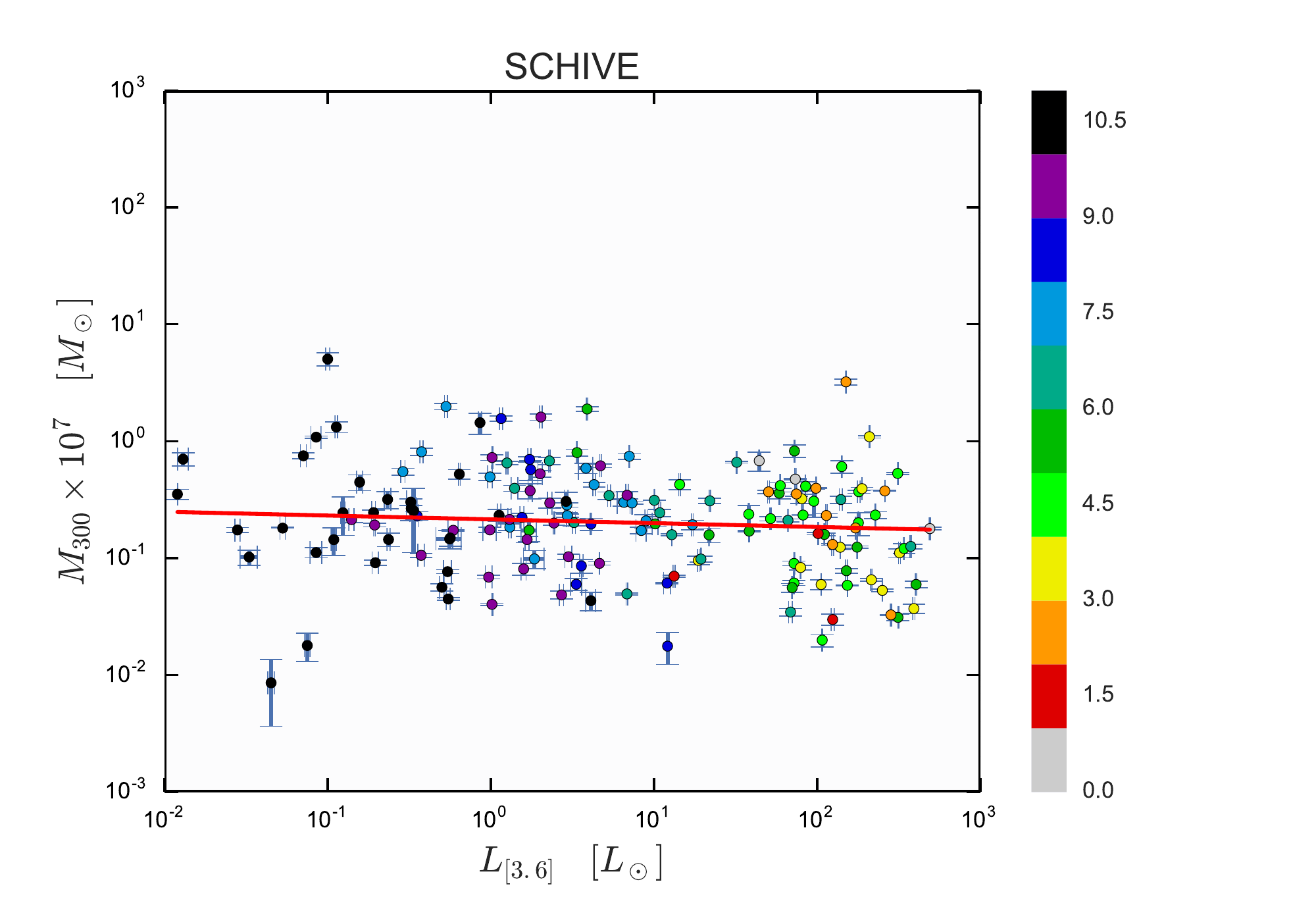}
\caption{Shown are the characteristic volume density, the scale length, the characteristic central surface density and the mass within 300 pc as function of luminosity at 3.6 $\mu$m. Schive model.
Galaxies are colored by Hubble type. In all panels, solid lines show linear fits. See caption of Fig 
\ref{fig:ModelsPISO_NFW}.}
\label{fig:ModelsSchive}
\end{figure}

\begin{figure*}
\begin{center}$
\begin{array}{ccc}
\includegraphics[width=3.25in]{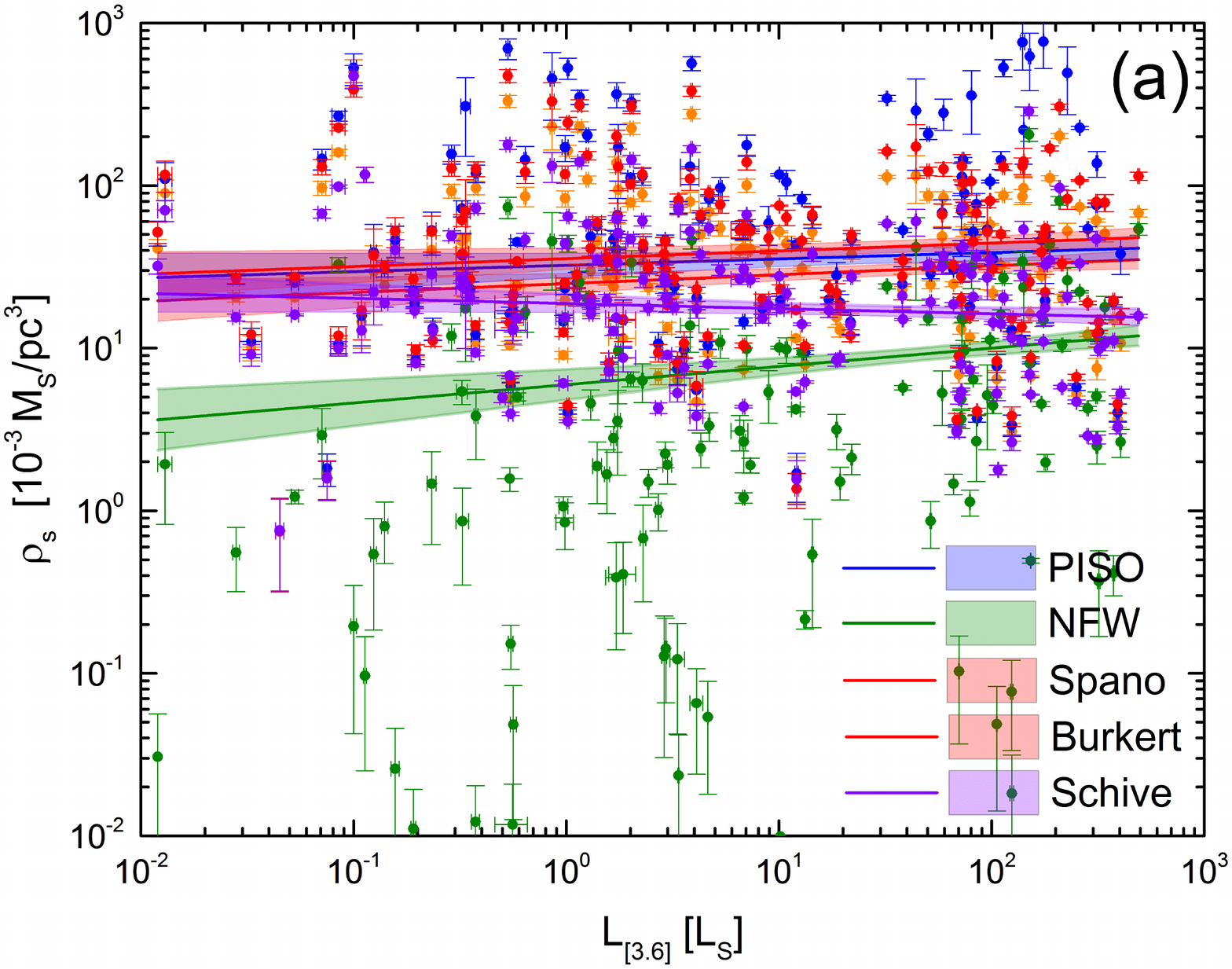}&
\includegraphics[width=3.25in]{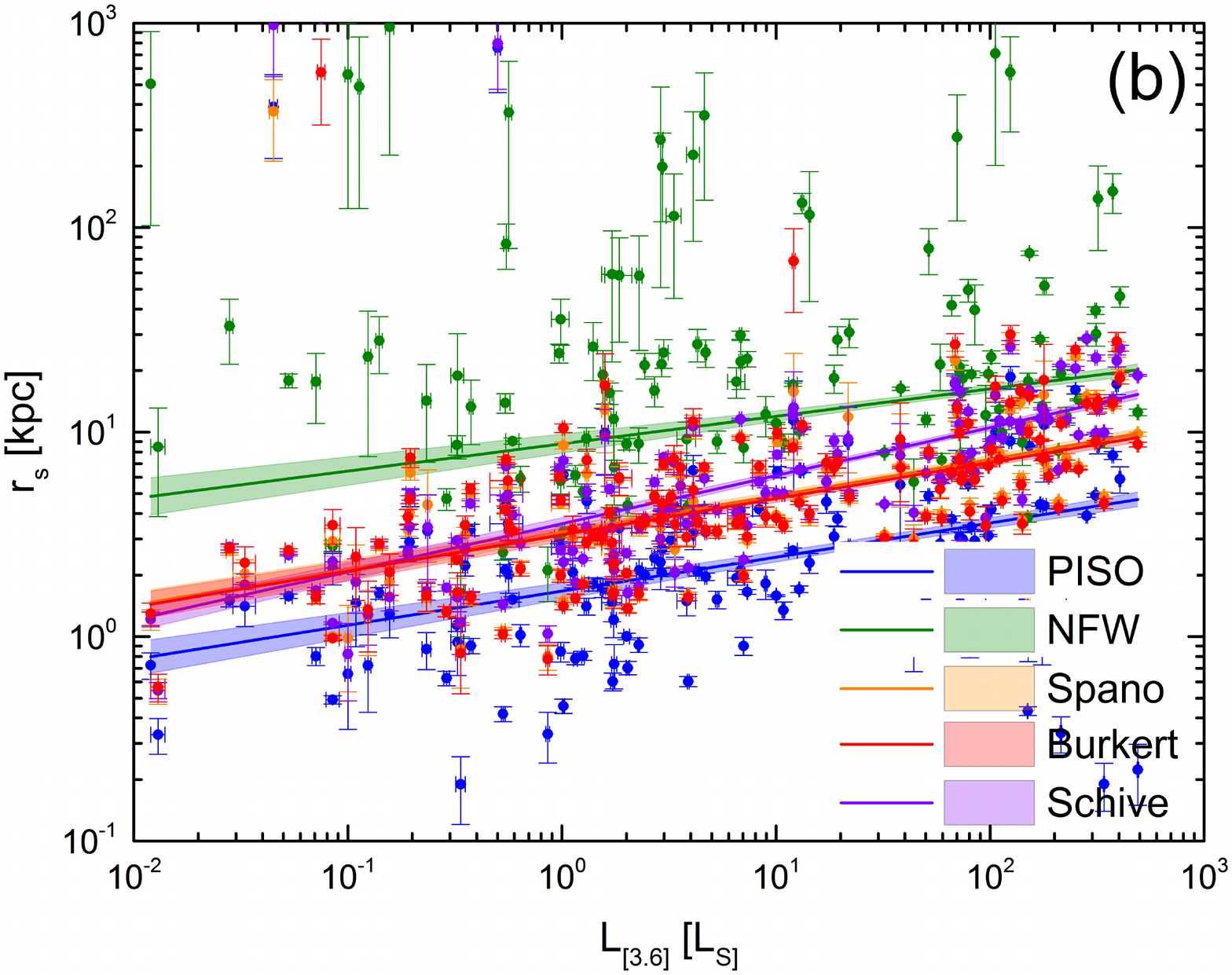}\\
\includegraphics[width=3.25in]{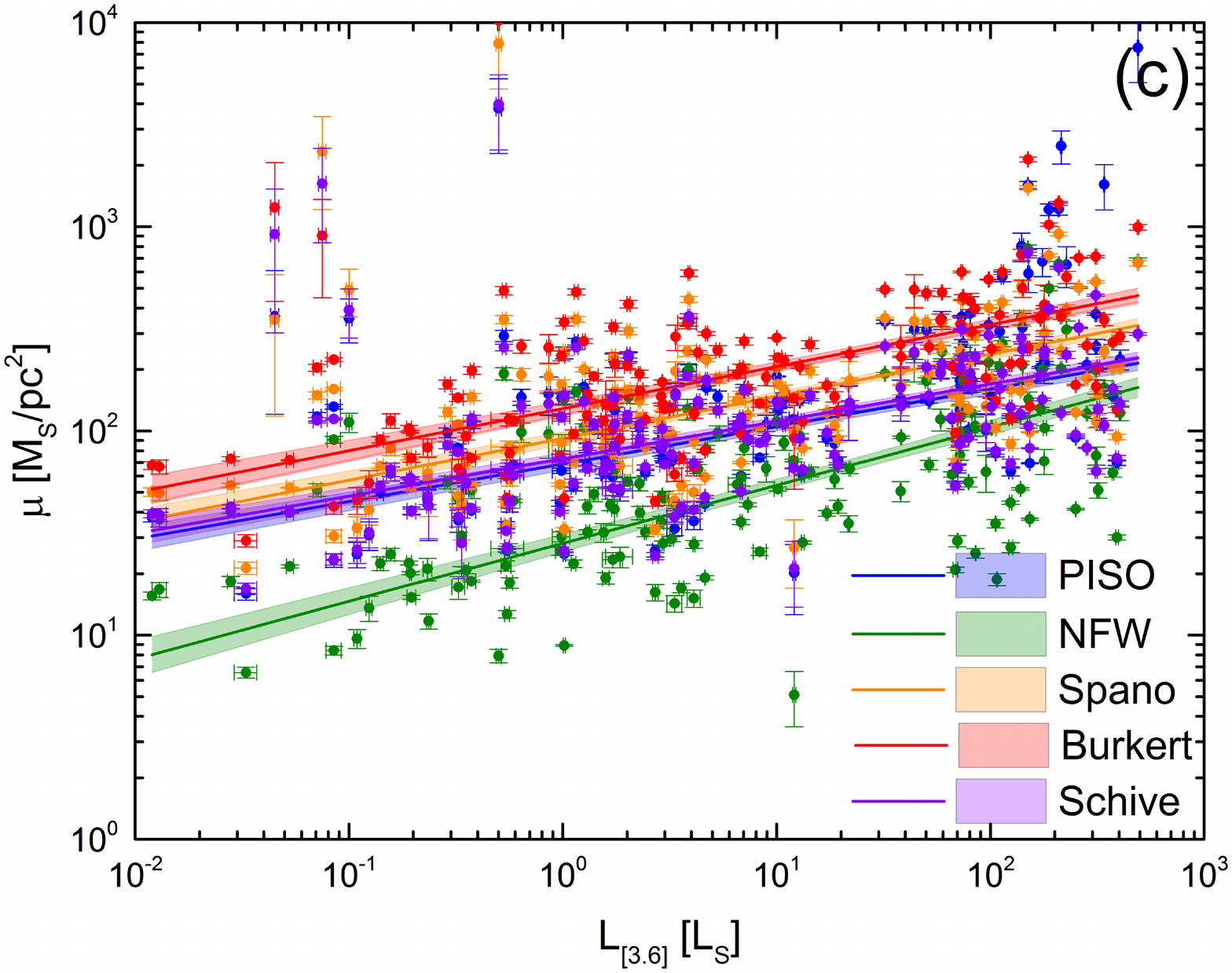}&
\includegraphics[width=3.25in]{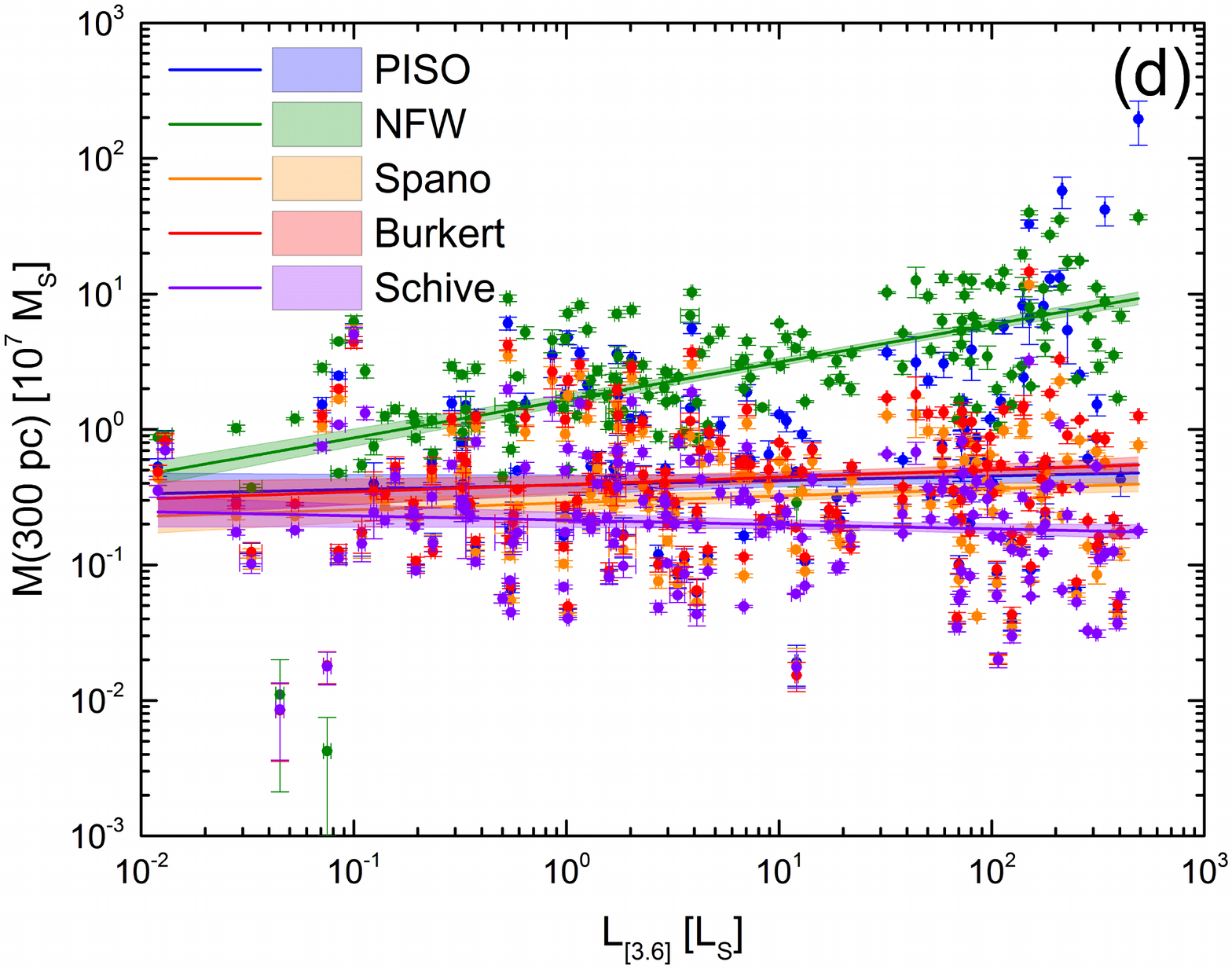} \\
\end{array}$
\end{center}
\caption{Same as Figs. \ref{fig:ModelsPISO_NFW}--\ref{fig:ModelsSchive} however we are illustrating all the models together and also displaying the fit and $1\sigma$ bands of the linear (in Log-Log scale) fit.}
\label{fig:allModels}
\end{figure*}

\begin{figure}
\begin{center}
(a)
\includegraphics[width=3.25in]{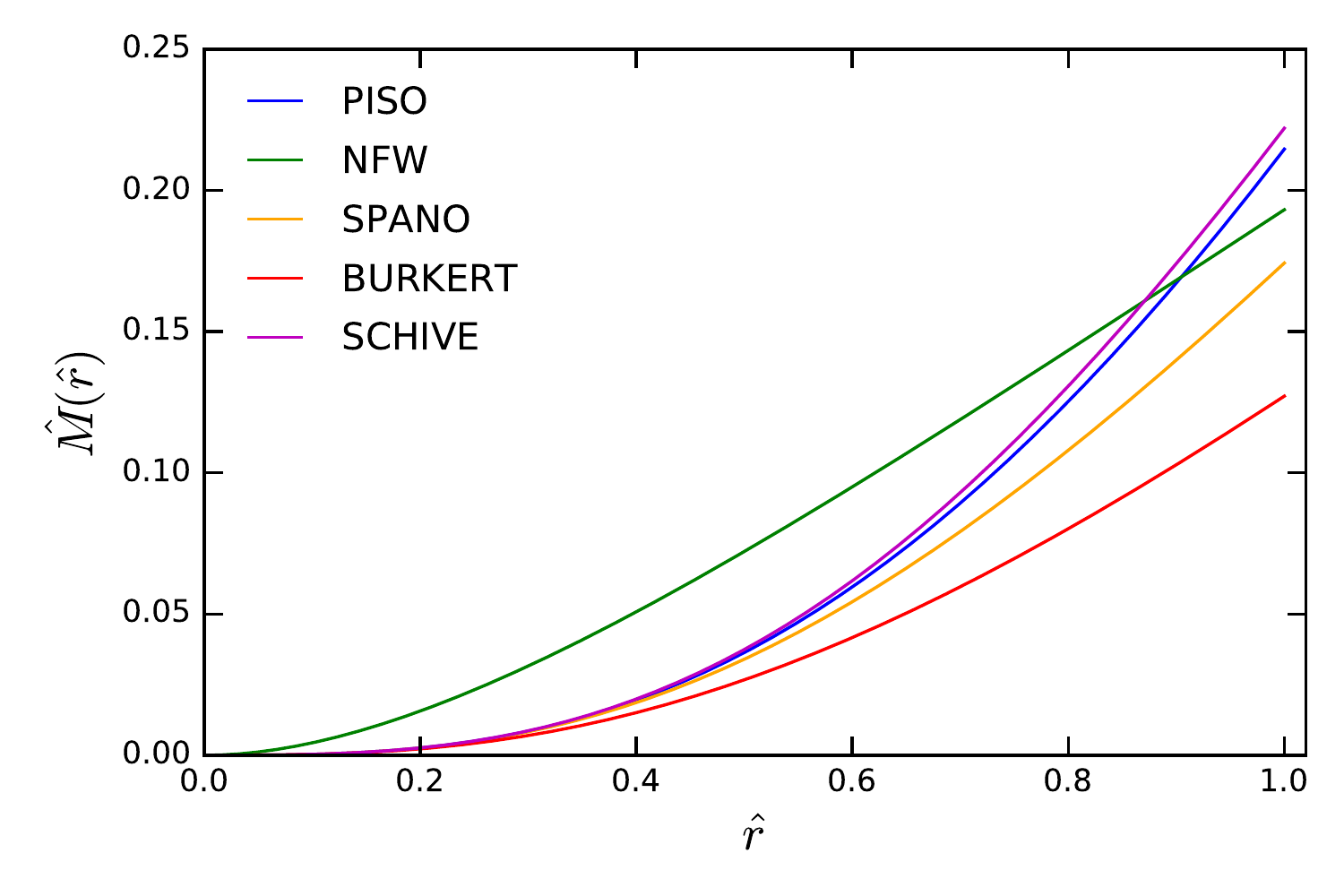} \\
(b)
\includegraphics[width=3.25in]{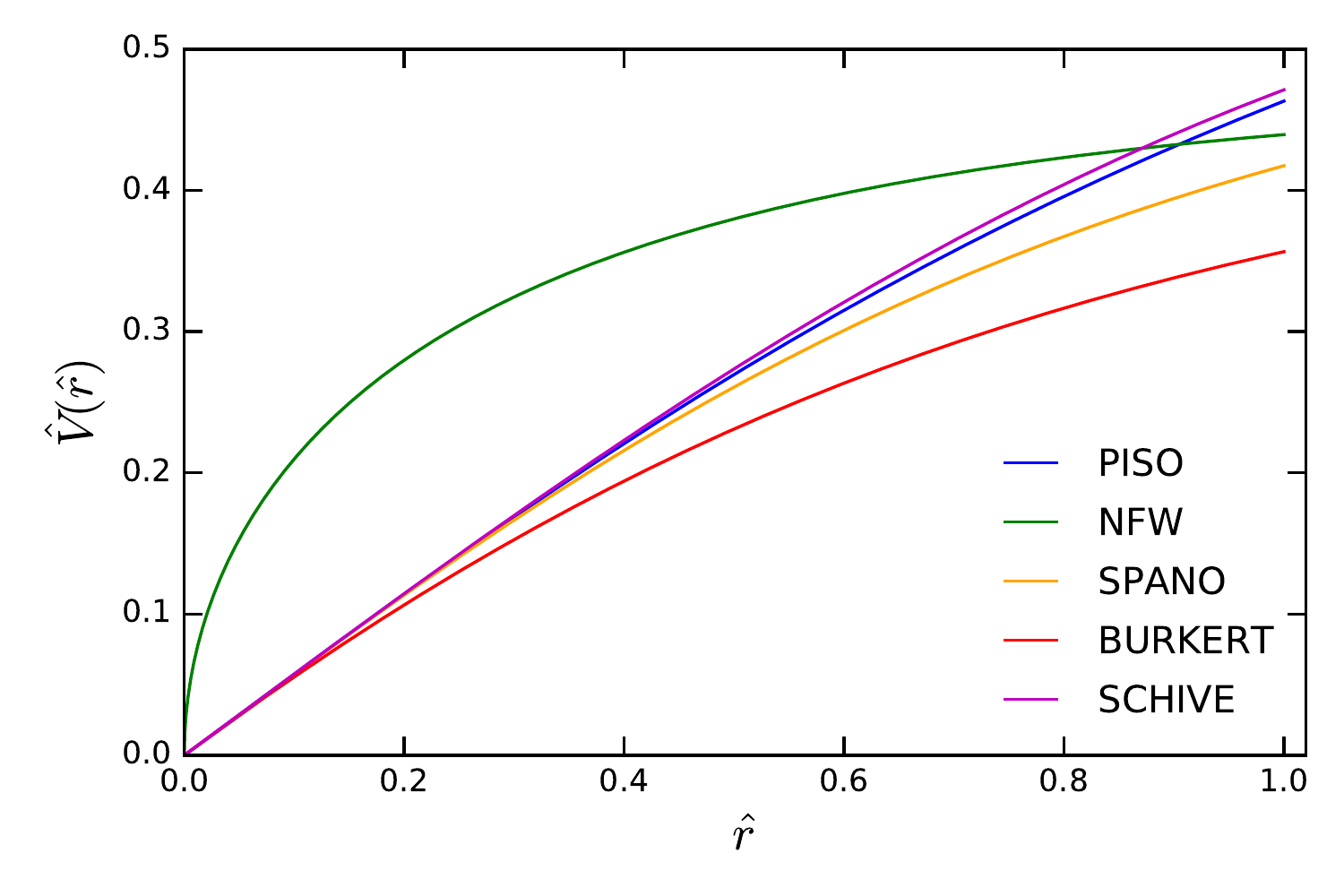}
\end{center}
\caption{In (a) mass model for PISO, NFW, Spano, Burkert and Schive DM models. $\hat{M}$ and $\hat{r}$ are dimensionless quantities defined in Appendix \ref{Ap}. In (b) velocity model for the same DM models as in (a). $\hat{V}$ is also a dimensionless velocity defined in Appendix \ref{Ap}.
}
\label{fig:allModels_massvel}
\end{figure}

%
%

\section{CONCLUSIONS} \label{Conclusions}
In this work we have performed a non parametric reconstruction of galactic rotation curves by using the LOESS+SIMEX technique 
as a method to discern between seven different DM models. Our main objective is to explain the DM contribution within the rotation 
curves of galaxies. In order to show the efficiency of LOESS+SIMEX and to report a quantitative result we have obtained the normalized area 
between the reconstructed curve and the fitted rotation curve, including also the area between 1$\sigma$ bands. Since we are comparing 
different models with different number of parameters, we use the AIC and BIC information criteria methods in order to assess the model fitness 
while penalizing the number of estimated parameters and, at the same time, giving support to LOESS+SIMEX results. 
Thus, to perform the model selection we use three conditions as a criterion to accept or reject models: 
\begin{itemize}
\item $B_{\rm BIC}$: the best BIC value, 
\item $B_{\rm DIST}$: the area between the reconstructed curve and the rotation curve obtained with the best fit parameters, and
\item $B_{\rm D1\sigma}$: the area between the 1$\sigma$ reconstruction band and the 1$\sigma$ model fitting band. 
\end{itemize}

We asign to each model a numerical code (1--4), which was ordered from the lowest value up to the highest one, according to any 
of the above requirements, indicating in the first place as the most preferred model by the data.

Furthermore, the DM models were separated in two groups (core + cusp and core + cusp + fuzzy models). In the first group 
we found that $\sim 44.32 \% $ of the galaxies satisfy the three conditions, where Spano is the most favored model with $ \sim 53.85 \% $ 
while the NFW model is the least favored one, because is appearing in the last place according to the three conditions 
with the $ \sim 59 \% $.
From the second group we found that Schive satisfies the three conditions in a $44\%$, 
while NFW turns out to be the rejected model with the $ 21.6\% $ of the cases.

By performing a comparison between these three conditions, we are able to do a Core and Cusp classification for some of the galaxies which 
we have worked with. The DDO 064, DDO 168, F 571-v1, KK 98-251, NGC 6789, UGC 05986 and UGC 06399 are 
core type galaxies. On the other hand, NGC 0247, NGC 3741, UGC 05716, UGC 08490 and UGC 12732 can be classified as cusp galaxies. 
In some cases, when the FUZZY models were selected by one or more of the criteria used here, the corresponding transition radius 
turned out an important factor for the classification. The cases when $r_{\epsilon} \sim r_{\rm min}$ pointing out to NFW as the best 
model to explain the structure of the galaxy; this showed that the LOESS+SIMEX method can be useful as technique of models selection.

To complement our analysis, by using a MCMC method, we have computed the haloes characteristic surface density, $\rho_s r_s$ for each one
of the DM models and also the DM mass within 300 pc. We obtained that there is a common mass for spiral galaxies 
of the order of $10^7$ M$_\odot$, result that is in agreement with ones of dSph Milky Way satellites. 
This would give a central density for the halo of $\sim 0.1$ M$_\odot$ pc$^{-2}$ regardless the dark matter model. One extra result consistent 
with our findings is that there is a constant characteristic volume density of DM haloes: $\rho_s$ is constant 
for the core models we analyzed. 
Moreover, we have found that characteristic central surface density is not constant for the SPARC galaxies, 
which is in tension with previous results, but in favor of core models which is in agreement with the non-parametric-parametric analysis.

Finally, we conclude that, in the study of galactic rotation curves, the non parametric LOESS+SIMEX method in combination with standard 
fitting methods turns out to be a reliable tool to perform DM models selection.


\section*{Acknowledgements}
L.M.F.H. acknowledges fruitful advice by S. McGaugh. All authors thank SNI-CONACYT for partial support. A.M. acknowledges support from postdoctoral grants from DGAPA-UNAM. M.A.R.M. acknowledges financial support from CONACyT project 283151, from ``Fondo Sectorial de Investigaci\'on para la Educaci\'on''. All authors also thank to Juan Carlos Hidalgo, Nora Bret\'on and Luisa Jaime  for reading the manuscript and for their helpful comments and suggestions.





\bibliography{biblio}
\bibliographystyle{mnras}



\appendix

\section{CORE, CUSP AND SFDM MODELS} \label{Ap}

Here we give a brief summary of the most common models in the literature reported in \cite{piso,NFW,Burkert,Spano,Schive1,2018MNRAS.475.1447B}. 
In general a rotation curve can be described by a general expression, see \cite{Aspeitia-Rodriguez} for more details
\begin{equation}
\label{GVelocity}
V^2_{model} (r) = 4\pi G r_{model} \mu_{model}\hat{V}^2_{model} (\hat{r}),
\end{equation}
where $\hat{V}(\hat{r})$ is a dimensionless function given by
\begin{equation}
\label{GVelocity2}
\hat{V}(\hat{r}) =\sqrt{\frac{M(\hat{r})}{\hat{r}}} ;
\end{equation}
$\hat{r}=r/r_{model}$ with $r_{model}$ being a length scale  parameter for every model and $\mu_{model} = \rho_{model} r_{model}$, which is almost a constant when we  fit observations but take a different value for each model 
\citep{Aspeitia-Rodriguez}.
The dimensionless mass at a given radius $\hat{r}$ is given by
\begin{equation}
M(\hat{r})= 4\pi \int_0^{\hat{r}}  dr' r'^2 \rho_{DM}(r') \label{massDM} .
\end{equation}

\begin{enumerate}[(a)]
\item {\bf Pseudo isothermal profile.}
The DM density profile is  written as:
\begin{equation}
\rho_{\rm PISO}(r)=\frac{\rho_p}{1+(r/r_p)^{2}}. \label{PIP}
\end{equation}
See \cite{piso} for more details. On the other hand, we can write the Eq. (\ref{GVelocity}), for Piso model as
\begin{equation}
\label{VPiso}
\hat{V}^2_{\rm p}(\hat{r}_p)=\frac{\hat{r}_p-\arctan(\hat{r}_p)}{\hat{r}_p}\, ,
\end{equation}
where $\hat{r}_p=r/r_p$.

\item {\bf Navarro-Frenk-White profile.}
Another interesting case (motivated by cosmological $N$-body simulations) is the cuspy NFW density profile which is given by \cite{NFW}:
\begin{equation}
\rho_{\rm NFW}(r)=\frac{\rho_n}{(r/r_n)(1 + r/r_n)^{2}}, \label{NFW}
\end{equation}
where the dimensionless velocity for this model is
\begin{equation}
\label{VNavarro}
\hat{V}^2_{\rm NFW}(\hat{r}_n)=\frac{1}{\hat{r}_n}\ln(1+\hat{r}_n) -   \frac{1}{1+\hat{r}_n}\, ,
\end{equation}
and $\hat{r}_n=r/r_n$ is defined using the scale radius $r_n$.

\item {\bf Burkert profile.}
Another core density profile proposed by \cite{Burkert} is:
\begin{equation}
\rho_{\rm Burk}=\frac{\rho_{b}}{(1+r/r_b)(1+(r/r_b)^{2})}, \label{Burk}
\end{equation}
with the above density profile we have
\begin{eqnarray}
\label{VBurkert}
\hat{V}^2_{\rm Burk}(\hat{r}_b)&=&\frac{1}{4\hat{r}_b}  
\left(-2\arctan(\hat{r}_b) 
\right.
\nonumber \\
&&
\left. + \ln\left[(1+\hat{r}_b^2)(1+\hat{r}_b)^2 \right]\right) \, .
\end{eqnarray}

\item{\bf Spano profile.}
The density profile proposed by \cite{Spano} is
\begin{equation}
\label{eq:Spano-density}
\rho_{\rm Spano} =\frac{\rho_{sp}}{\left(1+\left({r}/{r_{sp}}\right)^2\right)^{3/2}},
\end{equation}
we obtain a dimensionless quantity given by
\begin{equation}
\label{VSpano}
\hat{V}^2_{\rm sp}(\hat{r}_{sp})= - \frac{1}{\sqrt{1+\hat{r}^2_{sp}}} + \frac{1}{\hat{r}_{sp}}\arcsinh(\hat{r}_{sp})\, .
\end{equation}

\item {\bf Schive profile.}
Here we consider that DM density profile is given by WaveDM density profile, see \cite{Schive2,2018MNRAS.475.1447B}:
\begin{equation}
\rho_{\rm WDM}(r)=\frac{\rho_w}{(1+a_w(r/r_w)^2)^{8}}, \label{eq:WaveDM}
\end{equation}
where $a_w=0.091$. Then, if we define $\hat{r}_w=r/r_w$, it is obtained
\begin{eqnarray}
\label{VSchive}
\hat{V}^2_{\rm w}(\hat{r}_w)&=&\frac{1}{215040 a_w^{3/2}}\left[\frac{a_w^{1/2}}{\left(1 + a_w \hat{r}_w^2\right)^7}\left(-3465  \right. \right. \nonumber\\
&+&48580 a_w \hat{r}_w^2 + 92323 a_w ^2 \hat{r}_w^4 + 101376 a_w^3 \hat{r}_w^6  \nonumber\\
&+&\left. 65373 a_w^4 \hat{r}_w^8 + 231000 a_w^5 \hat{r}_w^{10} + 3465 a_w^6 \hat{r}_w ^{12} \right) \nonumber\\
&-&\left. 3465\frac{\arctan(\sqrt{a_w}  \hat{r}_w)}{\hat{r}_w}\right]\, .
\end{eqnarray}

\item{\bf Wave+NFW (SNFW) with  continuity condition in the density profile.}
In this subsection we consider the DM profile motivated by simulations where the DM can be modeled with a Scalar Field \citep{Schive2}, using the expression for density profile for the core Eq. (\ref{eq:WaveDM}) and the Eq. (\ref{NFW}) for the cuspy we obtain
\begin{eqnarray}
\label{WDM-density}
\rho_{\rm SNFWC}(r)&=& \Theta(r_\epsilon - r) \rho_{\rm WDM}(r)\nonumber\\
 &+& \Theta(r-r_\epsilon) \rho_{\rm NFW}(r), 
\end{eqnarray}
with the continuity condition on the density profiles at the matching radius $r_\epsilon$, $\rho_{\rm WDM} (r_\epsilon) = \rho_{\rm NFW} (r_\epsilon) $.   With this condition we have four free parameters ${r}_w$, ${r}_\epsilon$, ${r}_n$ and $\rho_w$, Eqs. (\ref{VNavarro}, \ref{VSchive}). Then we can write the velocity rotation curve as \cite{2018MNRAS.475.1447B}
\begin{multline}
\hat{V}^2_{\rm SNFWC}(\hat{r}_w, \hat{r}_\epsilon, \hat{r}_n) =  \\
= \begin{cases}
               \hat{V}_{WDM}(\hat{r}_w) & \text{if $\hat{r}_w \leq \frac{\hat{r}_w}{\hat{r}_\epsilon}$},
\nonumber\\
               \hat{V}_{WDM}(\frac{\hat{r}_w}{\hat{r}_\epsilon}) - \hat{V}_{NFW}(\frac{\hat{r}_n}{\hat{r}_\epsilon}) + \hat{V}_{NFW}(\hat{r}_n) & \text{if $\hat{r}_n > \frac{\hat{r}_n}{\hat{r}_\epsilon}$}
\end{cases} 
\end{multline}

\item{\bf Wave+NFW (SNFW) with  continuity condition on the density profile and on its derivative.}
This case was proposed in \cite{2018MNRAS.475.1447B}.
Here we consider the same density profile Eq. (\ref{WDM-density}), the continuity condition $\rho_{\rm WDM} (r_\epsilon) = \rho_{\rm NFW} (r_\epsilon)$ and a differentiable continuity condition at the transition radius $r_\epsilon$, $\rho '_{\rm WDM} (r_\epsilon) = \rho '_{\rm NFW} (r_\epsilon) $. With these conditions we have only three free parameters  ${r}_w$, ${r}_\epsilon$ and $\rho_w$, where 
\begin{multline}
\hat{V}^2_{\rm SNFWDC}(\hat{r}_w, \hat{r}_\epsilon) = \\
=\begin{cases}
               \hat{V}_{WDM}(\hat{r}_w) & \text{if $\hat{r}_w \leq \frac{\hat{r}_w}{\hat{r}_\epsilon}$}. \nonumber\\
               \hat{V}_{WDM}(\frac{\hat{r}_w}{\hat{r}_\epsilon}) - \hat{V}_{NFW}(\frac{\hat{r}_w}{\hat{r}_\epsilon}) + \hat{V}_{NFW}(\hat{r}_w) & \text{if $\hat{r}_n > \frac{\hat{r}_n}{\hat{r}_\epsilon}$}.
            \end{cases}
\end{multline}

\clearpage
\end{enumerate}


\section{Comparison parametric versus non-parametric tables} \label{ApTables}

\begin{center}
\begin{table*}
\ra{1.24}
\begin{center}
\begin{turn}{90}
\resizebox{!}{0.24\paperheight}{%
\begin{tabular}{@{}l | r | r | r | r | r | r | r | r | r | r | r | r | r | r | r | r | r | r | r | r | r | r | r | r | r | r | r| r  @ {}}\toprule
    \hline \hline   
    \multicolumn{29}{c}{SPARC Galaxies} \\
    \multicolumn{29}{c}{Statistics}\\
    \hline 
\multicolumn{1}{c|}{Galaxy} &
\multicolumn{6}{c|}{PISO (1) } &
\multicolumn{6}{c|}{NFW (2)}&
\multicolumn{6}{c|}{SPANO (3)}&
\multicolumn{6}{c|}{BURKERT (4)} &
\multicolumn{4}{c}{BEST MODEL}\\

\hline
    & $\chi^2_{red}$& $P$-Value & AIC & BIC & DIST & D$1\sigma$ & $\chi^2_{red}$ & $P$-Value & AIC & BIC & DIST & D$1\sigma$ & $\chi^2_{red}$ & $P$-Value & AIC & BIC & DIST & D$1\sigma$ & $\chi^2_{red}$ & $P$-Value & AIC & BIC & DIST & D$1\sigma$ & $B_{\rm BIC}$ &  $B_{\rm DIST}$ &  $B_{\rm D1\sigma}$ & $B_{\chi^2_{red}}$ \\ \hline

\cmidrule[0.4pt](r{0.125em}){1-1}
\cmidrule[0.4pt](lr{0.125em}){2-2}
\cmidrule[0.4pt](lr{0.125em}){3-3}
\cmidrule[0.4pt](lr{0.125em}){4-4}
\cmidrule[0.4pt](lr{0.125em}){5-5}
\cmidrule[0.4pt](lr{0.125em}){6-6}
\cmidrule[0.4pt](lr{0.125em}){7-7}
\cmidrule[0.4pt](lr{0.125em}){8-8}
\cmidrule[0.4pt](lr{0.125em}){9-9}
\cmidrule[0.4pt](lr{0.125em}){10-10}
\cmidrule[0.4pt](lr{0.125em}){11-11}
\cmidrule[0.4pt](lr{0.125em}){12-12}
\cmidrule[0.4pt](lr{0.125em}){13-13}
\cmidrule[0.4pt](lr{0.125em}){14-14}
\cmidrule[0.4pt](lr{0.125em}){15-15}
\cmidrule[0.4pt](lr{0.125em}){16-16}
\cmidrule[0.4pt](lr{0.125em}){17-17}
\cmidrule[0.4pt](lr{0.125em}){18-18}
\cmidrule[0.4pt](lr{0.125em}){19-19}
\cmidrule[0.4pt](lr{0.125em}){20-20}
\cmidrule[0.4pt](lr{0.125em}){21-21}
\cmidrule[0.4pt](lr{0.125em}){22-22}
\cmidrule[0.4pt](lr{0.125em}){23-23}
\cmidrule[0.4pt](lr{0.125em}){24-24}
\cmidrule[0.4pt](lr{0.125em}){25-25}
\cmidrule[0.4pt](lr{0.125em}){26-26}
\cmidrule[0.4pt](lr{0.125em}){27-27}
\cmidrule[0.4pt](lr{0.125em}){28-28}
\cmidrule[0.4pt](lr{0.125em}){29-29}
 

\multicolumn{1}{c|}{1} & 
\multicolumn{1}{c|}{2} & 
\multicolumn{1}{c|}{3} & 
\multicolumn{1}{c|}{4} & 
\multicolumn{1}{c|}{5} & 
\multicolumn{1}{c|}{6} & 
\multicolumn{1}{c|}{7} & 
\multicolumn{1}{c|}{8} & 
\multicolumn{1}{c|}{9} & 
\multicolumn{1}{c|}{10} & 
\multicolumn{1}{c|}{11} & 
\multicolumn{1}{c|}{12} & 
\multicolumn{1}{c|}{13} & 
\multicolumn{1}{c|}{14} & 
\multicolumn{1}{c|}{15} & 
\multicolumn{1}{c|}{16} & 
\multicolumn{1}{c|}{17} & 
\multicolumn{1}{c|}{18} & 
\multicolumn{1}{c|}{19} & 
\multicolumn{1}{c|}{20} & 
\multicolumn{1}{c|}{21} & 
\multicolumn{1}{c|}{22} & 
\multicolumn{1}{c|}{23} & 
\multicolumn{1}{c|}{24} & 
\multicolumn{1}{c|}{25} & 
\multicolumn{1}{c|}{26} & 
\multicolumn{1}{c|}{27} & 
\multicolumn{1}{c|}{28} & 
\multicolumn{1}{c}{29} \\
\hline

D 512-2 & 0.12 & 0.12  & 22.03 & 23.57 & 0.31 & 0.44 & 0.31 & 0.27  & 22.41 & 23.95 & 0.29 & 0.48 & 0.04 & 0.042  & 21.87 & 23.41 &  0.31 & 0.46 & 0.06 & 0.06  & 21.90 & 23.45 & 0.31 & 0.46 & 3412 & 2(134) & 1(34)2 & 3412 \\ 
D 564-8 & 0.06 & 0.007  & 21.90 & 25.07 & 0.24 & 0.28 & 0.91 & 0.55  & 25.30 & 28.47 & 0.22 & 0.36 & 0.04 & 0.003  & 21.80 & 24.97 &  0.25 & 0.27 & 0.07 & 0.009  & 21.92 & 25.09 & 0.24 & 0.27 & 3142 & 2(14)3 & (34)12 & 3142 \\ 
D 631-7 & 0.93 & 0.48  & 73.27 & 80.36 & 0.38 & 0.28 & 6.04 & 1.000  & 144.79 & 151.88 & 0.79 & 0.59 & 0.86 & 0.402  & 72.29 & 79.38 &  0.36 & 0.27 & 1.02 & 0.570  & 74.45 & 81.54 & 0.39 & 0.29 & 3142 & 3142 & 3142 & 3142 \\ 
DDO 064 & 0.33 & 0.0156  & 78.65 & 85.21 & 0.24 & 0.17 & 0.56 & 0.124  & 81.42 & 87.98 & 0.30 & 0.16 & 0.31 & 0.0115  & 78.39 & 84.94 &  0.23 & 0.16 & 0.31 & 0.0128  & 78.47 & 85.03 & 0.24 & 0.17 & 3412 & 3(14)2 & (23)(14) & (34)12 \\ 
DDO 154 & 2.48 & 0.994  & 34.12 & 40.06 & 0.94 & 0.62 & 11.77 & 1.000  & 127.05 & 132.99 & 1.86 & 1.55 & 0.82 & 0.390  & 17.53 & 23.47 &  0.55 & 0.35 & 1.24 & 0.743  & 21.78 & 27.72 & 0.74 & 0.44 & 3412 & 3412 & 3412 & 3412 \\ 
DDO 168 & 4.20 & 1.00  & 67.33 & 72.54 & 0.70 & 0.38 & 10.69 & 1.00  & 119.26 & 124.47 & 1.05 & 0.64 & 3.88 & 1.00  & 64.75 & 69.96 &  0.67 & 0.37 & 4.35 & 1.00  & 68.51 & 73.72 & 0.71 & 0.38 & 3142 & 3142 & 3(14)2 & 3142 \\ 
DDO 170 & 1.31 & 0.750  & 30.63 & 34.95 & 0.58 & 0.40 & 2.16 & 0.956  & 35.76 & 40.08 & 0.64 & 0.57 & 1.93 & 0.929  & 34.40 & 38.72 &  0.56 & 0.35 & 1.69 & 0.880  & 32.90 & 37.22 & 0.57 & 0.32 & 1432 & 3412 & 4312 & 1432 \\ 
ESO 116-G012 & 1.01 & 0.56  & 77.66 & 84.49 & 0.31 & 0.33 & 2.35 & 0.9961  & 95.06 & 101.89 & 0.60 & 0.28 & 0.83 & 0.37  & 75.28 & 82.11 &  0.22 & 0.42 & 0.78 & 0.314  & 74.61 & 81.44 & 0.22 & 0.42 & 4312 & (34)12 & 21(34) & 4312 \\ 
ESO 444-G084 & 1.12 & 0.66  & 36.25 & 40.04 & 0.30 & 0.16 & 0.67 & 0.354  & 33.98 & 37.76 & 0.35 & 0.24 & 2.36 & 0.963  & 42.44 & 46.23 &  0.27 & 0.19 & 1.64 & 0.85  & 38.82 & 42.60 & 0.27 & 0.16 & 2143 & (34)12 & (14)32 & 2143 \\ 
F 565-V2 & 0.06 & 0.0025  & 44.86 & 48.64 & 0.19 & 0.13 & 0.37 & 0.134  & 46.43 & 50.21 & 0.25 & 0.20 & 0.04 & 0.0009  & 44.75 & 48.54 &  0.19 & 0.12 & 0.06 & 0.0021  & 44.84 & 48.62 & 0.19 & 0.13 & 3412 & (134)2 & 3(14)2 & 3(14)2 \\ 
F 568-3 & 1.03 & 0.584  & 128.57 & 136.13 & 0.30 & 0.15 & 3.20 & 1.000  & 163.19 & 170.75 & 0.29 & 0.14 & 0.87 & 0.392  & 125.90 & 133.46 &  0.27 & 0.16 & 1.05 & 0.597  & 128.76 & 136.32 & 0.28 & 0.16 & 3142 & 3421 & 21(34) & 3142 \\ 
F 568-V1 & 0.14 & 0.00015  & 112.20 & 119.04 & 0.34 & 0.24 & 0.22 & 0.00154  & 113.19 & 120.02 & 0.35 & 0.24 & 0.06 & 6.0 $\times 10^{-7}$  & 111.09 & 117.92 &  0.21 & 0.16 & 0.05 & 4.0$\times 10^{-7}$  & 111.05 & 117.88 & 0.24 & 0.17 & 4312 & 3412 & 34(12) & 4312 \\ 
F 571-V1 & 0.07 & 0.0034  & 44.65 & 48.43 & 0.25 & 0.23 & 0.32 & 0.098  & 45.90 & 49.68 & 0.26 & 0.27 & 0.02 & 9.85$\times 10^{-5}$  & 44.38 & 48.17 &  0.26 & 0.23 & 0.03 & 0.0005  & 44.46 & 48.24 & 0.26 & 0.22 & 3412 & 1(234) & 4(13)2 & 3412 \\ 
F 574-1 & 0.28 & 0.0080  & 75.72 & 82.28 & 0.26 & 0.21 & 1.64 & 0.928  & 92.04 & 98.60 & 0.43 & 0.26 & 0.06 & 3.4 $\times 10^{-6}$  & 73.09 & 79.65 &  0.23 & 0.18 & 0.09 & 0.000028  & 73.45 & 80.00 & 0.24 & 0.20 & 3412 & 3412 & 3412 & 3412 \\ 
F 583-1 & 0.36 & 0.0020  & 153.12 & 161.99 & 0.33 & 0.21 & 1.60 & 0.9661  & 181.73 & 190.61 & 0.40 & 0.29 & 0.13 & 2.2 $\times 10^{-7}$   & 147.93 & 156.80 &  0.25 & 0.12 & 0.21 & 0.000016  & 149.62 & 158.50 & 0.28 & 0.15 & 3412 & 3412 & 3412 & 3412 \\ 
F 583-4 & 0.29 & 0.0172  & 66.96 & 72.90 & 0.36 & 0.25 & 0.15 & 0.00112  & 65.54 & 71.48 & 0.24 & 0.33 & 0.55 & 0.145  & 69.53 & 75.46 &  0.38 & 0.26 & 0.45 & 0.076  & 68.48 & 74.42 & 0.36 & 0.26 & 2143 & 2(14)3 & 1(34)2 & 2143 \\ 
IC 2574 & 2.84 & 1.000  & 207.18 & 217.29 & 0.13 & 0.12 & 33.57 & 1.0000  & 1190.55 & 1200.65 & 0.50 & 0.36 & 3.10 & 1.000  & 215.61 & 225.72 &  0.14 & 0.11 & 2.52 & 1.000  & 196.93 & 207.03 & 0.14 & 0.11 & 4132 & 1(34)2 & (34)12 & 4132 \\ 
KK 98-251 & 0.29 & 0.0062  & 56.09 & 62.92 & 0.26 & 0.08 & 1.82 & 0.9658  & 76.05 & 82.88 & 0.56 & 0.34 & 0.27 & 0.0043  & 55.82 & 62.66 &  0.26 & 0.08 & 0.30 & 0.0081  & 56.29 & 63.12 & 0.27 & 0.08 & 3142 & (13)42 & (134)2 & 3142 \\ 
NGC 0024 & 0.34 & 0.00049  & 180.86 & 190.33 & 0.26 & 0.19 & 0.93 & 0.43  & 196.93 & 206.40 & 0.31 & 0.24 & 1.0 & 0.53  & 198.75 & 208.22 &  0.59 & 0.43 & 0.78 & 0.221  & 192.94 & 202.40 & 0.53 & 0.37 & 1423 & 1243 & 1243 & 1423 \\ 
NGC 0055 & 0.62 & 0.104  & 105.85 & 114.03 & 0.29 & 0.19 & 2.97 & 1.000  & 150.50 & 158.68 & 0.45 & 0.35 & 0.32 & 0.0022  & 100.15 & 108.33 &  0.24 & 0.14 & 0.50 & 0.037  & 103.65 & 111.83 & 0.27 & 0.17 & 3412 & 3412 & 3412 & 3412 \\ 
NGC 0100 & 0.09 & 7. $\times 10^{-8}$  & 113.10 & 121.28 & 0.23 & 0.16 & 0.79 & 0.281  & 126.50 & 134.67 & 0.36 & 0.28 & 0.08 & 3.3 $\times 10^{-8}$  & 112.97 & 121.15 &  0.22 & 0.16 & 0.09 & 9.$\times 10^{-8}$  & 113.15 & 121.33 & 0.23 & 0.16 & 3142 & 3(14)2 & (134)2 & 3(14)2 \\ 
NGC 0247 & 2.12 & 0.9989  & 170.64 & 179.67 & 0.38 & 0.25 & 1.85 & 0.9931  & 164.15 & 173.18 & 0.22 & 0.14 & 5.66 & 1.000  & 255.54 & 264.57 &  0.66 & 0.50 & 4.26 & 1.000  & 221.91 & 230.95 & 0.57 & 0.41 & 2143 & 2143 & 2143 & 2143 \\ 
NGC 0300 & 0.41 & 0.0061  & 143.21 & 152.09 & 0.27 & 0.21 & 0.65 & 0.101  & 148.58 & 157.46 & 0.30 & 0.23 & 0.74 & 0.193  & 150.75 & 159.62 &  0.30 & 0.24 & 0.58 & 0.058  & 147.11 & 155.99 & 0.29 & 0.23 & 1423 & 14(23) & 1(24)3 & 1423 \\ 
NGC 3109 & 0.20 & 0.000013  & 100.37 & 109.24 & 0.16 & 0.1 & 8.55 & 1.0000  & 292.53 & 301.40 & 0.69 & 0.50 & 0.23 & 0.000039  & 100.96 & 109.83 &  0.18 & 0.12 & 0.18 & 4.0 $\times 10^{-6}$   & 99.85 & 108.72 & 0.15 & 0.11 & 4132 & 4132 & 1432 & 4132 \\ 
NGC 3741 & 1.06 & 0.61  & 97.54 & 105.72 & 0.29 & 0.22 & 0.34 & 0.0037  & 83.98 & 92.16 & 0.20 & 0.20 & 1.50 & 0.9255  & 105.92 & 114.10 &  0.38 & 0.23 & 1.21 & 0.77  & 100.51 & 108.69 & 0.33 & 0.22 & 2143 & 2143 & 2(14)3 & 2143 \\ 
NGC 6789 & 0.29 & 0.25  & 24.66 & 26.20 & 0.23 & 0.17 & 2.04 & 0.870  & 28.16 & 29.71 & 0.45 & 0.21 & 0.29 & 0.26  & 24.67 & 26.21 &  0.23 & 0.16 & 0.25 & 0.22  & 24.58 & 26.12 & 0.22 & 0.16 & 4132 & 4(13)2 & (34)12 & 4(13)2 \\ 
UGC 00191 & 1.35 & 0.78  & 40.42 & 45.21 & 0.70 & 0.58 & 3.58 & 0.999  & 56.09 & 60.87 & 0.93 & 0.78 & 5.57 & 1.00  & 70.01 & 74.80 &  0.59 & 0.36 & 4.03 & 1.00  & 59.21 & 64.00 & 0.63 & 0.39 & 1243 & 3412 & 3412 & 1243 \\ 
UGC 00634 & 1.52 & 0.780  & 19.87 & 21.42 & 1.10 & 1.0 & 3.20 & 0.959  & 23.24 & 24.79 & 1.20 & 0.88 & 0.16 & 0.15  & 17.16 & 18.71 &  1.00 & 1.09 & 0.36 & 0.30  & 17.55 & 19.10 & 1.04 & 1.06 & 3412 & 3412 & 2143 & 3412 \\ 
UGC 00891 & 0.23 & 0.127  & 14.33 & 16.77 & 0.25 & 0.22 & 3.10 & 0.974  & 22.94 & 25.38 & 0.71 & 0.55 & 0.17 & 0.083  & 14.14 & 16.58 &  0.17 & 0.22 & 0.22 & 0.115  & 14.28 & 16.72 & 0.23 & 0.23 & 3412 & 3412 & (13)42 & 3412 \\ 
UGC 02259 & 0.49 & 0.180  & 33.37 & 37.68 & 0.38 & 0.31 & 0.71 & 0.355  & 34.69 & 39.01 & 0.37 & 0.31 & 3.16 & 0.996  & 49.40 & 53.71 &  0.65 & 0.28 & 2.31 & 0.969  & 44.30 & 48.62 & 0.58 & 0.24 & 1243 & 2143 & 43(12) & 1243 \\ 
UGC 04325 & 1.46 & 0.812  & 45.32 & 49.64 & 0.37 & 0.36 & 2.87 & 0.991  & 53.77 & 58.08 & 0.42 & 0.39 & 0.23 & 0.034  & 37.95 & 42.27 &  0.35 & 0.40 & 0.41 & 0.128  & 39.03 & 43.35 & 0.33 & 0.42 & 3412 & 4312 & 1234 & 3412 \\ 
UGC 04499 & 0.32 & 0.056  & 42.10 & 46.89 & 0.27 & 0.31 & 1.09 & 0.64  & 47.49 & 52.28 & 0.28 & 0.43 & 0.26 & 0.029  & 41.63 & 46.42 &  0.28 & 0.28 & 0.27 & 0.033  & 41.71 & 46.50 & 0.28 & 0.29 & 3412 & 1(234) & 3412 & 3412 \\ 
UGC 05716 & 2.51 & 0.995  & 51.54 & 57.48 & 0.48 & 0.31 & 1.81 & 0.946  & 44.55 & 50.49 & 0.42 & 0.19 & 5.53 & 1.00  & 81.75 & 87.69 &  0.74 & 0.60 & 4.09 & 1.00  & 67.34 & 73.28 & 0.61 & 0.52 & 2143 & 2143 & 2143 & 2143 \\ 
UGC 05764 & 4.46 & 1.00  & 58.68 & 63.89 & 1.09 & 0.78 & 6.73 & 1.00  & 76.91 & 82.12 & 1.28 & 0.96 & 2.36 & 0.984  & 41.91 & 47.12 &  0.71 & 0.48 & 2.53 & 0.991  & 43.30 & 48.51 & 0.79 & 0.54 & 3412 & 3412 & 3412 & 3412 \\ 
UGC 05918 & 0.02 & 0.00002  & 40.70 & 45.02 & 0.22 & 0.18 & 0.15 & 0.0102  & 41.48 & 45.80 & 0.19 & 0.21 & 0.07 & 0.0014  & 41.03 & 45.35 &  0.23 & 0.18 & 0.04 & 0.0003  & 40.86 & 45.18 & 0.23 & 0.18 & 1432 & 21(34) & (134)2 & 1432 \\ 
UGC 05986 & 1.60 & 0.9219  & 86.06 & 92.89 & 0.61 & 0.35 & 5.98 & 1.000  & 142.99 & 149.82 & 0.96 & 0.67 & 0.57 & 0.123  & 72.77 & 79.60 &  0.39 & 0.25 & 0.92 & 0.47  & 77.27 & 84.10 & 0.47 & 0.27 & 3412 & 3412 & 3412 & 3412 \\ 
UGC 06399 & 0.09 & 0.0011  & 49.74 & 54.52 & 0.26 & 0.17 & 0.66 & 0.292  & 53.73 & 58.52 & 0.32 & 0.27 & 0.03 & 0.00003  & 49.35 & 54.13 &  0.25 & 0.18 & 0.05 & 0.00014  & 49.46 & 54.25 & 0.26 & 0.17 & 3412 & 3(14)2 & (14)32 & 3412 \\ 
UGC 06446 & 0.17 & 0.00017  & 90.99 & 98.32 & 0.22 & 0.21 & 0.21 & 0.00062  & 91.59 & 98.93 & 0.23 & 0.21 & 0.61 & 0.132  & 97.55 & 104.88 &  0.22 & 0.23 & 0.45 & 0.036  & 95.12 & 102.45 & 0.19 & 0.25 & 1243 & 4(13)2 & (12)34 & 1243 \\ 
UGC 06667 & 0.22 & 0.020  & 43.99 & 48.78 & 0.28 & 0.31 & 1.31 & 0.76  & 51.57 & 56.36 & 0.41 & 0.31 & 0.12 & 0.0028  & 43.25 & 48.04 &  0.31 & 0.28 & 0.09 & 0.0014  & 43.08 & 47.87 & 0.30 & 0.29 & 4312 & 1432 & 34(12) & 4312 \\ 
UGC 07399 & 0.22 & 0.0128  & 47.85 & 53.06 & 0.21 & 0.11 & 1.01 & 0.577  & 54.19 & 59.40 & 0.33 & 0.15 & 1.36 & 0.792  & 56.97 & 62.18 &  0.38 & 0.25 & 0.89 & 0.472  & 53.16 & 58.37 & 0.30 & 0.22 & 1423 & 1423 & 1243 & 1423 \\ 
UGC 07524 & 0.31 & 0.000127  & 146.29 & 156.03 & 0.16 & 0.13 & 0.81 & 0.250  & 160.93 & 170.67 & 0.34 & 0.15 & 0.27 & 0.000033  & 145.22 & 154.96 &  0.19 & 0.11 & 0.23 & 4.6 $\times 10^{-6}$  & 143.96 & 153.69 & 0.16 & 0.12 & 4312 & (14)32 & 3412 & 4312 \\ 
UGC 07603 & 0.33 & 0.0268  & 54.83 & 60.77 & 0.27 & 0.17 & 1.51 & 0.873  & 66.67 & 72.61 & 0.43 & 0.21 & 0.14 & 0.00081  & 52.93 & 58.87 &  0.24 & 0.17 & 0.20 & 0.0039  & 53.56 & 59.50 & 0.24 & 0.19 & 3412 & (34)12 & (13)42 & 3412 \\ 
UGC 08286 & 0.86 & 0.39  & 74.73 & 82.06 & 0.34 & 0.27 & 2.55 & 0.9992  & 100.05 & 107.39 & 0.45 & 0.34 & 1.63 & 0.9425  & 86.26 & 93.59 &  0.45 & 0.26 & 1.27 & 0.79  & 80.77 & 88.11 & 0.41 & 0.25 & 1432 & 14(23) & 4312 & 1432 \\ 
UGC 08490 & 0.19 & 7.9 $\times 10^{-7}$ & 142.75 & 152.36 & 0.21 & 0.17 & 0.11 & 1.5 $\times 10^{-9}$  & 140.62 & 150.23 & 0.21 & 0.15 & 0.98 & 0.51  & 164.91 & 174.52 &  0.35 & 0.23 & 0.66 & 0.088  & 156.00 & 165.61 & 0.29 & 0.22 & 2143 & (12)43 & 2143 & 2143 \\ 
UGC 08550 & 0.37 & 0.049  & 44.00 & 49.59 & 0.25 & 0.45 & 0.76 & 0.350  & 47.58 & 53.17 & 0.19 & 0.51 & 1.31 & 0.77  & 52.50 & 58.09 &  0.40 & 0.37 & 0.94 & 0.51  & 49.17 & 54.77 & 0.33 & 0.40 & 1243 & 2143 & 3412 & 1243 \\ 
UGC 12632 & 0.13 & 0.000094  & 70.16 & 76.99 & 0.20 & 0.15 & 0.34 & 0.0143  & 72.87 & 79.70 & 0.28 & 0.24 & 0.10 & 0.000024  & 69.80 & 76.63 &  0.20 & 0.12 & 0.07 & 1.5 $\times 10^{-6}$  & 69.30 & 76.13 & 0.20 & 0.11 & 4312 & (134)2 & 4312 & 4312 \\ 
UGC 12732 & 0.47 & 0.049  & 80.96 & 88.05 & 0.32 & 0.1 & 0.21 & 0.00075  & 77.30 & 84.39 & 0.22 & 0.15 & 1.79 & 0.966  & 99.48 & 106.57 &  0.46 & 0.24 & 1.34 & 0.824  & 93.12 & 100.21 & 0.41 & 0.21 & 2143 & 2143 & 1243 & 2143 \\ 
UGC A442 & 0.80 & 0.432  & 33.34 & 37.66 & 0.26 & 0.26 & 2.05 & 0.944  & 40.82 & 45.14 & 0.45 & 0.22 & 0.76 & 0.400  & 33.10 & 37.42 &  0.30 & 0.22 & 0.62 & 0.285  & 32.25 & 36.56 & 0.26 & 0.25 & 4312 & (14)32 & (23)41 & 4312 \\  
\bottomrule
\hline \hline
\end{tabular}
}
\end{turn}
\end{center}
\caption{In this table we report the different statistical values for Piso (1), NFW (2), Spano (3) and Burkert (4) models. In the Columns 26--29 we associated the number code, mentioned in the text for every model, to select in ascendent order  the best BIC, best distance between model and LOESS+SIMEX, best 1$\sigma$ bands distance from the model and LOESS+SIMEX and best $\chi^2_{red}$ value. The numbers inside parenthesis mean that these models have the same value. }\label{tab:PNSB}
\end{table*}
\end{center}


\setcounter{table}{0}
%
\begin{center}
\begin{table*}
\ra{1.24}
\begin{center}
\begin{turn}{90}
\resizebox{!}{0.2\paperheight}{%
\begin{tabular}{@{}l | r | r | r | r | r | r | r | r | r | r | r | r | r | r | r | r | r | r | r | r | r | r | r | r | r | r | r| r  @ {}}\toprule
    \hline \hline
    \multicolumn{29}{c}{SPARC Galaxies} \\
    \multicolumn{29}{c}{Statistics}\\
    \hline 
\multicolumn{1}{c|}{Galaxy} &
\multicolumn{6}{c|}{PISO (1) } &
\multicolumn{6}{c|}{NFW (2)}&
\multicolumn{6}{c|}{SPANO (3)}&
\multicolumn{6}{c|}{BURKERT (4)} &
\multicolumn{4}{c}{BEST MODEL}\\

\hline
    & $\chi^2_{red}$& $P$-Value & AIC & BIC & DIST & D$1\sigma$ & $\chi^2_{red}$ & $P$-Value & AIC & BIC & DIST & D$1\sigma$ & $\chi^2_{red}$ & $P$-Value & AIC & BIC & DIST & D$1\sigma$ & $\chi^2_{red}$ & $P$-Value & AIC & BIC & DIST & D$1\sigma$ & $B_{\rm BIC}$ &  $B_{\rm DIST}$ &  $B_{\rm D1\sigma}$ & $B_{\chi^2_{red}}$ \\ \hline

\cmidrule[0.4pt](r{0.125em}){1-1}
\cmidrule[0.4pt](lr{0.125em}){2-2}
\cmidrule[0.4pt](lr{0.125em}){3-3}
\cmidrule[0.4pt](lr{0.125em}){4-4}
\cmidrule[0.4pt](lr{0.125em}){5-5}
\cmidrule[0.4pt](lr{0.125em}){6-6}
\cmidrule[0.4pt](lr{0.125em}){7-7}
\cmidrule[0.4pt](lr{0.125em}){8-8}
\cmidrule[0.4pt](lr{0.125em}){9-9}
\cmidrule[0.4pt](lr{0.125em}){10-10}
\cmidrule[0.4pt](lr{0.125em}){11-11}
\cmidrule[0.4pt](lr{0.125em}){12-12}
\cmidrule[0.4pt](lr{0.125em}){13-13}
\cmidrule[0.4pt](lr{0.125em}){14-14}
\cmidrule[0.4pt](lr{0.125em}){15-15}
\cmidrule[0.4pt](lr{0.125em}){16-16}
\cmidrule[0.4pt](lr{0.125em}){17-17}
\cmidrule[0.4pt](lr{0.125em}){18-18}
\cmidrule[0.4pt](lr{0.125em}){19-19}
\cmidrule[0.4pt](lr{0.125em}){20-20}
\cmidrule[0.4pt](lr{0.125em}){21-21}
\cmidrule[0.4pt](lr{0.125em}){22-22}
\cmidrule[0.4pt](lr{0.125em}){23-23}
\cmidrule[0.4pt](lr{0.125em}){24-24}
\cmidrule[0.4pt](lr{0.125em}){25-25}
\cmidrule[0.4pt](lr{0.125em}){26-26}
\cmidrule[0.4pt](lr{0.125em}){27-27}
\cmidrule[0.4pt](lr{0.125em}){28-28}
\cmidrule[0.4pt](lr{0.125em}){29-29}
 
\multicolumn{1}{c|}{1} & 
\multicolumn{1}{c|}{2} & 
\multicolumn{1}{c|}{3} & 
\multicolumn{1}{c|}{4} & 
\multicolumn{1}{c|}{5} & 
\multicolumn{1}{c|}{6} & 
\multicolumn{1}{c|}{7} & 
\multicolumn{1}{c|}{8} & 
\multicolumn{1}{c|}{9} & 
\multicolumn{1}{c|}{10} & 
\multicolumn{1}{c|}{11} & 
\multicolumn{1}{c|}{12} & 
\multicolumn{1}{c|}{13} & 
\multicolumn{1}{c|}{14} & 
\multicolumn{1}{c|}{15} & 
\multicolumn{1}{c|}{16} & 
\multicolumn{1}{c|}{17} & 
\multicolumn{1}{c|}{18} & 
\multicolumn{1}{c|}{19} & 
\multicolumn{1}{c|}{20} & 
\multicolumn{1}{c|}{21} & 
\multicolumn{1}{c|}{22} & 
\multicolumn{1}{c|}{23} & 
\multicolumn{1}{c|}{24} & 
\multicolumn{1}{c|}{25} & 
\multicolumn{1}{c|}{26} & 
\multicolumn{1}{c|}{27} & 
\multicolumn{1}{c|}{28} & 
\multicolumn{1}{c}{29} \\
\hline

DDO 161 
& 0.33 & 0.00030 & 113.83 & 123.56 & 0.28 & 0.22  
& 1.15  & 0.73  & 137.44  & 147.18  & 0.56  & 0.50  
& 0.25 	& 0.000015  & 111.46  & 121.20  & 0.25  & 0.19  
& 0.27  & 0.000026 	& 111.81 & 121.55  & 0.27  & 0.21  
& 3412  & 3412  & 3412  & 3412 \\
F 563-1 
& 0.56 & 0.095 & 18726.84 & 18734.17 & 0.29 & 0.22 
& 1.07 & 0.62 & 16230.84 & 16238.17 & 0.30 & 0.23 
& 0.55 & 0.088 & 23172.08 & 23179.41 & 0.28 & 0.23 
& 0.56 & 0.095 & 22021.41 & 22028.75 & 0.28 & 0.22 
& 2143 & (34)12 & (14)32 & 3(14)2 \\
F 563v2 
& 0.29 & 0.0310 & 51926.01 & 51931.22 & 0.30 & 0.21 
& 1.23 & 0.725 & 59093.79 & 59099.00 & 0.44 & 0.33 
& 0.11 & 0.001 & 42960.47 & 42965.68 & 0.20 & 0.15 
& 0.18 & 0.0062 & 44883.91 & 44889.12 & 0.23 & 0.17 
& 3412 & 3412 & 3412 & 3412 \\
F 568-1 
& 0.14 &0.00082 &59395.17 &59401.11 &0.17 &0.07 
&0.86 &0.429 &63319.03 &63324.97 &0.27 &0.13 
&0.07 &0.000023 &55216.03 &55221.97 &0.15 &0.10 
&0.10 &0.00018 &56246.50 &56252.44 &0.15 &0.1 
&3412 &(34)12 &1(34)2 &3412 \\
F 571-8 & 10.75 &1.0000 &61560.53 &61566.79 &0.88 &0.65 &19.17 &1.0000 &54951.23 &54957.49 &1.20 &0.93 &10.22 &1.00 &61939.64 &61945.90 &0.80 &0.62 &10.84 &1.0000 &61554.34 &61560.60 &0.86 &0.64 &2413 &3412 &3412 &3142 \\
F 579v1 
& 0.05 & $5.0\times 10^{-7}$
&33553.33 &33559.89 &0.18 &0.06 
&0.19 &0.00108 &33342.86 &33349.41 &0.18 &0.12 
&0.37  &0.027 &30594.21 &30600.77 &0.26 &0.16 
&0.25 &0.0049 &31124.25 &31130.80 &0.22 &0.15 
&3421 &(12)43 &1243 &1243 \\
NGC 1003 & 3.34 &1.000 &255.13 &265.46 &0.46 &0.28 &2.57 &1.000 &229.02 &239.36 &0.45 &0.31 &6.56 &1.000 &364.76 &375.10 &0.78 &0.55 &5.54 &1.0000 &330.25 &340.58 &0.69 &0.47 		&2143 &2143 &1243 &2143 \\
NGC 2915 & 0.70 &0.120 &202.07 &211.68 &0.36 &0.21 &0.96 &0.47 &209.31 &218.92 &0.39 &0.27 &0.53 &0.021 &197.48 &207.08 &0.32 &0.20 &0.56 &0.031 &198.28 &207.88 &0.33 &0.21 		&3412 &3412 &3(14)2 &3412 \\
NGC 2976  & 0.30 &0.00024 &127.00 &136.18 &0.23 &0.22 &1.90 &0.9957 &167.08 &176.27 &0.54 &0.24 &0.29 &0.00023 &126.95 &136.14 &0.23 &0.22 &0.30 &0.00032 &127.21 &136.40 &0.24 &0.22 &3142 &(13)42 &(134)2 &3(14)2 \\
NGC 4068 & 0.21 &0.069 &28.08 &31.24 &0.34 &0.21 &1.86 &0.886 &34.68 &37.85 &0.64 &0.42 &0.21 &0.069 &28.08 &31.24 &0.34 &0.21 &0.22 &0.072 &28.10 &31.26 &0.34 &0.20 &(13)42 		&(134)2 &4(13)2 &(13)42 \\
NGC 5585 & 5.56 &1.000 &203.50 &212.21 &0.38 &0.24 &7.17 &1.0000 &239.11 &247.83 &0.83 &0.50 &5.12 &1.00 &193.84 &202.55 &0.32 &0.26 &5.18 &1.000 &195.23 &203.95 &0.34 &0.25 		&3412 &3412 &1432 &3412 \\
NGC 7793 &0.97 &0.47 &305.76 &317.07 &0.31 &0.20 &0.89 &0.32 &302.45 &313.76 &0.31 &0.21 &1.03 &0.580 &308.44 &319.76 &0.27 &0.17 &0.96 &0.45 &305.47 &316.78 &0.27 &0.17 			&2413 &(34)(12) &(34)12 &2413 \\
UGC 00731 &0.16 &0.00126 &52.08 &58.02 &0.21 &0.16 &0.34 &0.0295 &53.92 &59.86 &0.20 &0.12 &0.96 &0.521 &60.09 &66.03 &0.34 &0.25 &0.57 &0.160 &56.23 &62.17 &0.26 &0.23 			&1243 &2143 &2143 &1243 \\
UGC 01281 &0.24 &0.000068 &133.07 &141.95 &0.21 &0.15 &1.30 &0.8478 &157.46 &166.34 &0.37 &0.32 &0.20 &0.00001 &132.08 &140.95 &0.20 &0.15 &0.24 &0.000082 &133.19 &142.06 &0.21 &0.15 &3142 &3(14)2 &(134)2 &3(14)2 \\
UGC 04278 &0.73 &0.181 &150.26 &159.13 &0.28 &0.22 &1.47 &0.9324 &167.28 &176.16 &0.43 &0.37 &0.74 &0.193 &150.50 &159.37 &0.28 &0.22 &0.68 &0.133 &149.17 &158.04 &0.27 &0.22 	&4132 &4(13)2 &(134)2 &4132 \\
UGC 04483 &0.27 &0.049 &29.15 &33.47 &0.28 &0.23 &0.59 &0.264 &31.09 &35.41 &0.32 &0.32 &0.23 &0.034 &28.93 &33.25 &0.28 &0.23 &0.25 &0.040 &29.02 &33.34 &0.29 &0.22 				&3412 &(13)42 &4(13)2 &3412 \\
UGC 05005 
&0.03 & $3.0\times 10^{-6}$ 
&78.18 &83.77 &0.19 &0.12 
&0.22 &0.0079 &79.85 &85.44 &0.18 &0.11 
&0.01 & $4.0\times 10^{-8}$  
&78.00 &83.59 &0.21 &0.14 
&0.01 & $5.0\times 10^{-8}$  
&78.01 &83.60 &0.21 &0.13 
&3412 &21(34) &2143 &(34)12 \\
UGC 05414 
&0.09 &0.014 &26.15 &29.32 &0.30 &0.28 
&1.05 &0.619 &29.98 &33.14 &0.37 &0.36 		
&0.13 &0.028 &26.30 &29.47 &0.31 &0.27 
&0.09 &0.014 &26.14 &29.30 &0.30 &0.28 		
&4132 &(14)32 &3(14)2 &(14)32 \\
UGC 05721 &0.81 &0.290 &137.23 &145.77 &0.36 &0.20 &0.96 &0.49 &140.45 &148.99 &0.37 &0.20 	&0.42 &0.01 &129.04 &137.58 &0.22 &0.13 &0.39 &0.0060 &128.46 &137.00 &0.21 &0.14 	&4312 &4312 &34(12) &4312 \\
UGC 05750 &0.17 &0.0031 &71.06 &76.65 &0.16 &0.18 &1.00 &0.57 &78.57 &84.16 &0.25 &0.16 		&0.07 &0.00011 &70.20 &75.79 &0.15 &0.18 &0.12 &0.00071 &70.59 &76.18 &0.15 &0.19 	&3412 &(34)12 &2(13)4 &3412 \\
UGC 05829 &0.15 &0.0018 &61.23 &66.82 &0.21 &0.16 &0.09 &0.00021 &60.67 &66.27 &0.19 &0.16 	&0.26 &0.015 &62.22 &67.81 &0.22 &0.18 &0.20 &0.0054 &61.65 &67.25 &0.20 &0.17 		&2143 &2413 &(12)43 &2143 \\
UGC 06818 &1.17 &0.681 &51.84 &56.15 &0.34 &0.28 &3.55 &0.998 &66.12 &70.44 &0.65 &0.36 		&1.16 &0.678 &51.80 &56.12 &0.34 &0.28 &1.21 &0.703 &52.08 &56.40 &0.34 &0.27 		&3142 &(134)2 &4(13)2 &3142 \\ 
UGC 06917 &0.27 &0.0163 &57.68 &63.27 &0.21 &0.17 &0.80 &0.387 &62.52 &68.11 &0.30 &0.14 		&0.19 &0.005 &57.01 &62.60 &0.21 &0.19 &0.20 &0.0053 &57.05 &62.64 &0.19 &0.20 		&3412 &4(13)2 &2134 &3412 \\
UGC 06923 &0.51 &0.273 &37.03 &40.19 &0.32 &0.34 &0.95 &0.57 &38.80 &41.96 &0.36 &0.35 		&0.42 &0.209 &36.67 &39.84 &0.33 &0.33 &0.48 &0.251 &36.91 &40.07 &0.33 &0.32 		&3412 &1(34)2 &4312 &3412 \\
UGC 06983 &0.46 &0.041 &101.41 &108.75 &0.28 &0.27 &0.60 &0.120 &103.40 &110.73 &0.27 &0.27 	&0.47 &0.044 &101.52 &108.85 &0.18 &0.29 &0.42 &0.027 &100.80 &108.13 &0.20 &0.28 	&4132 &3421 &(12)43 &4132 \\
UGC 07089 &0.16 &0.00139 &66.49 &72.43 &0.21 &0.11 &0.27 &0.0119 &67.57 &73.51 &0.22 &0.16 	&0.19 &0.003 &66.75 &72.69 &0.22 &0.12 &0.15 &0.00115 &66.42 &72.36 &0.21 &0.11 		&4132 &(14)(23) &(14)32 &4132 \\
UGC 07125 &0.46 &0.072 &57.34 &63.60 &0.35 &0.19 &0.85 &0.41 &61.67 &67.93 &0.42 &0.24 &0.23 	&0.004 &54.77 &61.03 &0.23 &0.12 &0.25 &0.0060 &54.99 &61.25 &0.25 &0.13 			&3412 &3412 &3412 &3412 \\
UGC 07151 &1.05 &0.60 &54.29 &59.88 &0.37 &0.27 &2.86 &0.998 &70.63 &76.23 &0.47 &0.29 		&0.76 &0.347 &51.71 &57.30 &0.29 &0.31 &0.85 &0.43 &52.54 &58.14 &0.30 &0.32 		&3412 &3412 &1234 &3412 \\
UGC 07232 &0.11 &0.11 &21.13 &22.67 &0.22 &0.26 &2.41 &0.911 &25.73 &27.28 &0.54 &0.27 		&0.11 &0.107 &21.13 &22.67 &0.22 &0.26 &0.12 &0.11 &21.14 &22.68 &0.23 &0.26 			&(13)42 &(13)42 &(134)2 &(13)42 \\
UGC 07261 &0.09 &0.006 &41.58 &45.37 &0.21 &0.15 &0.05 &0.0014 &41.38 &45.16 &0.20 &0.13 		&0.50 &0.223 &43.63 &47.41 &0.28 &0.21 &0.37 &0.132 &43.00 &46.78 &0.26 &0.20 		&2143 &2143 &2143 &2143 \\
UGC 07323 &0.32 &0.042 &52.01 &57.22 &0.21 &0.10 &0.71 &0.316 &55.11 &60.32 &0.37 &0.26 		&0.33 &0.045 &52.09 &57.30 &0.21 &0.11 &0.30 &0.0321 &51.80 &57.01 &0.21 &0.10 		&4132 &(134)2 &(14)32 &4132 \\
UGC 07559 &0.10 &0.008 &33.79 &37.57 &0.21 &0.17 &0.43 &0.171 &35.42 &39.20 &0.25 &0.24 		&0.08 &0.005 &33.69 &37.48 &0.21 &0.17 &0.10 &0.008 &33.79 &37.57 &0.21 &0.17 		&3(14)2 &(134)2 &3(14)2 &3(14)2 \\
UGC 07577 &0.08 &0.0008 &41.87 &46.65 &0.22 &0.08 &0.20 &0.0153 &42.74 &47.53 &0.26 &0.10 	&0.08 &0.0008 &41.87 &46.66 &0.22 &0.08 &0.08 &0.0009 &41.89 &46.68 &0.22 &0.08 		&1342 &(134)2 &(134)2 &(134)2 \\
UGC 07690 &0.40 &0.154 &39.34 &43.12 &0.23 &0.07 &0.34 &0.108 &39.00 &42.78 &0.20 &0.08 		&0.08 &0.005 &37.72 &41.51 &0.18 &0.12 &0.11 &0.010 &37.87 &41.66 &0.18 &0.11 		&3421 &(34)21 &1243 &3421 \\
UGC 07866 &0.05 &0.0012 &36.75 &40.53 &0.21 &0.16 &0.04 &0.0011 &36.73 &40.52 &0.19 &0.18 	&0.09 &0.006 &36.95 &40.73 &0.21 &0.17 &0.07 &0.0032 &36.86 &40.64 &0.21 &0.17 		&2143 &2(134) &1(34)2 &2143 \\
UGC 08837 &0.67 &0.325 &37.29 &41.61 &0.26 &0.27 &6.33 &1.00 &71.27 &75.59 &0.88 &0.52 &0.68 	&0.332 &37.34 &41.66 &0.26 &0.27 &0.77 &0.408 &37.91 &42.23 &0.29 &0.25 			&1342 &(13)42 &4(13)2 &1342 \\
UGC 09992 &0.008 &0.001 &30.55 &32.98 &0.20 &0.19 &0.01 &0.002 &30.56 &33.00 &0.21 &0.19 		&0.03 &0.007 &30.62 &33.05 &0.22 &0.18 &0.02 &0.005 &30.60 &33.03 &0.22 &0.18 		&1243 &12(34) &(34)(12) &1243 \\
UGC 10310 &0.27 &0.069 &40.90 &44.68 &0.22 &0.10 &0.57 &0.276 &42.41 &46.19 &0.25 &0.09 		&0.06 &0.002 &39.85 &43.63 &0.18 &0.11 &0.09 &0.007 &40.03 &43.82 &0.18 &0.12 		&3412 &(34)12 &2134 &3412 \\
UGC 11820 &5.95 &1.00 &84.50 &89.71 &2.36 &2.04 &1.33 &0.778 &47.54 &52.75 &1.98 &1.79 &8.67 &1.00 &106.21 &111.42 &2.64 &2.33 &7.56 &1.00 &97.34 &102.55 &2.54 &2.24 				&2143 &2143 &2143 &2143 \\
UGC A444 &0.20 & $9.0\times 10^{-8}$ 
&172.24 &182.58 &0.17 &0.12 &0.06 & $1.7\times 10^{-15}$ 
&167.64 &177.97 &0.16 &0.09 &0.26 & $5.2\times 10^{-6}$  
&174.57 &184.91 &0.18 &0.13 &0.22 & $3.4\times 10^{-7}$ 
&172.90 &183.23 &0.17 &0.13 &2143 &2(14)3 &21(34) &2143  \\

\bottomrule
\hline \hline
\end{tabular}
}
\end{turn}
\end{center}
\caption{Continue. }\label{tab:PNSBc}
\end{table*}

\end{center}
%

%
%
\begin{center}
\begin{table*}
\ra{1.24}
\begin{center}
\begin{turn}{90}
\resizebox{!}{0.24\paperheight}{%
\begin{tabular}{@{}l | r | r | r | r | r | r | r | r | r | r | r | r | r | r | r | r | r | r | r | r | r | r | r | r | r | r | r | r | r  @ {}}\toprule
    \hline \hline
    \multicolumn{30}{c}{SPARC Galaxies} \\
    \multicolumn{30}{c}{Statistics}\\
    \hline 
\multicolumn{1}{c|}{Galaxy} &
\multicolumn{1}{c|}{$\delta A_V$} &
\multicolumn{6}{c|}{SCHIVE (1) } &
\multicolumn{6}{c|}{NFW (2)} &
\multicolumn{6}{c|}{SNFWDC (3)}&
\multicolumn{6}{c|}{SNFWC (4)} &
\multicolumn{4}{c}{BEST MODEL}\\

\hline
   & & $\chi^2_{red}$& $P$-Value & AIC & BIC & DIST & D$1\sigma$ & $\chi^2_{red}$ & $P$-Value & AIC & BIC & DIST & D$1\sigma$ & $\chi^2_{red}$ & $P$-Value & AIC & BIC & DIST & D$1\sigma$ & $\chi^2_{red}$ & $P$-Value & AIC & BIC &  DIST & D$1\sigma$ & $B_{\rm BIC}$ &  $B_{\rm DIST}$ &  $B_{\rm D1\sigma}$ & $B_{\chi^2_{red}}$  \\ \hline

\cmidrule[0.4pt](r{0.125em}){1-1}
\cmidrule[0.4pt](lr{0.125em}){2-2}
\cmidrule[0.4pt](lr{0.125em}){3-3}
\cmidrule[0.4pt](lr{0.125em}){4-4}
\cmidrule[0.4pt](lr{0.125em}){5-5}
\cmidrule[0.4pt](lr{0.125em}){6-6}
\cmidrule[0.4pt](lr{0.125em}){7-7}
\cmidrule[0.4pt](lr{0.125em}){8-8}
\cmidrule[0.4pt](lr{0.125em}){9-9}
\cmidrule[0.4pt](lr{0.125em}){10-10}
\cmidrule[0.4pt](lr{0.125em}){11-11}
\cmidrule[0.4pt](lr{0.125em}){12-12}
\cmidrule[0.4pt](lr{0.125em}){13-13}
\cmidrule[0.4pt](lr{0.125em}){14-14}
\cmidrule[0.4pt](lr{0.125em}){15-15}
\cmidrule[0.4pt](lr{0.125em}){16-16}
\cmidrule[0.4pt](lr{0.125em}){17-17}
\cmidrule[0.4pt](lr{0.125em}){18-18}
\cmidrule[0.4pt](lr{0.125em}){19-19}
\cmidrule[0.4pt](lr{0.125em}){20-20}
\cmidrule[0.4pt](lr{0.125em}){21-21}
\cmidrule[0.4pt](lr{0.125em}){22-22}
\cmidrule[0.4pt](lr{0.125em}){23-23}
\cmidrule[0.4pt](lr{0.125em}){24-24}
\cmidrule[0.4pt](lr{0.125em}){25-25}
\cmidrule[0.4pt](lr{0.125em}){26-26}
\cmidrule[0.4pt](lr{0.125em}){27-27}
\cmidrule[0.4pt](lr{0.125em}){28-28}
\cmidrule[0.4pt](lr{0.125em}){29-29}
\cmidrule[0.4pt](lr{0.125em}){30-30}

\multicolumn{1}{c|}{1} & 
\multicolumn{1}{c|}{2} & 
\multicolumn{1}{c|}{3} & 
\multicolumn{1}{c|}{4} & 
\multicolumn{1}{c|}{5} & 
\multicolumn{1}{c|}{6} & 
\multicolumn{1}{c|}{7} & 
\multicolumn{1}{c|}{8} & 
\multicolumn{1}{c|}{9} & 
\multicolumn{1}{c|}{10} & 
\multicolumn{1}{c|}{11} & 
\multicolumn{1}{c|}{12} & 
\multicolumn{1}{c|}{13} & 
\multicolumn{1}{c|}{14} & 
\multicolumn{1}{c|}{15} & 
\multicolumn{1}{c|}{16} & 
\multicolumn{1}{c|}{17} & 
\multicolumn{1}{c|}{18} & 
\multicolumn{1}{c|}{19} & 
\multicolumn{1}{c|}{20} & 
\multicolumn{1}{c|}{21} & 
\multicolumn{1}{c|}{22} & 
\multicolumn{1}{c|}{23} & 
\multicolumn{1}{c|}{24} & 
\multicolumn{1}{c|}{25} & 
\multicolumn{1}{c|}{26} & 
\multicolumn{1}{c|}{27} & 
\multicolumn{1}{c|}{28} & 
\multicolumn{1}{c|}{29} & 
\multicolumn{1}{c}{30} \\
\hline
 
D 512-2* & 46.3 & 0.08 & 0.07 & 21.93 & 23.48 & 0.32 & 0.47 & 0.31 & 0.27 & 22.41 & 23.95 & 0.29 & 0.48 & 0.09 & 0.24 & 23.87 & 26.19 & 0.31 & 0.73 & -- & -- & 25.87 & 28.96 & 0.31 & 0.77 & 1234 & 2(34)1 & 1234 & 132 \\ 
D 564-8 & 43.9 & 0.04 & 0.003 & 21.82 & 24.98 & 0.26 & 0.26 & 0.91 & 0.55 & 25.30 & 28.47 & 0.22 & 0.36 & 0.04 & 0.009 & 23.75 & 28.50 & 0.25 & 0.35 & 0.08 & 0.08 & 25.82 & 32.15 & 0.26 & 0.31 & 1234 & 23(14) & 1432 & (13)42 \\ 
D 631-7 & 49.1 & 0.75 & 0.280 & 70.75 & 77.84 & 0.33 & 0.25 & 6.04 & 1.000 & 144.79 & 151.88 & 0.79 & 0.59 & 0.81 & 0.35 & 72.75 & 83.38 & 0.33 & 0.25 & 0.60 & 0.156 & 71.38 & 85.56 & 0.25 & 0.25 & 1342 & 4(13)2 & (134)2 & 4132 \\ 
DDO 064 & 49.7 & 0.27 & 0.0066 & 77.97 & 84.52 & 0.22 & 0.16 & 0.56 & 0.124 & 81.42 & 87.98 & 0.30 & 0.16 & 0.30 & 0.0132 & 79.97 & 89.80 & 0.22 & 0.16 & 0.33 & 0.0255 & 81.97 & 95.08 & 0.22 & 0.16 & 1234 & (134)2 & (1234) & 1342 \\ 
DDO 154 & 41.9 & 2.00 & 0.971 & 29.33 & 35.27 & 0.38 & 0.26 & 11.77 & 1.000 & 127.05 & 132.99 & 1.86 & 1.55 & 1.24 & 0.73 & 22.49 & 31.40 & 0.43 & 0.47 & 1.39 & 0.807 & 24.49 & 36.37 & 0.43 & 0.49 & 3142 & 1(34)2 & 1342 & 3412 \\ 
DDO 168 & 56.0 & 3.32 & 0.999 & 60.30 & 65.51 & 0.61 & 0.35 & 10.69 & 1.00 & 119.26 & 124.47 & 1.05 & 0.64 & 3.79 & 1.00 & 62.30 & 70.11 & 0.61 & 0.35 & 4.43 & 1.00 & 64.30 & 74.72 & 0.61 & 0.35 & 1342 & (134)2 & (134)2 & 1342 \\ 
DDO 170 & 49.4 & 6.20 & 1.00 & 60.01 & 64.33 & 0.75 & 0.42 & 2.16 & 0.956 & 35.76 & 40.08 & 0.64 & 0.57 & 1.45 & 0.80 & 32.06 & 38.54 & 0.55 & 0.41 & 0.48 & 0.246 & 28.69 & 37.33 & 0.47 & 0.54 & 4321 & 4321 & 3142 & 4321 \\ 
ESO 116-G012 & 50.0 & 1.42 & 0.86 & 82.96 & 89.79 & 0.34 & 0.33 & 2.35 & 0.9961 & 95.06 & 101.89 & 0.60 & 0.28 & 1.03 & 0.584 & 78.89 & 89.14 & 0.24 & 0.46 & 1.13 & 0.66 & 80.89 & 94.56 & 0.24 & 0.48 & 3142 & (34)12 & 2134 & 3412 \\ 
ESO 444-G084 & 34.7 & 3.87 & 0.998 & 50.00 & 53.78 & 0.36 & 0.23 & 0.67 & 0.354 & 33.98 & 37.76 & 0.35 & 0.24 & 0.17 & 0.047 & 33.31 & 38.99 & 0.35 & 0.19 & 0.23 & 0.123 & 35.31 & 42.88 & 0.35 & 0.28 & 2341 & (234)1 & 3124 & 3421 \\ 
F 565-V2 & 36.0 & 0.05 & 0.0017 & 44.82 & 48.60 & 0.18 & 0.12 & 0.37 & 0.133 & 46.42 & 50.21 & 0.25 & 0.20 & 0.03 & 0.002 & 46.68 & 52.36 & 0.19 & 0.17 & 0.02 & 0.004 & 48.62 & 56.19 & 0.17 & 0.31 & 1234 & 4132 & 1324 & 4312 \\  
F 568-3 & 46.2 & 0.90 & 0.435 & 126.48 & 134.04 & 0.34 & 0.20 & 3.20 & 1.000 & 163.19 & 170.75 & 0.29 & 0.14 & 0.76 & 0.276 & 125.41 & 136.76 & 0.26 & 0.14 & 0.71 & 0.232 & 125.94 & 141.06 & 0.23 & 0.13 & 1342 & 4321 & 4(23)1 & 4312 \\ 
F 568-V1 & 36.4 & 0.43 & 0.042 & 115.98 & 122.81 & 0.29 & 0.20 & 0.22 & 0.00154 & 113.19 & 120.02 & 0.35 & 0.24 & 0.06 & 2.4 $\times 10^{-6}$ & 113.08 & 123.33 & 0.23 & 0.15 & 0.07 & 1.0  $\times 10^{-5}$ & 115.08 & 128.74 & 0.23 & 0.13 & 2134 & (34)12 & 4312 & 3421 \\ 
F 571-V1 & 45.7 & 0.09 & 0.006 & 44.73 & 48.52 & 0.23 & 0.24 & 0.32 & 0.098 & 45.90 & 49.68 & 0.26 & 0.27 & 0.01 & 0.0002 & 46.35 & 52.02 & 0.25 & 0.31 & 0.008 & 0.0009 & 48.32 & 55.89 & 0.25 & 0.36 & 1234 & 1(34)2 & 1234 & 4312 \\ 
F 574-1 & 42.7 & 0.78 & 0.330 & 81.70 & 88.26 & 0.31 & 0.24 & 1.64 & 0.928 & 92.04 & 98.60 & 0.43 & 0.26 & 0.03 & 2. 0  $\times 10^{-7}$ & 74.65 & 84.49 & 0.23 & 0.21 & 0.02 & 7.0  $\times 10^{-8}$ & 76.51 & 89.63 & 0.23 & 0.24 & 3142 & (34)12 & 3(14)2 & 4312 \\ 
F 583-1 & 40.1 & 0.28 & 0.00030 & 151.42 & 160.29 & 0.27 & 0.18 & 1.60 & 0.9661 & 181.73 & 190.61 & 0.40 & 0.29 & 0.11 & 6.0   $\times 10^{-8}$ & 149.29 & 162.61 & 0.20 & 0.08 & 0.11 & 1.7  $\times 10^{-7}$ & 151.28 & 169.03 & 0.20 & 0.11 & 1342 & (34)12 & 3412 & (34)12 \\ 
F 583-4 & 49.6 & 1.04 & 0.591 & 74.39 & 80.33 & 0.30 & 0.29 & 0.15 & 0.00112 & 65.54 & 71.48 & 0.24 & 0.33 & 0.14 & 0.0014 & 67.28 & 76.19 & 0.24 & 0.34 & 0.16 & 0.0039 & 69.28 & 81.15 & 0.25 & 0.41 & 2314 & (23)41 & 1234 & 3241 \\ 
IC 2574 & 49.8 & 3.49 & 1.000 & 228.11 & 238.22 & 0.17 & 0.1 & 33.54 & 1.0000 & 1189.77 & 1199.88 & 0.49 & 0.36 & 1.31 & 0.8845 & 158.98 & 174.14 & 0.24 & 0.14 & 3.40 & 1.000 & 222.26 & 242.47 & 0.14 & 0.14 & 3142 & 4132 & 1(34)2 & 3412 \\ 
KK 98-251 & 61.4 & 0.23 & 0.0022 & 55.39 & 62.23 & 0.25 & 0.09 & 1.82 & 0.9658 & 76.05 & 82.88 & 0.56 & 0.34 & 0.25 & 0.0047 & 57.39 & 67.64 & 0.25 & 0.09 & 0.28 & 0.0097 & 59.39 & 73.06 & 0.25 & 0.09 & 1342 & (134)2 & (134)2 & 1342 \\ 
NGC 0024 & 28.3 & 3.34 & 1.000 & 262.10 & 271.57 & 1.12 & 0.92 & 0.93 & 0.43 & 196.93 & 206.40 & 0.31 & 0.24 & 0.44 & 0.0057 & 185.14 & 199.34 & 0.36 & 0.24 & 0.21 & 8.8  $\times 10^{-6}$  & 181.02 & 199.96 & 0.11 & 0.20 & 3421 & 4231 & 4(23)1 & 4321 \\ 
NGC 0055 & 61.2 & 0.13 & 2.1 $\times 10^{-6}$ & 96.56 & 104.73 & 0.22 & 0.12 & 2.97 & 1.000 & 150.50 & 158.68 & 0.45 & 0.35 & 0.13 & 4.5  $\times 10^{-6}$  & 98.48 & 110.74 & 0.21 & 0.09 & 0.12 & 3.3  $\times 10^{-6}$ & 100.09 & 116.44 & 0.20 & 0.07 & 1342 & 4312 & 4312 & 4(13)2 \\ 
NGC 0100 & 52.4 & 0.18 & 0.000035 & 114.92 & 123.10 & 0.22 & 0.17 & 0.77 & 0.250 & 126.00 & 134.18 & 0.36 & 0.28 & 0.06 & 6.5  $\times 10^{-9}$ & 114.52 & 126.79 & 0.22 & 0.13 & 0.06 & 2.1 $\times 10^{-8}$  & 116.49 & 132.85 & 0.21 & 0.11 & 1342 & 4(13)2 & 4312 & (34)12 \\ 
NGC 0247 & 49.3 & 14.10 & 1.0000 & 458.10 & 467.13 & 0.96 & 0.78 & 1.85 & 0.9931 & 164.15 & 173.18 & 0.22 & 0.14 & 1.27 & 0.83 & 150.95 & 164.50 & 0.20 & 0.16 & 1.32 & 0.856 & 152.78 & 170.84 & 0.20 & 0.16 & 3421 & (34)21 & 2(34)1 & 3421 \\ 
NGC 0300 & 44.7 & 1.80 & 0.9896 & 175.20 & 184.07 & 0.42 & 0.35 & 0.65 & 0.101 & 148.58 & 157.46 & 0.30 & 0.23 & 0.41 & 0.0068 & 144.72 & 158.03 & 0.28 & 0.19 & 0.41 & 0.0075 & 146.22 & 163.97 & 0.22 & 0.12 & 2341 & 4321 & 4321 & (34)21 \\ 
NGC 3109 & 36.2 & 0.37 & 0.0027 & 104.33 & 113.20 & 0.22 & 0.16 & 8.55 & 1.0000 & 292.53 & 301.40 & 0.69 & 0.50 & 0.22 & 0.000047 & 102.62 & 115.93 & 0.17 & 0.1 & 0.23 & 0.000101 & 104.62 & 122.37 & 0.17 & 0.13 & 1342 & (34)12 & 3412 & 3412 \\ 
NGC 3741 & 32.2 & 2.29 & 0.9989 & 120.87 & 129.05 & 0.51 & 0.30 & 0.34 & 0.0037 & 83.98 & 92.16 & 0.20 & 0.20 & 0.36 & 0.0057 & 85.84 & 98.11 & 0.19 & 0.21 & 0.37 & 0.0083 & 87.65 & 104.01 & 0.20 & 0.23 & 2341 & 3(24)1 & 2341 & 2341 \\ 
NGC 6789* & 40.6 & 0.30 & 0.26 & 24.68 & 26.23 & 0.23 & 0.16 & 2.04 & 0.870 & 28.16 & 29.71 & 0.45 & 0.21 & 0.18 & 0.33 & 26.26 & 28.58 & 0.24 & 0.36 & -- & --& 28.68 & 31.77 & 0.23 & 0.34 & 1324 & (14)32 & 1243 & 312 \\ 
UGC 00191 & 49.2 & 27.75 & 1.000 & 225.28 & 230.06 & 1.03 & 0.54 & 3.58 & 0.999 & 56.09 & 60.87 & 0.93 & 0.78 & 1.89 & 0.922 & 44.35 & 51.53 & 0.74 & 0.56 & 1.16 & 0.67 & 40.78 & 50.36 & 0.36 & 0.41 & 4321 & 4321 & 4132 & 4321 \\ 
UGC 00634* & 43.7 & 1.34 & 0.737 & 19.51 & 21.06 & 1.01 & 1.07 & 3.20 & 0.959 & 23.24 & 24.79 & 1.20 & 0.88 & 0.68 & 0.59 & 19.52 & 21.84 & 0.99 & 1.50 & -- & -- & 21.52 & 24.61 & 0.99 & 1.46 & 1342 & (34)12 & 2143 & 312 \\ 
UGC 00891 & 47.0 & 0.61 & 0.392 & 15.47 & 17.90 & 0.21 & 0.14 & 3.10 & 0.974 & 22.94 & 25.38 & 0.71 & 0.55 & 0.27 & 0.24 & 16.17 & 19.83 & 0.18 & 0.31 & 0.36 & 0.45 & 17.99 & 22.87 & 0.15 & 0.73 & 1342 & 4312 & 1324 & 3412 \\ 
UGC 02259 & 40.7 & 12.50 & 1.00 & 105.48 & 109.80 & 1.20 & 0.74 & 0.71 & 0.355 & 34.69 & 39.01 & 0.37 & 0.31 & 0.85 & 0.484 & 36.69 & 43.17 & 0.37 & 0.34 & 0.91 & 0.54 & 38.08 & 46.72 & 0.44 & 0.40 & 2341 & (23)41 & 2341 & 2341 \\ 
UGC 04325 & 43.7 & 1.94 & 0.929 & 48.19 & 52.51 & 0.59 & 0.33 & 2.87 & 0.991 & 53.77 & 58.08 & 0.42 & 0.39 & 0.13 & 0.013 & 39.18 & 45.66 & 0.36 & 0.45 & 0.04 & 0.003 & 40.71 & 49.34 & 0.35 & 0.53 & 3412 & 4321 & 1234 & 4312 \\ 
UGC 04499 & 54.7 & 0.73 & 0.355 & 44.97 & 49.76 & 0.29 & 0.30 & 1.09 & 0.64 & 47.49 & 52.28 & 0.28 & 0.43 & 0.24 & 0.038 & 43.31 & 50.49 & 0.28 & 0.34 & 0.21 & 0.041 & 44.89 & 54.47 & 0.27 & 0.38 & 1324 & 4(23)1 & 1342 & 4312 \\ 
UGC 05716 & 41.1 & 18.53 & 1.000 & 211.76 & 217.70 & 1.57 & 1.18 & 1.81 & 0.946 & 44.55 & 50.49 & 0.42 & 0.19 & 2.01 & 0.966 & 46.54 & 55.45 & 0.42 & 0.25 & 2.26 & 0.979 & 48.54 & 60.42 & 0.42 & 0.25 & 2341 & (234)1 & 2(34)1 & 2341 \\ 
UGC 05764 & 36.5 & 4.30 & 1.00 & 57.45 & 62.66 & 0.40 & 0.53 & 6.73 & 1.00 & 76.91 & 82.12 & 1.28 & 0.96 & 2.86 & 0.995 & 45.08 & 52.90 & 0.69 & 0.49 & 3.34 & 0.997 & 47.08 & 57.50 & 0.69 & 0.52 & 3412 & 1(34)2 & 3412 & 3412 \\ 
 UGC 05918 & 39.1 & 0.37 & 0.104 & 42.84 & 47.16 & 0.26 & 0.21 & 0.15 & 0.0102 & 41.48 & 45.80 & 0.19 & 0.21 & 0.02 & 9.3 $\times 10^{-5}$& 42.68 & 49.16 & 0.22 & 0.23 & 8.9 $\times 10^{-3}$   & 0.0002 & 44.64 & 53.27 & 0.22 & 0.30 & 2134 & 2(34)1 & (12)34 & 4321 \\ 
UGC 05986 & 47.5 & 0.1 & 0.000014 & 66.55 & 73.38 & 0.19 & 0.17 & 5.98 & 1.000 & 142.99 & 149.82 & 0.96 & 0.67 & 0.09 & 0.000018 & 68.36 & 78.61 & 0.23 & 0.17 & 0.11 & 0.00015 & 70.55 & 84.21 & 0.19 & 0.17 & 1342 & (14)32 & (134)2 & 3142 \\ 
UGC 06399 & 46.6 & 0.14 & 0.0048 & 50.11 & 54.89 & 0.21 & 0.21 & 0.66 & 0.292 & 53.73 & 58.52 & 0.32 & 0.27 & 0.04 & 0.0003 & 51.39 & 58.57 & 0.24 & 0.24 & 0.05 & 0.0017 & 53.39 & 62.97 & 0.24 & 0.30 & 1234 & 1(34)2 & 1324 & 3412 \\ 
UGC 06446 & 41.4 & 2.01 & 0.9889 & 118.60 & 125.93 & 0.48 & 0.24 & 0.21 & 0.00062 & 91.59 & 98.93 & 0.23 & 0.21 & 0.16 & 0.000190 & 92.66 & 103.66 & 0.21 & 0.23 & 0.13 & 0.000095 & 94.09 & 108.76 & 0.24 & 0.24 & 2341 & 3241 & 23(14) & 4321 \\ 
UGC 06667 & 29.8 & 0.66 & 0.298 & 47.07 & 51.86 & 0.40 & 0.24 & 1.31 & 0.76 & 51.57 & 56.36 & 0.41 & 0.31 & 0.21 & 0.028 & 45.71 & 52.89 & 0.30 & 0.34 & 0.26 & 0.064 & 47.71 & 57.29 & 0.30 & 0.40 & 1324 & (34)12 & 1234 & 3412 \\ 
UGC 07399 & 35.0 & 6.03 & 1.00 & 94.33 & 99.54 & 0.85 & 0.64 & 1.01 & 0.577 & 54.19 & 59.40 & 0.33 & 0.15 & 0.30 & 0.048 & 50.21 & 58.02 & 0.23 & 0.15 & 0.35 & 0.089 & 52.17 & 62.59 & 0.23 & 0.15 & 3241 & (34)21 & (234)1 & 3421 \\ 
UGC 07524 & 47.3 & 0.60 & 0.042 & 154.63 & 164.36 & 0.34 & 0.17 & 0.81 & 0.250 & 160.93 & 170.67 & 0.34 & 0.15 & 0.34 & 0.00048 & 148.96 & 163.57 & 0.18 & 0.15 & 0.36 & 0.00083 & 150.96 & 170.44 & 0.18 & 0.15 & 3142 & (34)(12) & (234)1 & 3412 \\ 
UGC 07603 & 44.3 & 0.42 & 0.064 & 55.76 & 61.70 & 0.26 & 0.21 & 1.51 & 0.873 & 66.67 & 72.61 & 0.43 & 0.21 & 0.09 & 0.00029 & 54.38 & 63.29 & 0.24 & 0.20 & 0.09 & 0.00047 & 56.22 & 68.10 & 0.23 & 0.25 & 1342 & 4312 & 3(12)4 & (34)12 \\ 
UGC 08286 & 40.2 & 7.68 & 1.000 & 176.95 & 184.28 & 1.08 & 0.71 & 2.55 & 0.9992 & 100.05 & 107.39 & 0.45 & 0.34 & 0.66 & 0.180 & 72.96 & 83.96 & 0.30 & 0.25 & 0.18 & 0.00055 & 68.13 & 82.80 & 0.26 & 0.30 & 4321 & 4321 & 3421 & 4321 \\ 
UGC 08490 & 41.4 & 4.49 & 1.000 & 263.07 & 272.68 & 0.80 & 0.57 & 0.11 & 1.5  $\times 10^{-9}$  & 140.62 & 150.23 & 0.21 & 0.15 & 0.11 & 1.4  $\times 10^{-9}$ & 142.33 & 156.74 & 0.21 & 0.15 & 0.11 & 4.5  $\times 10^{-9}$  & 144.33 & 163.54 & 0.21 & 0.15 & 2341 & (234)1 & (234)1 & (234)1 \\ 
UGC 08550 & 41.8 & 4.49 & 1.00 & 81.11 & 86.70 & 0.71 & 0.38 & 0.76 & 0.350 & 47.58 & 53.17 & 0.19 & 0.51 & 0.40 & 0.080 & 45.92 & 54.31 & 0.24 & 0.49 & 0.23 & 0.021 & 46.29 & 57.47 & 0.27 & 0.52 & 2341 & 2341 & 1324 & 4321 \\ 
UGC 12632 & 43.7 & 0.94 & 0.50 & 80.72 & 87.55 & 0.40 & 0.21 & 0.34 & 0.0143 & 72.87 & 79.70 & 0.28 & 0.24 & 0.07 & 3.8  $\times 10^{-6}$ & 71.23 & 81.48 & 0.19 & 0.17 & 0.07 & 0.000014 & 73.23 & 86.89 & 0.19 & 0.21 & 2341 & (34)21 & 3(14)2 & (34)21 \\ 
UGC 12732 & 42.1 & 4.38 & 1.000 & 135.67 & 142.76 & 0.75 & 0.51 & 0.21 & 0.00075 & 77.30 & 84.39 & 0.22 & 0.15 & 0.20 & 0.00104 & 79.04 & 89.68 & 0.23 & 0.15 & 0.21 & 0.00189 & 80.92 & 95.10 & 0.24 & 0.16 & 2341 & 2341 & (23)41 & 3(24)1 \\ 
UGC A442 & 43.9 & 1.15 & 0.672 & 35.45 & 39.77 & 0.44 & 0.15 & 2.05 & 0.944 & 40.82 & 45.14 & 0.45 & 0.22 & 1.09 & 0.64 & 35.96 & 42.44 & 0.28 & 0.32 & 1.36 & 0.755 & 37.96 & 46.60 & 0.28 & 0.38 & 1324 & (34)12 & 1234 & 3142 \\  \\ 
\bottomrule
\hline \hline
\end{tabular}
}
\end{turn}
\end{center}
\caption{In this table we report the different statistical values for Schive (1), NFW (2), SNFWDC (3) and SNFWC (4) models. The galaxies with * have four data points. In the Columns 27--30 we associated the number code, mentioned in the text for each model, to select in ascendent order  the best BIC, best distance between model and LOESS+SIMEX, the best 1$\sigma$ bands distance from the model and LOESS+SIMEX and the best $\chi^2_{red}$ value. The numbers inside of parenthesis mean the same value (or BIC).}\label{tab:SNSS}
\end{table*}
\end{center}

\setcounter{table}{1}
%

\begin{center}
\begin{table*}
\ra{1.24}
\begin{center}
\begin{turn}{90}
\resizebox{!}{0.2\paperheight}{%
\begin{tabular}{@{}l | r | r | r | r | r | r | r | r | r | r | r | r | r | r | r | r | r | r | r | r | r | r | r | r | r | r | r | r | r  @ {}}\toprule
    \hline \hline
    \multicolumn{30}{c}{SPARC Galaxies (Continue)} \\
    \multicolumn{30}{c}{Statistics}\\
    \hline 
\multicolumn{1}{c|}{Galaxy} &
\multicolumn{1}{c|}{$\delta A_V$} &
\multicolumn{6}{c|}{SCHIVE (1) } &
\multicolumn{6}{c|}{NFW (2)} &
\multicolumn{6}{c|}{SNFWDC (3)}&
\multicolumn{6}{c|}{SNFWC (4)} &
\multicolumn{4}{c}{BEST MODEL}\\

\hline
   & & $\chi^2_{red}$& $P$-Value & AIC & BIC & DIST & D$1\sigma$ & $\chi^2_{red}$ & $P$-Value & AIC & BIC & DIST & D$1\sigma$ & $\chi^2_{red}$ & $P$-Value & AIC & BIC & DIST & D$1\sigma$ & $\chi^2_{red}$ & $P$-Value & AIC & BIC &  DIST & D$1\sigma$ & $B_{\rm BIC}$ &  $B_{\rm DIST}$ &  $B_{\rm D1\sigma}$ & $B_{\chi^2_{red}}$  \\ \hline

\cmidrule[0.4pt](r{0.125em}){1-1}
\cmidrule[0.4pt](lr{0.125em}){2-2}
\cmidrule[0.4pt](lr{0.125em}){3-3}
\cmidrule[0.4pt](lr{0.125em}){4-4}
\cmidrule[0.4pt](lr{0.125em}){5-5}
\cmidrule[0.4pt](lr{0.125em}){6-6}
\cmidrule[0.4pt](lr{0.125em}){7-7}
\cmidrule[0.4pt](lr{0.125em}){8-8}
\cmidrule[0.4pt](lr{0.125em}){9-9}
\cmidrule[0.4pt](lr{0.125em}){10-10}
\cmidrule[0.4pt](lr{0.125em}){11-11}
\cmidrule[0.4pt](lr{0.125em}){12-12}
\cmidrule[0.4pt](lr{0.125em}){13-13}
\cmidrule[0.4pt](lr{0.125em}){14-14}
\cmidrule[0.4pt](lr{0.125em}){15-15}
\cmidrule[0.4pt](lr{0.125em}){16-16}
\cmidrule[0.4pt](lr{0.125em}){17-17}
\cmidrule[0.4pt](lr{0.125em}){18-18}
\cmidrule[0.4pt](lr{0.125em}){19-19}
\cmidrule[0.4pt](lr{0.125em}){20-20}
\cmidrule[0.4pt](lr{0.125em}){21-21}
\cmidrule[0.4pt](lr{0.125em}){22-22}
\cmidrule[0.4pt](lr{0.125em}){23-23}
\cmidrule[0.4pt](lr{0.125em}){24-24}
\cmidrule[0.4pt](lr{0.125em}){25-25}
\cmidrule[0.4pt](lr{0.125em}){26-26}
\cmidrule[0.4pt](lr{0.125em}){27-27}
\cmidrule[0.4pt](lr{0.125em}){28-28}
\cmidrule[0.4pt](lr{0.125em}){29-29}
\cmidrule[0.4pt](lr{0.125em}){30-30}

\multicolumn{1}{c|}{1} & 
\multicolumn{1}{c|}{2} & 
\multicolumn{1}{c|}{3} & 
\multicolumn{1}{c|}{4} & 
\multicolumn{1}{c|}{5} & 
\multicolumn{1}{c|}{6} & 
\multicolumn{1}{c|}{7} & 
\multicolumn{1}{c|}{8} & 
\multicolumn{1}{c|}{9} & 
\multicolumn{1}{c|}{10} & 
\multicolumn{1}{c|}{11} & 
\multicolumn{1}{c|}{12} & 
\multicolumn{1}{c|}{13} & 
\multicolumn{1}{c|}{14} & 
\multicolumn{1}{c|}{15} & 
\multicolumn{1}{c|}{16} & 
\multicolumn{1}{c|}{17} & 
\multicolumn{1}{c|}{18} & 
\multicolumn{1}{c|}{19} & 
\multicolumn{1}{c|}{20} & 
\multicolumn{1}{c|}{21} & 
\multicolumn{1}{c|}{22} & 
\multicolumn{1}{c|}{23} & 
\multicolumn{1}{c|}{24} & 
\multicolumn{1}{c|}{25} & 
\multicolumn{1}{c|}{26} & 
\multicolumn{1}{c|}{27} & 
\multicolumn{1}{c|}{28} & 
\multicolumn{1}{c|}{29} & 
\multicolumn{1}{c}{30} \\
\hline
 
DDO 161 		&58.4 &0.31 &0.000118 &112.99 &122.73 &0.23 &0.18 &1.15 		&0.73 &137.44 &147.18 &0.56 &0.50 &0.26 		&0.000027 &113.36 &127.97 &0.23 &0.14 &0.27 &0.000051 &115.33 &134.80 &0.23 &0.12 	&1342 &(134)2 &4312 &3412 \\
F 563-1 		&32.8 &1.33 &0.83 &30413.27 &30420.61 &0.40 &0.32 &1.07 		&0.62 &16230.84 &16238.17 &0.30 &0.23 &0.47 	&0.051 &21101.19 &21112.19 &0.27 &0.16 &0.40 &0.028 &21933.37 &21948.04 &0.26 &0.16 	&2341 &4321 &(34)21 &4321 \\
F 563v2 		&34.5 &0.08 &0.00036 &33188.33 &33193.54 &0.11 &0.06 &1.23 	&0.725 &59093.79 &59099.00 &0.44 &0.33 &0.04 	&0.00009 &38828.65 &38836.46 &0.13 &0.09 &0.0034 &0.0 &39623.65 &39634.07 &0.12 &0.12 &1342 &1432 &1342 &4312 \\
F 568-1 		&36.1 &0.16 &0.00143 &49615.22 &49621.16 &0.24 &0.12 &0.86 &0.429 &63319.03 &63324.97 &0.27 &0.13 &0.04 
& $5.0\times 10^{-6}$
&55300.01 &55308.92 &0.15 &0.14 &0.01 
& $2.0\times 10^{-7}$
&59085.67 &59097.55 &0.18 &0.18 	&1342 &3412 &1234 &4312 \\
F 571-8 		&49.2 &9.63 &1.000 &63191.52 &63197.78 &0.74 &0.61 &19.17 &1.0000 &54951.23 &54957.49 &1.20 &0.93 &10.59 &1.00 &63193.52 &63202.91 &0.74 &0.61 &10.80 &1.00 &62474.05 &62486.57 &0.62 &0.67 	&2413 &4(13)2 &(13)42 &1342 \\
F 579v1 		&46.5 &1.78 &0.955 &29628.76 &29635.32 &0.52 &0.37 &0.19 &0.00108 &33342.86 &33349.41 &0.18 &0.12 &0.07 &0.000013 &32730.12 &32739.95 &0.17 &0.15 &0.02 
& $9.0\times 10^{-8}$
&32863.31 &32876.42 &0.17 &0.12  &1342 &(34)21 &(24)31 &4321 \\
NGC 1003 	&45.6 &12.02 &1.0000 &550.57 &560.90 &1.20 &0.93 &2.57 		&1.000 &229.02 &239.36 &0.45 &0.31 &2.65 		&1.000 &231.03 &246.53 &0.45 &0.28 &1.17 &0.767 &183.19 &203.86 &0.33 &0.18 		&4231 &4(23)1 &4321 &4231 \\
NGC 2915 	&33.5 &0.95 &0.46 &209.11 &218.71 &0.39 &0.20 &0.96 			&0.47 &209.31 &218.92 &0.39 &0.27 &0.49 		&0.0119 &197.70 &212.11 &0.30 &0.21 &0.35 &0.00090 &195.63 &214.84 &0.26 &0.26 	&3412 &43(12) &1342 &4312 \\
NGC 2976 	&75.3 &0.29 &0.00021 &126.89 &136.08 &0.23 &0.22 &1.90 		&0.9957 &167.08 &176.27 &0.54 &0.24 &0.30 		&0.00041 &128.89 &142.67 &0.23 &0.22 &0.31 &0.00063 &130.73 &149.09 &0.22 &0.28 	&1342 &4(13)2 &(13)24 &1342 \\
NGC 4068 	&72.3 &60.21 &0.069 &28.08 &31.24 &0.34 &0.21 &1.86 			&0.886 &34.68 &37.85 &0.64 &0.42 &0.28 		&0.164 &30.08 &34.83 &0.34 &0.23 &0.43 &0.35 &32.08 &38.41 &0.34 &0.23 			&1324 &(134)2 &1(34)2 &1342 \\
NGC 5585 	&54.1 &5.11 &1.000 &193.74 &202.45 &0.44 &0.32 &7.17 		&1.0000 &239.11 &247.83 &0.83 &0.50 &5.27 		&1.000 &193.98 &207.04 &0.38 &0.31 &5.53 &1.000 &195.98 &213.40 &0.38 &0.30 		&1342 &(34)12 &4312 &1342 \\
NGC 7793 	&66.2 &1.05 &0.62 &309.46 &320.78 &0.21 &0.14 &0.89 			&0.32 &302.45 &313.76 &0.31 &0.21 &0.81 		&0.195 &300.12 &317.10 &0.30 &0.20 &0.81 &0.199 &301.35 &323.98 &0.30 &0.20 		&2314 &1(34)2 &1(34)2 &(34)21 \\
UGC 00731 	&38.5 &4.36 &1.00 &94.11 &100.05 &0.75 &0.56 &0.34 			&0.0295 &53.92 &59.86 &0.20 &0.12 &0.11 		&0.00053 &53.51 &62.42 &0.20 &0.12 &0.12 &0.0017 &55.51 &67.39 &0.20 &0.16 		&2341 &(234)1 &(23)41 &3421 \\
UGC 01281 	&44.0 &0.14 
& $4.6\times 10^{-7}$
&130.83 &139.71 &0.20 &0.13 &1.30 &0.8478 &157.46 &166.34 &0.37 &0.32 &0.15 
& $1.25\times 10^{-6}$
&132.83 &146.14 &0.20 &0.13 &0.10 
& $7.2\times 10^{-8}$
&133.73 &151.48 &0.20 &0.08 	&1342 &(134)2 &4(13)2 &4132 \\
UGC 04278 	&44.8 &0.75 &0.207 &150.77 &159.64 &0.27 &0.22 &1.47 &0.9324 &167.28 &176.16 &0.43 &0.37 &0.53 &0.038 &147.19 &160.51 &0.28 &0.21 &0.80 &0.279 &154.29 &172.04 &0.27 &0.21 			&1342 &(14)32 &(34)12 &3142 \\
UGC 04483 	&58.1 &0.32 &0.071 &29.42 &33.74 &0.28 &0.22 &0.59 &0.264 &31.09 &35.41 &0.32 &0.32 &0.25 &0.059 &30.77 &37.25 &0.28 &0.29 &0.26 &0.094 &32.55 &41.19 &0.26 &0.39 					&1234 &4(13)2 &1324 &3412 \\
UGC 05005 	&47.1 &0.05 &0.000017 &78.33 &83.92 &0.24 &0.18 &0.22 &0.0079 &79.85 &85.44 &0.18 &0.11 &0.02 
&   $2.0\times 10^{-6}$
&80.06 &88.45 &0.22 &0.13 &0.03 &0.00002 &82.06 &93.25 &0.22 &0.13 	&1234 &2(34)1 &2(34)1 &3412 \\
UGC 05414 	&60.5 &0.23 &0.079 &26.71 &29.88 &0.31 &0.27 &1.05 			&0.619 &29.98 &33.14 &0.37 &0.36 &0.11 	&0.044 &28.11 &32.86 &0.30 &0.37 &0.15 &0.14 &30.10 &36.43 &0.30 &0.49 				&1324 &(34)12 &1234 &3412 \\
UGC 05721 	&39.0 &1.63 &0.9655 &154.41 &162.95 &0.48 &0.35 &0.96 		&0.49 &140.45 &148.99 &0.37 &0.20 &0.46 	&0.0205 &131.48 &144.29 &0.22 &0.13 &0.41 &0.0119 &132.06 &149.14 &0.19 &0.13 		&3241 &4321 &(34)21 &4321 \\
UGC 05750 	&45.1 &0.05 &0.000021 &69.98 &75.58 &0.18 &0.22 &1.00 		&0.57 &78.57 &84.16 &0.25 &0.16 &0.04 		&0.00004 &71.89 &80.27 &0.19 &0.21 &0.05 &0.00013 &73.85 &85.04 &0.18 &0.15 			&1324 &(14)32 &4231 &3(14)2 \\
UGC 05829 	&48.1 &0.47 &0.105 &64.12 &69.71 &0.27 &0.16 &0.09 			&0.00021 &60.67 &66.27 &0.19 &0.16 &0.04 	&0.000024 &62.21 &70.60 &0.20 &0.17 &0.02 &0.00001 &64.05 &75.23 &0.20 &0.22 		&2134 &2(34)1 &(12)34 &4321 \\
UGC 06818 	&58.7 &1.15 &0.672 &51.74 &56.06 &0.33 &0.29 &3.55 			&0.998 &66.12 &70.44 &0.65 &0.36 &1.39 	&0.77 &53.74 &60.22 &0.33 &0.29 &1.32 &0.739 &54.08 &62.72 &0.29 &0.45 				&1342 &4(13)2 &(13)24 &1432 \\
UGC 06917 	&51.6 &0.78 &0.364 &62.30 &67.89 &0.32 &0.21 &0.80 			&0.387 &62.52 &68.11 &0.30 &0.14 &0.16 	&0.0046 &58.60 &66.99 &0.20 &0.23 &0.07 &0.0005 &59.78 &70.97 &0.21 &0.27 			&3124 &3421 &2134 &4312 \\
UGC 06923 	&66.2 &0.31 &0.126 &36.20 &39.37 &0.33 &0.31 &0.95 			&0.57 &38.80 &41.96 &0.36 &0.35 &0.41 		&0.253 &38.20 &42.95 &0.33 &0.33 &0.18 &0.17 &39.34 &45.67 &0.28 &0.59 				&1234 &4(13)2 &1324 &4132 \\
UGC 06983 	&46.6 &1.59 &0.9319 &118.28 &125.61 &0.34 &0.21 &0.60 		&0.120 &103.40 &110.73 &0.27 &0.27 &0.36 	&0.0156 &101.55 &112.55 &0.23 &0.28 &0.34 &0.0133 &102.81 &117.47 &0.23 &0.30 		&2341 &(34)21 &1234 &4321 \\
UGC 07089 	&62.0 &0.24 &0.0074 &67.27 &73.21 &0.23 &0.13 &0.27 			&0.0119 &67.57 &73.51 &0.22 &0.16 &0.12 	&0.00090 &68.02 &76.93 &0.20 &0.15 &0.19 &0.0070 &70.38 &82.26 &0.21 &0.14 			&1234 &3421 &1432 &3412 \\
UGC 07125 	&64.1 &0.55 &0.129 &58.32 &64.58 &0.25 &0.15 &0.85 			&0.41 &61.67 &67.93 &0.42 &0.24 &0.28 		&0.0146 &57.10 &66.49 &0.23 &0.13 &0.31 &0.029 &59.10 &71.62 &0.23 &0.14 			&1324 &(34)12 &3412 &3412 \\
UGC 07151 	&58.2 &1.83 &0.942 &61.32 &66.91 &0.39 &0.31 &2.86 			&0.998 &70.63 &76.23 &0.47 &0.29 &0.64 	&0.255 &51.98 &60.37 &0.28 &0.36 &0.44 &0.125 &51.97 &63.15 &0.27 &0.41 			&3412 &4312 &2134 &4312 \\
UGC 07232 	&62.4 &0.11 &0.11 &21.13 &22.67 &0.22 &0.26 &2.41 			&0.911 &25.73 &27.28 &0.54 &0.27 &0.09 	&0.23 &22.99 &25.31 &0.19 &0.52 &-- &-- &25.13 &28.22 &0.22 &0.43 					&1324 &3(14)2 &1243 &312 \\ 
UGC 07261 	&52.2 &1.57 &0.84 &48.98 &52.77 &0.40 &0.24 &0.05			&0.0014 &41.38 &45.16 &0.20 &0.13 &0.06 	&0.007 &43.38 &49.05 &0.20 &0.14 &0.03 &0.006 &45.21 &52.78 &0.19 &0.27 			&2314 &4(23)1 &2314 &4231 \\
UGC 07323 	&64.2 &0.35 &0.053 &52.22 &57.43 &0.20 &0.11 &0.71 			&0.316 &55.11 &60.32 &0.37 &0.26 &0.33 	&0.061 &53.78 &61.59 &0.27 &0.11 &0.46 &0.161 &56.20 &66.62 &0.20 &0.18 			&1234 &(14)32 &(13)42 &3142 \\
UGC 07559 	&62.2 &0.05 &0.0018 &33.55 &37.33 &0.22 &0.16 &0.43 			&0.171 &35.42 &39.20 &0.25 &0.24 &0.07 	&0.008 &35.55 &41.22 &0.22 &0.17 &0.01 &0.002 &37.32 &44.89 &0.21 &0.31 			&1234 &4(13)2 &1324 &4132 \\
UGC 07577 	&86.4 &0.08 &0.0008 &41.86 &46.65 &0.22 &0.08 &0.20 			&0.0153 &42.74 &47.53 &0.26 &0.10 &0.09 	&0.0029 &43.86 &51.05 &0.22 &0.08 &0.11 &0.010 &45.86 &55.44 &0.22 &0.08 			&1234 &(134)2 &(134)2 &1342 \\
UGC 07690 	&59.8 &0.45 &0.188 &39.58 &43.36 &0.26 &0.19 &0.34 			&0.108 &39.00 &42.78 &0.20 &0.08 &0.05 	&0.004 &39.51 &45.19 &0.19 &0.17 &0.0022 &0.0 &41.32 &48.89 &0.18 &0.25 			&2134 &4321 &2314 &4321 \\
UGC 07866 	&56.7 &0.18 &0.031 &37.43 &41.21 &0.21 &0.18 &0.04 			&0.0011 &36.73 &40.52 &0.19 &0.18 &0.03 	&0.002 &38.65 &44.33 &0.20 &0.21 &0.01 &0.002 &40.56 &48.12 &0.21 &0.34 			&2134 &23(14) &(12)34 &4321 \\
UGC 08837 	&67.5 &0.66 &0.322 &37.27 &41.59 &0.26 &0.27 &6.33 			&1.00 &71.27 &75.59 &0.88 &0.52 &0.79 		&0.446 &39.25 &45.73 &0.25 &0.27 &0.93 &0.56 &41.01 &49.64 &0.25 &0.46 				&1342 &(34)12 &(13)42 &1342 \\
UGC 09992 	&66.1 &0.20 &0.103 &31.12 &33.56 &0.24 &0.20 &0.01 			&0.002 &30.56 &33.00 &0.21 &0.19 &0.02 	&0.018 &32.56 &36.22 &0.21 &0.32 &0.02 &0.1 &34.54 &39.42 &0.21 &0.68 				&2134 &(234)1 &2134 &2(34)1 \\
UGC 10310 	&51.0 &0.25 &0.062 &40.83 &44.61 &0.22 &0.16 &0.57 			&0.276 &42.41 &46.19 &0.25 &0.09 &0.04 	&0.003 &41.74 &47.41 &0.19 &0.17 &0.02 &0.004 &43.63 &51.19 &0.18 &0.25 			&1234 &4312 &2134 &4312 \\
UGC 11820 	&53.0 &14.85 &1.000 &155.68 &160.89 &2.97 &2.65 &1.33 		&0.778 &47.54 &52.75 &1.98 &1.79 &1.55 	&0.85 &49.74 &57.56 &1.98 &1.82 &0.25 &0.041 &42.40 &52.82 &1.46 &1.26 				&2431 &4(23)1 &4231 &4231 \\
UGC A444 	&48.2 &0.37 &0.00033 &178.28 &188.61 &0.22 &0.12 &0.06 
&   $1.7\times 10^{-15}$
&167.64 &177.97 &0.16 &0.09 &0.05 
&  $5.0\times 10^{-16}$
&169.32 &184.82 &0.16 &1.10 &0.08 
&   $4.9\times 10^{-13}$
&172.07 &192.74 &0.15 &0.11 	&2314 &4(23)1 &2413 &3241 \\


\bottomrule
\hline \hline
\end{tabular}
}
\end{turn}
\end{center}
\caption{Continue.}\label{tab:SNSSc}
\end{table*}
\end{center}

%


\bsp	
\label{lastpage}
\end{document}